\documentclass{cambridge6A}
\usepackage{natbib}

\setcitestyle{square}

\usepackage{rotating}
\usepackage{floatpag}
\usepackage{epsfig}
\usepackage{array}

\rotfloatpagestyle{empty}

\usepackage{graphicx}
\usepackage{amssymb}

\usepackage[usenames]{color}

\newcommand{\url}{}

\newcounter{mnotecount}[section]

\renewcommand{\theequation}{\thesection.\arabic{equation}}
\begin{document}

\title[Chapter 7: Probing Strong Field Gravity Through Numerical Simulations]
{General Relativity and Gravitation: A Centennial Perspective}

\author{Matthew W. Choptuik\\
        Luis Lehner\\
        Frans Pretorius}

\frontmatter
\maketitle
\tableofcontents
\mainmatter

\addtocounter{chapter}{6}
\chapter{Probing Strong Field Gravity Through Numerical Simulations}
\label{intro}
This chapter  describes what has been learned about the dynamical, 
strong field regime of general relativity via numerical methods. 
There is no rigorous way to identify this regime, in particular
since notions of energies, velocities, length and timescales 
are observer dependent at best, and at worst are not well-defined
locally or even globally. Loosely
speaking, however, dynamical strong field phenomena exhibit the following
properties: there is at least one region of spacetime of characteristic
size $R$ containing energy $E$ where the compactness $2G E/c^4 R$ approaches unity,
local velocities approach the speed of light $c$, and luminosities (of
gravitational or matter fields) can approach the Planck luminosity $c^5/G$.
A less physical characterization, though one better suited to classifying
solutions, are spacetimes where even in ``well-adapted'' coordinates 
the non-linearities of the field equations are strongly manifest.
In many of the cases where these conditions are met, numerical methods are the 
only option available to solve the Einstein field equations, and such scenarios
are the subject of this chapter. 

Mirroring trends in the growth and efficacy of computation,
numerical solutions have had greatest impact on the field in the decades
following the 1987 volume~\citep{300yrs} celebrating the $300^{\tt th}$ anniversary of 
Newton's {\em Principia}. However, several pioneering studies laying the foundation
for subsequent advances were undertaken before this, and they are briefly
reviewed in section~\ref{sec_historical} below. 
Though this review focuses
on the physics that has been gleaned from computational solutions, there are some unique challenges
in numerical evolution of the Einstein equations; these as well as the basic 
computational strategies that are currently dominant in numerical relativity
are discussed in section~\ref{sec_NR}.
As important as computational
science has become in uncovering details of solutions too complex to model analytically,
it is a rare moment when qualitatively new physics is uncovered. The standout example
in general relativity is the discovery of critical phenomena in gravitation
collapse (section~\ref{sec_CP}); another noteworthy example is the formation
of so-called spikes in the approach to cosmological singularities (section~\ref{sec_CS}).
A significant motivation for obtaining solutions in the dynamical strong field
has been to support the upcoming field of gravitational wave astronomy, which
requires predictions of emitted waveforms for optimal detection and parameter
extraction. The expected primary sources are compact object mergers, where numerical
methods are crucial in the modeling of the final stages of the events. 
Binary black hole systems are discussed
in section~\ref{sec_BBH}, black hole-neutron star and binary neutron stars systems in 
section~\ref{sec_BHNS_BNS}. 
Though not of astrophysical or experimental relevance---barring the existence
of an unexpectedly small Planck energy scale---the ultra-relativistic limit of the
two body problem is of considerable theoretical interest, and this is discussed
in section~\ref{sec_URC}. Spurred by the gauge-gravity dualities of string theory,
the study of higher dimensional gravity has been very active in the past decade; 
related numerical discoveries are presented in section~\ref{sec_HDG}.
Some miscellaneous topics are mentioned in section~\ref{sec_misc}, and
we conclude the review in section~\ref{sec_unsolved} with a discussion
of open problems for the coming years.

Regarding notation, for the most part we report results in geometric units
where Newton's constant $G$ and the speed of light $c$ are set to unity,
though for clarity some expressions will explicitly include these constants.
In referring to the dimensionality of a manifold, metric or tensor field, we will use lower case
``d" for spacetime dimensions, and upper case ``D" for purely spacelike dimensions; e.g.,
``4d" refers to $3+1$ spacetime dimensions (this latter ``$n+1$'' form we will also use),
and ``3D" means three spatial dimensions.

\section{Historical Perspective}\label{sec_historical}

The similarly oriented book released in 1987~\citep{300yrs} 
gave a snapshot
of the various interesting subjects and problems in gravitational research.
However, there was no chapter on numerical solution
of the Einstein equations, even though the subfield of numerical relativity had
been in active development for over a decade by then. The discipline
was still coming into its own, and the breadth and scope of works within its purview was still
limited. Nevertheless, these incipient studies did provide a hint of developments
to come as the know-how, computational resources and experience improved.
It is thus important to set some perspective by describing a subset of works leading
to the current status of the field. 

The particular topics we review later in this chapter are weighted
towards developments that have occurred within the past decade or two. This is
natural as numerical relativity has been a rapidly growing field during this time.
However, as mentioned, the foundations for building a mature field were initiated before, and here we
briefly discuss, loosely organized by subject, some of these more important early results. 
Unfortunately due to space limitations we cannot mention all the relevant works,
nor discuss those we do mention in any detail. Also, we do not include results,
in particular the more recent ones, that are discussed elsewhere in this review.

{\bf Binary black hole mergers.}
The first attempt at a numerical solution of the binary black hole merger problem
was made by Hahn and Lindquist~\citep{Hahn64} (1964). At
that time the term ``black hole'' had not yet been coined, and the full significance
of the problem, in particular with regards to gravitational wave emission
and black hole mergers in the universe, was not recognized. 
Using Gaussian normal coordinates, \index{Misner wormhole initial data}
Misner's ``wormhole'' initial
data~\citep{1960PhRv..118.1110M}, representing two black holes initially at 
rest, was evolved until $t \approx m/2$ (with
$m=\sqrt{A/16\pi}$, $A$ being the area of each throat). At that point numerical
errors had grown too large to warrant further evolution but it was nonetheless possible
to measure the mutual attraction between the holes, and the fact that the throats were
beginning to pinch off. Smarr~\citep{Smarr75,Smarr79a} and Eppley~\citep{Eppley75}
independently revisited the head-on collision problem a decade later, now with
a profound new understanding of black holes gained in the preceding years, both from theory, 
and from observations
suggesting that they likely exist in the universe. These works used the same initial data
as Hahn and Lindquist, {\v C}ade{\v z} 
coordinates \index{{\v C}ade{\v z} coordinates} to simultaneously conform to the throats and approach the usual spherical polar coordinates
at large distances~\citep{1976PhRvD..14.2443S}, and maximal slicing \index{maximal slicing}. 
The culmination of these studies showed that the collision emitted radiation of order $0.1\%$ of the total
mass, and that the waveform was very similar to that computed from a perturbative calculation 
(the first indications of the ``relative simplicity'' of black hole merger waveforms discussed
in Sec.~\ref{sec_BBH}). 

In anticipation of construction of the LIGO gravitational wave
detectors, and the recognized need for waveform models to enable detection, the head-on
collision calculations were reinitiated by the NCSA group in the early '90s~\citep{Anninos:1993zj}.
The new simulations offered improved treatment of the {\v C}ade{\v z} coordinate singularity and radiation extraction,
but essentially confirmed previous results. With hindsight, it is amusing that in~\citep{Smarr79a}
the status of this field was summarized as ``The two black hole collision problem has been largely completed.''
This was the prevailing opinion through the mid-'90s, with the consensus being 
that the most significant impediment to solving the full 3D
merger problem was simply lack of available computational power.
This turned out not to be the case, and a tremendous effort by the community
was expended in going from the first short-lived  grazing collision simulations 
reported in 1997~\citep{Bruegmann:1997uc}
to the breakthroughs in 2005~\citep{Pretorius:2005gq,Campanelli:2005dd,Baker:2005vv} that
facilitated stable evolution of the full problem and the impressive results
that have followed
(see~\citep{Pretorius:2007nq} for more discussion of this development).
As briefly discussed in Sec.~\ref{sec_NR},
some of the key stumbling blocks were related to the underlying mathematical character of the 
Einstein field equations and the existence of geometric singularities inside black holes.
This is not to say that limited computational power was not an issue; in fact it did hamper
the effort to rapidly find solutions to the more fundamental problems, as numerous
attempts to isolate and solve issues in a symmetry-reduced (or similar) setting that
could be tackled more quickly with available computational resources, failed when
carried over to the full problem.

{\bf Gravitational collapse.}
Numerical studies of the gravitational collapse of stars
began with the work of May and White~\citep{1966PhRv..141.1232M}, who looked at
the collapse of ideal fluid spheres with a $\Gamma$-law
equation of state (specifically $\Gamma=5/3$). They found, depending on
the initial conditions, that collapse would continue to black hole formation,
or halt and then bounce (a necessary condition for an eventual supernova).
The ``second generation'' of codes were developed over the next
couple of decades, with the pioneering efforts of, among others, Wilson~\citep{1971ApJ...163..209W,Wilson79a},
Shapiro and Teukolsky~\citep{Shapiro80}, Stark and Piran~\citep{Stark:1985da},
Nakamura~\citep{Nakamura81, Nakamura83} and Evans~\citep{Evans86}. Advances
included evolution of axisymmetric models to study the effects of
rotation and asymmetries, solution of the
the hydrodynamic equations written in conservative form, improvement in the
handling of axis coordinate singularities, 
development of moving mesh methods, incorporation of 
effects of neutrino emission, and exploration of a variety
of slicing and spatial coordinate conditions.
The more recent studies of stellar collapse are reviewed in
Sec.~\ref{sec_collapse}.

Although many studies of gravitational collapse are motivated
by application to the wide variety of observed phenomena attributed
to stellar collapse, there has been considerable work on more
theoretical scenarios, in particular critical collapse, reviewed
in Sec.~\ref{sec_CP}. A notable work we mention here
is the evolution of collapsing, axisymmetric configurations of collisionless
matter
by Shapiro and Teukolsky~\citep{Shapiro:1991zza}. In all cases the
formation of a geometric singularity was observed, but, intriguingly,
for sufficiently prolate distributions no apparent horizon
was found when the time-slice ran into the singularity.
This could be a slicing issue in that a horizon could
still form at a later time; however the threshold of prolateness
above which no horizons were found is consistent with the hoop conjecture~\citep{Thorne72a},\index{hoop conjecture}
suggesting these cases are examples of cosmic censorship conjecture violation \index{cosmic censorship conjecture}
in asymptotically flat spacetimes. 
Another work of theoretical interest that had significant
impact on the foundations of the field was the numerical study of cylindrical
gravitational wave spacetimes \index{cylindrical
gravitational wave spacetimes} by Piran~\citep{Piran80} (here
``collapse'' of non-linear waves always leads to naked singularities, \index{naked singularities}
though the spacetime is not asymptotically flat). In particular,
the modern notions of free vs constrained evolution were introduced, and 
the utility of using coordinate conditions to modify the structure of the numerical scheme
was demonstrated.

{\bf Binary neutron star, black hole/neutron star mergers.}
Unlike the binary black hole problem \index{binary black hole problem} which featured extensive early development 
around the head-on collision case, relatively little work on the full general relativistic
modeling of binary neutron star or black hole/neutron star mergers, head-on or otherwise,
was undertaken until the early 2000's, as reviewed in Sec.~\ref{sec_BHNS_BNS}. A notable
exception is the head-on collision study done by Wilson in the late 1970s~\citep{Wilson79a}; he
found (by applying the quadrupole formula \index{quadrupole formula} to the matter dynamics) that, similar
to the black hole case, $\sim 0.1\%$ of the rest mass of the spacetime is emitted
in gravitational waves.

{\bf Initial Data.}
The numerical initial data problem in general relativity deserves an entire chapter by itself,
and unfortunately we are unable to devote space to it in this article (for reviews
see~\citep{Cook:2000LR,Gourgoulhon:2007tn,Pfeiffer:2005zm}, and the books mentioned
below). We would however be remiss not to mention the York formalism \index{York formalism}  for the construction
of initial data~\citep{York79,York-Piran-1982-in-Schild-lectures}. This has become
the standard method for producing generic initial data for a wide range
of problems. Moreover, it provides the framework in which modern ADM-based~\citep{Arnowitt:1962hi} Cauchy
evolution schemes are written, in particular being the starting point to develop the now commonly employed BSSN formalism \index{BSSN formalism} 
discussed in Sec.~\ref{sec_NR}. A few other notable initial data-related works
include the Bowen-York closed-form solutions \index{Bowen-York data} to the momentum constraints for
black hole initial data~\citep{Bowen:1980yu}, the ``puncture'' initial data~\citep{Brandt:1997tf} \index{puncture initial data} 
(which has had significant influence beyond initial data, leading
to the stable evolution of black hole spacetimes without the need for excision) and the use of
apparent horizons \index{apparent horizons} to provide boundary conditions and implement singularity excision~\citep{Thornburg87}.

{\bf Miscellaneous.}
We conclude this historical review by listing a few other developments of import to the growth
of the field in the 80's and 90's.
\begin{itemize}
\item {\em Cosmology.} We discuss work related to cosmological singularities in
Sec. \ref{sec_CS}, and numerical studies of cosmic bubble collisions and local
inhomogeneities in cosmology in Sec.~\ref{sec_misc}, but mention here that much 
pioneering work on numerical cosmologies and spacetimes with related
symmetries began with the works of Centrella, Anninos, Wilson, Kurki-Suonio, Laguna and 
Matzner, Berger and Moncrief~\citep{Centrella:1980np,Centrella:1983cj,Anninos91a,Kurki-Suonio93,Berger:1993ff}
\item {\em Boson stars.} \index{boson stars} The numerical study of these self-gravitating soliton-like
configurations of scalar fields has a
long history that begins with calculations in the late 1960's 
by Kaup~\citep{PhysRev.172.1331} and by Ruffini and Bonnazola~\citep{PhysRev.187.1767}, 
who found spherically-symmetric, static solutions in the Einstein-Klein-Gordon model.
A major resurgence of interest in 
the subject was sparked by Colpi, Shapiro and Wassermann's discovery that the addition
of a non-linear self-interaction could lead to boson star masses that, in contrast
to those originally constructed, were
in an astrophysically interesting range~\citep{PhysRevLett.57.2485}.
Much subsequent work investigating a wide variety of types of boson stars and related
objects has been carried out since, and we touch on some representative calculations
in Secs.~\ref{sec_CP} and \ref{sec_URC}.  Once more, however, space limitations preclude a thorough
coverage of this topic and we direct the interested reader to reviews such as~\citep{Liebling:2012fv}.
\item {\em Excision.} The first successful simulation incorporating the use
of black hole excision \index{excision} to eliminate  geometric singularities from 
the computational domain was presented by Seidel and Suen~\citep{Seidel:1992vd}.
\item {\em Hyperbolic evolution schemes.} One of the influential
efforts predating the wave of activity searching for stable
hyperbolic evolution schemes discussed in Sec.~\ref{sec_NR} was the 
formulation of Bona and Masso~\citep{Bona92b}.
\item {\em The Grand Challenge (1993-1998)}. \index{Grand Challenge} This
large scale NSF-funded project, aimed at solving the black hole inspiral and merger problem,
involved essentially all US-based numerical relativists and, crucially, many computer scientists.
The notable results culminating from this effort include
propagation of a single Schwarzschild black hole through a 3D mesh~\citep{Cook:1997na},
early efforts in refined gravitational wave extraction methods and improved outer boundary conditions~\citep{Abrahams:1997ut-etal},
and the development of a characteristic code \index{characteristic code} that could stably evolve even highly perturbed single black 
hole spacetimes~\citep{Gomez:1998uj}.
\end{itemize}

\section{Numerical Relativity: Current State of the Art}\label{sec_NR}
Before beginning our review of the important physics garnered from numerical solutions of the Einstein 
field equations over the past few decades, we describe some of the key
insights obtained along both formal and numerical fronts that have made these ventures possible. Of course, it
is impossible to exhaustively cover all of them; we thus choose particularly relevant ones that have
had a strong influence on the field.

\subsection{Mathematical Formalism}
Any numerical study of a dynamical process requires solving a suitably formulated initial value (or initial boundary-value)
problem. That is, provided a set of evolution equations, together
with a given state of the system at an initial time, its future evolution can be
obtained by a numerical integration. 
At face value any covariant theory is at odds with this requirement,
unless a suitable `time' foliation of the spacetime is introduced. In the case of the Einstein equations,
projections (tangential and normal to each leaf of the foliation) provide a natural hierarchy
of evolution and constraint equations. The latter are tied to the fact that the Einstein equations are
overdetermined with respect to the physical degrees of freedom, and allow one to
chose different combinations of equations to solve for the spacetime to the future of the initial hypersurface.
As a result, one can distinguish {\em free evolution} approaches---where only evolution equations are employed to
this effect---from {\em constrained (partially constrained)} approaches where (some of the) constraints are
used to solve for a subset of variables~\citep{Piran80}. 
We also note that the freedom in choosing a foliation
gives rise to Cauchy (or D+1), characteristic or hyperboloidal formulations, in the case of spacelike,
null or spacelike-but-asymptotically-null foliations \index{Cauchy formulation} \index{characteristic formulation} \index{hyperboloidal formulation}
respectively. For a review of related numerical approaches,  see e.g., ~\citep{Lehner:2001wq};
 a more pedagogical exposition of the basic concepts can be found in recent textbooks
on the subject~\citep{Bona05,Alcubierre08,2010nure.book.....B,2012LNP...846.....G}.

On the Cauchy front, early efforts employed the so-called York-ADM formulation~\citep{York79} (a
reformulation of the standard Hamiltonian-based ADM approach). \index{ADM formulation} \index{ADM-York formulation} This 
formulation is geometrically appealing in that it provides evolution equations for the intrinsic and 
extrinsic curvatures of the foliation.
However, beyond spacetimes where symmetries allow for reducing the dimensionality
of the problem, numerical evolution with the York-ADM method exhibits instabilities. 
In the early 2000s~\citep{Kreiss:2001cu}, it was recognized that such a formulation is only weakly hyperbolic,
implying that at the analytical level the system of PDEs lack properties required to achieve robust 
numerical implementation~\citep{Gustafsson95}. 
A flurry of activity in the following years provided much insight on how to deal with this issue and construct 
(desirable) symmetric or strongly hyperbolic formulations by suitable
modifications of the equations (primarily via the addition of constraints and the use of appropriate coordinate 
conditions---see e.g.~\citep{Sarbach:2012pr,Friedrich:2000qv,Reula:1998LR} for reviews on the subject). 
It turns out that arbitrarily many formulations could be defined with these desirable properties and all are, of course,
equivalent at the analytical level.

At the discrete level however, this is not the case. In particular in free evolution schemes
it is challenging to control the magnitude of the truncation error associated with the constraint equations (which are not explicitly
imposed except at the initial time). Such errors compound at different rates in different formulations, and
the practical physical time that accurate results can be achieved in simulations thus varies 
significantly. Two formulations have 
been shown empirically to display the robustness needed to construct a large class of 
4-dimensional spacetimes (without symmetries)
that are of particular relevance to the contemporary astrophysical and theoretical physics 
problems discussed here.
These are the generalized harmonic evolution with constraint damping~\citep{Pretorius:2005gq} (closely related to the Z4 formalism, see e.g.~\citep{Alic:2011gg}),
and the BSSN (or BSSNOK) approach~\citep{Nakamura:1987zz,Shibata:1995we,Baumgarte:1998te}. \index{BSSN formalism} 

Harmonic coordinates \index{harmonic coordinates} have a history even older than the field equations themselves---having
been used by Einstein in his search for a relativistic theory of gravity as early as 
1912~\citep{1999ewgr.book...87R}---and have played an important role in many key discoveries 
of properties of the field equations
since (see the introduction in~\citep{Lindblom:2005qh}). 
Harmonic coordinates can be defined
as the requirement that each spacetime coordinate obeys a homogeneous scalar wave equation.
Enforcing this at the level of the Einstein equations converts the latter
to a form that is manifestly symmetric hyperbolic. 
This desirable property is maintained if 
freely specifiable functions are added as source terms in the wave equations, resulting
in {\em generalized} harmonic coordinates. \index{generalized harmonic coordinates} In principle, the source functions
allow arbitrary gauges to be implemented within a harmonic
evolution scheme~\citep{Garfinkle:2001ni}.
Constraint damping terms~\citep{Gundlach:2005eh} \index{constraint damping} are further added to tame 
what otherwise would result in exponential growth of truncation error\footnote{The idea
of constraint damping via the addition of terms to the equations that are homogeneous in
the constrained variables (hence are zero for continuum solutions) dates back several
years earlier for the Einstein equations~\citep{Brodbeck:1998az}, and since has
also effectively been applied to other systems of PDEs that have internal constraints,
in particular Maxwell's equations~\citep{Palenzuela:2009hx}.}. This formulation has also
been particularly useful in finding stationary black hole solutions in higher dimensions where,
due to the stationary nature of the spacetime, the coordinate freedom can be exploited to define a 
convenient, strictly elliptic problem~\citep{Headrick:2009pv}.
The BSSN formulation is an extension of the York-ADM approach which introduces several additional (constrained)
variables to remove particular offending terms from the equations, and this, 
together with a judicious choice of coordinates, ensures strong hyperbolicity of the underlying equations.

When black holes are evolved, the geometric singularities inside the horizons need to be dealt
with in some manner to avoid numerical problems with infinities. One such method is excision~\citep{Thornburg87}, \index{excision}
where an inner {\em excision boundary} is placed inside each apparent horizon to remove the singular region
from the computational domain. Due to
the causal structure of the spacetime, the characteristics of the evolution equations
ostensibly all point out of the computational domain at the excision boundary, and no boundary conditions
are placed there. Excision is commonly used in harmonic evolution schemes.
For evolution within the BSSN approach, an 
alternative {\em moving puncture} method \index{moving puncture method} has proven successful~\citep{Campanelli:2005dd,Baker:2005vv}.
This is an extension of puncture initial data, where the puncture point inside the horizon
that formally represents spatial infinity on the other side of a ``wormhole'' is now evolved in time.
With the typical gauges employed during evolution the interior geometry evolves to a so-called 
``trumpet'' slice, \index{trumpet slice} where the puncture asymptotes to the future timelike infinity of the other 
universe~\citep{Hannam:2006vv,Hannam:2008sg}. 
Effectively then, the puncture also excises the singularity from the computational domain.

On the characteristic front, the structure of the equations is significantly different from 
the Cauchy problem, as
the foliation is defined using characteristic surfaces. The system of equations displays
a natural hierarchy of evolution equations, constraints, and a set of hypersurface equations for variables 
that asymptotically are intimately connected to the physical degrees of freedom of the 
theory~\citep{Winicour:1998LR}. Beyond spherically symmetric applications, numerical codes employing
this formulation show a remarkable degree of robustness, and can stably evolve
highly distorted single black hole spacetimes~\citep{Gomez:1997pd,Gomez:1998uj}. However, the rigidity 
in the choice of 
coordinates (being tied to characteristics) imply that difficulties arise when caustics and crossovers 
develop. 
For astrophysical purposes, the main role of the
characteristic approach has been to provide a clean
gravitational wave extraction procedure~\citep{Bishop:1996gt,Reisswig:2009rx} and to study
isolated black holes. \index{gravitational wave extraction}
Outside of the astrophysical domain, 
it has been convenient in studies of black hole interiors (e.g.~\citep{1995PhRvL..75.1256B}) and has become 
the predominant approach used to exploit the AdS/CFT (Anti-deSitter/Conformal Field Theory) 
duality \index{ADS/CFT duality} of string theory (see~\citep{Chesler:2013lia} and references therein). Here gravity
in asymptotically AdS spacetimes is used to study field theory problems,
some examples of which are described in section \ref{ADSCFT}.

Finally, the hyperboloidal formulation~\citep{Friedrich:2002xz} \index{hyperboloidal formulation} adopts a Cauchy approach in a conformally
related spacetime where the physical spacetime is recovered as a subset of a larger one. This 
allows studying, within a single framework, both the local and global structure of spacetime (as the larger manifold covers
spacelike, null and timelike infinity in a natural way).
However, it has received considerably less attention than the other approaches, though
some interesting first steps have been carried out (e.g.~\citep{Frauendiener:1997zc,Husa:2002kk}).

\subsection{Numerical Methods}
The subject of numerical analysis as it pertains to solutions of problems in
applied mathematics is, of course, vast.
Even when restricting to what is relevant for
numerical relativity applications, the breadth of
methods employed is considerable and, naturally,
depends on the particular goal one has in mind. From constructing initial data and evolving
the solution to the future of an initial hypersurface, to extracting 
physical information from the
numerical results,
an abundance of different techniques and methods have been used. Here we 
provide some brief comments with an aim to impart a basic understanding of 
the available options.

Any numerical implementation ultimately renders the problem of interest into an
algebraic problem for a discrete number of variables that describe the sought-after solution.
For gravitational studies, this involves devising approximate methods to solve
the relevant partial differential equations. 
Such methods can be conveniently visualized as providing
a way to discretize the underlying variables that describe the problem as well as the spatial
derivatives within the equations, and providing a recipe to advance the solution in discrete time.
 
The technique most commonly used is the {\em finite difference} (FD) method, \index{finite difference method}  with the solution represented
by its value at discrete grid-points covering the manifold of interest. Discrete spatial derivatives are defined
through suitable Taylor expansions, which can allow for high order accurate approximations for smooth solutions. Further refinements  can be achieved through the use of discrete derivative approximations that 
satisfy summation by parts. This
property is a direct analog of integration by parts often exploited to obtain estimates about the behavior
of general solutions at the analytical level~\citep{Gustafsson95,Sarbach:2012pr}. 
What has proven especially useful for many problems is the adoption 
of {\em adaptive mesh refinement} (AMR) \index {adaptive mesh refinement} to efficiently resolve a large range of relevant spatio-temporal 
lengthscales, and without {\em a priori} knowledge of the development of small scale 
features~\citep{Berger:1984zza,Choptuik89,Lehner:2005vc}.
Another discretization approach used is the {\em pseudo-spectral} \index{pseudo spectral method}
method~\citep{Boyd89a,Grandclement:2009LR}:
here the solution is expanded in 
terms of a suitably chosen basis (e.g. Chebyshev or Fourier), which also provides a simple
way to compute spatial derivatives. The coefficients of the expansion provide
the sought after solution.
Pseudo-spectral methods provide a highly efficient way to achieve high accuracy results for 
smooth solutions. As with AMR for FD, adaptive, multi-domain decomposition methods
can be employed for efficiently resolving the length scales in the problem (see e.g.~\citep{Szilagyi:2009qz}).

Advancing the solution in time requires integrating the discrete values of the FD solution, or
spectral coefficients, via suitable operators. One common approach is the method of lines, \index{method of lines} where
having discretized the spatial part of the problem, 
accurate methods devised for ordinary differential equations are used to integrate the
variables in time.

The above discretization approaches are also well-suited to evolving additional fields coupled 
to gravity that are
smooth and, in particular, do not develop discontinuities (such as scalar or electromagnetic fields). 
However, when matter sources such as neutron stars are incorporated, the equations 
of general-relativistic 
hydrodynamics (or magnetohydrodynamics) must also be solved.
These can be expressed
in a way fully consistent with the approaches employed to integrate Einstein equations (for a review on
this topic see~\citep{2008LRR....11....7F}). Nevertheless, as the solution to the hydrodynamic equations can induce
discontinuities even for smooth initial conditions, 
{\em finite volume} methods~\citep{Leveque92} are \index{finite volume methods} most often adopted as they are especially suited 
to accurately handle such features.

\section{Strong Field Gravity}
As mentioned, numerical simulations are often the only way to gain insights into the behavior
of gravity in the strongly non-linear regime. This arises naturally in gravitational collapse,
systems involving black holes and neutron stars, ultra-relativistic collisions and cosmology. Here
questions range from fundamental explorations of the theory itself,
to the resolution of questions of astrophysical relevance, such as the characteristics of gravitational wave signals produced in compact
binary mergers, or the effect of highly dynamical and strong gravitational fields on matter/gas/plasma and their
role in powering spectacular phenomena like gamma ray bursts. 
Complex numerical simulations have been developed over the past decades to start answering these long-standing questions,
and have produced results that often raise new questions.
In what follows, we discuss some of the more important findings and open questions, organizing the presentation by
subject area. These examples
are necessarily limited in scope and presentational depth, but serve as illustrations of the breadth of problems
addressed with simulations. For a complimentary review article on numerical relativity and its
applications, see~\citep{upcomingNRLR}.

\subsection{Critical Phenomena in Gravitational Collapse}\label{sec_CP}
\def\mbh{M_{\rm BH}}
\subsubsection{Overview}
Gravitational collapse, \index{gravitational collapse} including the process of black hole formation, is one of the 
hallmarks of general relativity. As has already been noted in Sec.\ ~\ref{sec_historical}, although 
simulations play an increasingly dominant role in advancing our 
understanding of collapse scenarios that we believe play out in the universe, they also 
provide the means to perform detailed studies of more fundamental aspects of 
the process. Albeit reflective of more than a little theorist's conceit, we can view 
computer programs as numerical laboratories which---paralleling real experiments in 
nonlinear science---are endowed with one or more control parameters that are 
varied in order to unearth and elucidate the phenomenology exhibited by the setup.

Critical phenomena \index{critical phenomena} are concerned with families of solutions to the coupled
Einstein-matter equations (including the vacuum case), where a continuous parameter 
$p$ labels the family members, and it is assumed that the spacetimes are constructed
dynamically---usually via simulation---starting from prescribed initial data that 
depends on $p$.  The initial data 
typically represents some bounded distribution of initially imploding matter and $p$ is 
chosen to control the maximal strength of the gravitational interaction that ensues.  
For $p$ sufficiently small, gravity remains weak during the evolution, and the spacetime 
is regular everywhere (if the matter is massless radiation, for example, the radiation 
will disperse to infinity, leaving flat spacetime in its wake).  For $p$ sufficiently
large, gravity becomes strong enough to trap some of the matter in a black hole, with 
mass $\mbh$, and within
which a singularity forms.  For some critical value $p^\star$ lying between
the very-weak and very-strong limits, the solution corresponds to the 
threshold of black hole formation, and is known as a critical spacetime for the given 
model. Collectively, the properties of these special configurations, as well as 
the features associated with the spacetimes close in solution space to 
the precisely critical solution, comprise what is meant by critical behaviour.
Evidence to date suggests that virtually any collapse 
model that admits black hole formation will contain critical solutions.

It transpires that $\mbh$ can be formally viewed as an order parameter, in the 
statistical mechanical sense, and most of the critical solutions identified to date 
can be sorted into two basic classes based on the 
behaviour of $\mbh$ at threshold.  Specifically, solution-space behaviour corresponding to
both first- and second-order phase transitions is seen, defining what are called
Type I and Type II critical solutions, respectively.  \index{Type I critical solutions} \index{Type II critical solutions}
Thus, in the Type I case $\mbh$ is {\em finite} at the threshold, so $\mbh(p)$
exhibits a gap (jump) at $p=p^\star$.  Conversely, in the Type II instance $\mbh$ becomes 
{\em infinitesimal} as $p\to p^\star$ from above, and there is no gap. 
It is also crucial to observe that the precisely critical solution for $p=p^\star$ does {\em not} contain a black hole.

There are three key features associated with both types of black hole critical phenomena:
universality, symmetries and scaling.

Concerning the first property, most, if not all, Type II solutions, and some 
Type I solutions, exhibit a type of universality \index{universality} in the sense that one finds the 
same critical configuration through numerical experimentation as sketched above, 
irrespective of the specific way $p$ parameterizes the initial data.  This implies a 
certain type of uniqueness, or at least isolation, of the critical spacetimes in 
solution space, analogous to the uniqueness of the Schwarzschild solution as
the endpoint of black hole formation in spherical symmetry.

Secondly, Type I critical solutions generically possess a time-translational symmetry,
which is either continuous (so the solution is static or stationary), or discrete 
(so the solution is periodic).  In the discrete case, the oscillation frequency forms
part of the precise description of the critical solution, and is
usually determined by an eigenvalue problem.
Type II critical solutions, on the other hand, typically have 
a scale, or homothetic, symmetry and are therefore scale invariant.  Once again, this 
symmetry can be either continuous or discrete (CSS/DSS for continuously/discretely
self-similar). \index{continuously self-similar; CSS} \index{discretely self-similar; DSS}  Continuously self-similar solutions have 
long been studied in relativity as well as in many other areas of science, frequently
arising in situations where the underlying 
physics has no intrinsic length scale.  On the other hand, it is 
safe to say that the observation of discrete self-similarity in the earliest numerical
calculations of critical collapse came as a complete surprise.  For DSS solutions, 
the analogue of the frequency of periodic Type I configurations is known as the 
echoing exponent, $\Delta$.  When expressed in coordinates 
adapted to the self-similarity, a DSS solution is oscillatory with period $\Delta$;
each complete oscillation represents a shrinking of the scale of the dynamics by
a factor of $e^\Delta$.
In relation to the collapse
process, self-similar behaviour of either type is particularly interesting because 
it means that at criticality
the strong field regime propagates to arbitrarily small spatiotemporal scales.  Indeed,
as the self-similar solution ``focuses in'' to an accumulation event at the center of the 
collapse, curvature quantities grow without bound, but with no formation of an 
event horizon.  Thus, Type II critical solutions possess naked singularities \index{naked singularities} and 
have significant relevance to the issue of cosmic censorship. \index{cosmic censorship conjecture}  However, in terms of their
origin from collapse, it is imperative to note that these naked singularities are produced 
from (infinite) fine tuning of the initial data, so are therefore not generic with 
respect to initial conditions.  

The third characteristic of critical collapse is the emergence of scaling laws.  \index{scaling laws}
Empirically, these are determined by studying the behaviour of certain physical
quantities as a function of the family parameter $p$ as $p\to p^\star$.  For 
Type I behaviour a typical scaling law measures the time interval, $\tau$, during which
the dynamically-evolving configuration is close to the precisely critical solution.
One finds
\begin{equation}
   \tau \sim - \sigma \ln | p - p^\star |
   \label{lifetime-scaling}
\end{equation}
where $\sigma$ is called the time-scaling exponent.  In Type II collapse
the black hole mass scales according to
\begin{equation}
   \mbh \sim C |p - p^\star|^\gamma
   \label{mass-scaling}
\end{equation}
where $\gamma$ is known as the mass-scaling exponent, and $C$ is a family dependent constant.  
For both types of behaviour, if the critical
solution is universal with respect to the initial data, then so is the corresponding 
scaling exponent.   Again, this is the case for all known Type II solutions, but not 
so for most Type I transitions where, as discussed in more detail below, a particular critical 
configuration is typically one member of an entire branch of unstable solutions.

The scaling laws can be understood in terms of perturbation theory.  
The key observation~\citep{Evans:1994pj,Koike:1995jm,Maison:1995cc} is that the appearance of critical solutions 
through a fine tuning process---wherein one of two distinct
end states characterizes the long-time dynamics---suggests that they have a {\em single} unstable 
perturbative mode. The inverse of the Lyapunov exponent associated with 
that mode is then precisely the scaling exponent.  Furthermore, 
leading subdominant modes can give arise to additional scaling laws, for 
example charge and angular momentum~\citep{Gundlach:2007gc}.

The scaling relations~(\ref{lifetime-scaling})--(\ref{mass-scaling}) underscore the 
fact that near criticality, there is exponentially sensitive dependence on 
initial conditions, and that irrespective of the original choice of 
initial data parametrization, it is the transformed quantity 
$\ln|p-p^\star|$ which is most natural in describing the phenomenology.
In simulations one wants to compute with $|p-p^\star|$ as small as possible in
order to most accurately determine the threshold solutions and their 
associated exponents.  This is especially true for the Type II case since
the asymptotically flat boundary conditions that are normally adopted
are incompatible with self-similarity, so one relies on calculations 
which probe as small scales as possible to ensure that boundary effects 
are minimized.  In practice, investigation of the critical regime is 
ultimately limited by the fact that $p$ can only be fine-tuned to machine 
precision. For spherical and axisymmetric calculations, this can be 
accomplished with current technologies, but, for the Type II scenarios, only 
if the numerical algorithm provides sufficient spatio-temporal dynamic range
using a technique such as adaptive mesh-refinement (AMR)~\cite{Berger:1984zza}.
Indeed, at this point it should be emphasized that almost all known critical 
solutions have been computed in models that impose symmetry restrictions.  
Spherical symmetry has been most 
commonly adopted. There have been some axisymmetric calculations, but very few fully
3D studies.

As already mentioned, the appearance of critical solutions seems completely 
generic, irrespective of the matter content of the model.  However, details 
of the phenomenology are dependent on a number of factors including the following: the 
type of matter and the specifics of any self-interaction terms, the imposed
symmetries, the spacetime dimensionality and the asymptotic boundary conditions.
The remainder of this section is devoted to a summary of a necessarily
incomplete selection of the 
many numerical studies performed to date, organized by the type of matter employed. 
In order to highlight the state of the art
in the subject, there is some bias towards more recent calculations, and an attempt
has been made to impart some sense of the wide variety 
of scenarios that have been explored.
Those interested in more information are directed to the 
excellent comprehensive reviews of the 
subject~\citep{Gundlach:1997wm,Gundlach:2007gc}.

\subsubsection{Scalar Fields}
\begin{figure}
\includegraphics[width=5.0in]{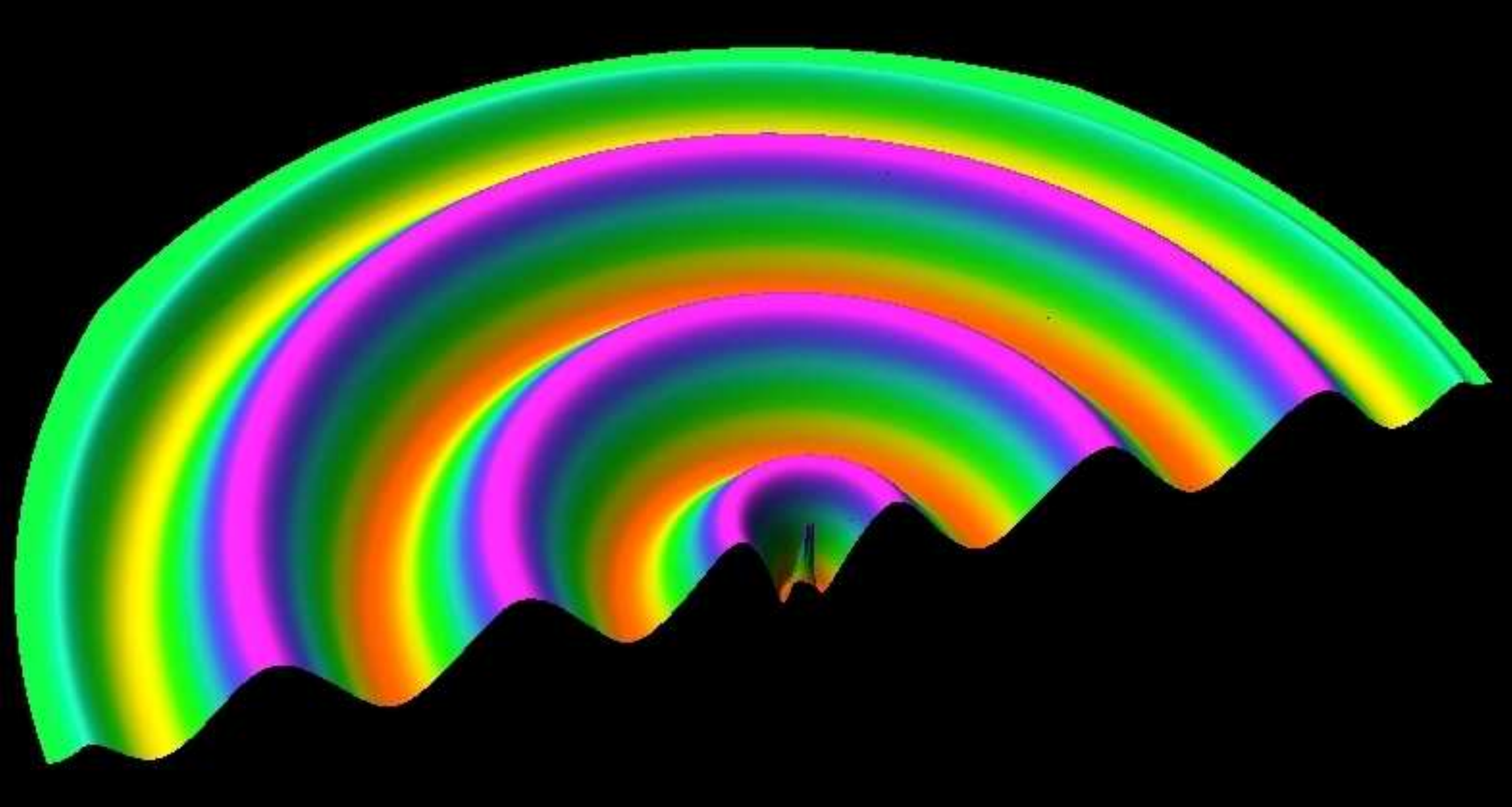}
\caption[Scalar field critical collapse solution]
{Type II discretely self-similar critical solution computed 
from the collapse of a spherically symmetric distribution of massless 
scalar field.
The figure shows the late time configuration of the scalar field from 
a marginally subcritical evolution 
where the family parameter has been tuned
to approximately a part in $10^{15}$.  The radial coordinate is logarithmic,
making the discretely self-similar (echoing) nature of the solution evident: each
successive echo represents a change in scale of $e^\Delta\approx31$.  The data
were generated using the axisymmetric code described in~\citep{Choptuik:2003ac}.
}
\label{fig:vt_phi_lnr}
\end{figure}

Critical collapse was first studied in the model of a spherically symmetric
minimally coupled massless scalar field~\cite{Choptuik:1992jv}.
Using 
several families of initial data,  a single Type II DSS solution with 
$\gamma \approx 0.37$ and $\Delta \approx 3.44$ was found (see Fig.~\ref{fig:vt_phi_lnr}).  
Due to the 
extreme dynamical range required to fully resolve the critical solution,  
use of adaptive mesh refinement was crucial.  Remarkably, the mass-scaling 
relation~(\ref{mass-scaling}) provided a good fit for $\mbh$ even when 
$|p-p^\star|$ was large enough that the final black hole contained 
most of the total mass of the spacetime.
The main results of~\cite{Choptuik:1992jv}  have been confirmed many times 
since using a 
variety of different algorithms and coordinate systems.  
Assuming the existence of a spacetime with 
a discrete homotheticity, certain regularity conditions and a tailored
numerical approach, the critical
solution and associated exponents were computed to very high 
accuracy in~\cite{Gundlach:1996eg}.  Additionally,
analysis of the implications of the discrete self-similarity led to 
the prediction and observation of a
modulation of the mass-scaling law~(\ref{mass-scaling}) with period
$\Delta/(2\gamma)$~\cite{Gundlach:1996eg,Hod:1996ar}.  Finally, an 
important analysis in~\cite{MartinGarcia:1998sk} found that all non-spherical
modes of the critical solution decay, strongly suggesting that the 
same threshold configuration would appear if the symmetry restriction
were relaxed.

Critical solutions from axisymmetric massless scalar collapse 
using multiple initial data families were constructed in~\cite{Choptuik:2003ac}.
For the most part the threshold configurations could be described as the spherical solution
plus perturbations (measured values for the scaling exponents were $\gamma\approx
0.28$--$0.41$ and $\Delta \approx 2.9$--$3.5$), but there were also indications 
of a single asymmetric mode which, as it grew, produced two separated regions 
(on axis) within which the solution locally resembled the spherical one.  This 
observation is in conflict with~\cite{MartinGarcia:1998sk}, but the accuracy 
of the results was insufficient to convincingly demonstrate that 
the growth was genuine and not a reflection of limitations in the simulations.
Adaptive mesh refinement was again crucial.

Very recently, a study of massless scalar collapse using a fully 3D code has been 
carried out~\cite{Healy:2013xia} and, in fact, represents the first calculations 
of Type II general-relativistic critical phenomena in 4 spacetime dimensions
without symmetry restrictions.
Four initial data families defining a spherical matter distribution deformed
to varying degrees with a $Y_{21}$ spherical harmonic 
anisotropy were considered. Even though AMR was used, compute-time limitations
kept the tuning of $p/p^\star$ to about a part in $10^{4}$. Nonetheless, evidence for the
emergence of the spherically symmetric critical solution with $\gamma\approx0.37$--$0.38$ 
was found.  There were also preliminary indications of echoing with 
$\Delta\approx3.1$--$3.3$.

The massless scalar model has no intrinsic length scale so, in retrospect, 
the appearance of a Type II solution at threshold is natural.  Introduction of 
a mass, $\mu$, breaks scale invariance and, as shown in~\cite{Brady:1997fj}, 
complicates the picture of criticality.   For initial data with a length 
scale $\lambda$ the massless behaviour is recovered 
when $\lambda\mu\ll1$.  However, for $\lambda\mu\gtrsim1$, a Type I transition
is seen with a critical solution which is one of the periodic, starlike 
configurations (oscillons) \index{oscillons} admitted by the model and constructed
in~\citep{Seidel:1991zh}. 
As with relativistic perfect fluid stars, the oscillons comprise a one-parameter 
family that can be labeled by the central density.  As the central density
increases the stellar mass also increases, but only up to a point, whereafter dynamical
instability sets in and the stars reside on the so-called unstable branch---it 
is precisely one of these unstable solutions that sits at the Type I transition.
This latter type of behaviour was also observed in~\citep{Hawley:2000dt} using a 
massive complex scalar field whose static solutions, known as boson stars, \index{boson stars} also have
stable and unstable branches.  In this instance stable stars were driven to a 
Type I threshold via an imploding pulse of massless scalar field, whose overall
amplitude was used as the tuning parameter.

Investigation of circularly symmetric massless scalar collapse in $2+1$ AdS
spacetime~\citep{Husain:2000vm,Pretorius:2000yu} represents one of the few instances 
where critical behaviour 
in a non-asymptotically flat setting has been seen (but also see the discussion 
of the turbulent instability of $3+1$ AdS~\citep{Bizon:2011gg} in Sec.~\ref{ads_stab}).
Evidence for a Type II transition with a CSS solution was found---with 
a mass-scaling exponent $\gamma\approx1.2$---but a thorough understanding of the 
picture of criticality here is still lacking.
In particular, an analytic CSS solution that shows good agreement with the numerical 
results has been found~\citep{Garfinkle:2000br},
but seems to have additional unstable modes. Its existence also seems paradoxical
in the sense that, heuristically, the cosmological constant should be irrelevant 
on the small scales pertinent to scale-invariance, yet is essential in the 
construction of the solution.
\subsubsection{Vacuum} 
Historically, the second example of black hole critical phenomena discovered 
was in the collapse of pure gravitational waves~\cite{Abrahams:1993wa} in axisymmetry.
The study employed one family of initial data representing initially incoming 
pulses of gravitational radiation with quadrupolar angular dependence, and with 
an overall amplitude factor serving as the control parameter.  Evidence 
for a Type II transition was found, with a discretely self-similar critical 
solution that was centred in the collapsing energy.
The computations yielded an estimated mass-scaling exponent
$\gamma\approx0.37$ and an echoing factor $\Delta\approx0.6$.  The calculations 
did not use AMR, but due to the use of spherical polar coordinates, increased 
central resolution could be achieved with a moving mesh technique.  Nonetheless,
the dynamic range of the code was very limited relative to that used 
in~\cite{Choptuik:1992jv},  so it was quite fortuitous that $\Delta$ in this case 
was quite small.

It is truly remarkable that in the two decades that have elapsed since the 
publication of~\citep{Abrahams:1993wa}, and despite several additional assaults
on the problem and a vast increase in the available amount of computer resources, 
little progress has been made in reproducing and extending these
early results.  One notable exception is~\cite{Sorkin:2010tm} in which the collapse 
of axisymmetric Brill waves was studied, using several different families of 
data with varying degrees of anisotropy.  Once more, evidence for a Type II 
transition was found in all of the experiments, with a scaling exponent 
$\gamma$---measured in this instance through the scaling of a curvature
invariant in subcritical collapse~\citep{Garfinkle:1998va}---in the range
$0.37$--$0.4$. However, in stark contrast to the observations in~\citep{Abrahams:1993wa},
most of the computed critical solutions showed accumulation on rings at finite distances
from the origin, rather than at the origin itself.
Additionally, indications 
of echoing were seen, but with an estimated $\Delta\approx1.1$ significantly 
different from that reported in~\citep{Abrahams:1993wa}. Development of a more
complete understanding of the critical behaviour of collapsing gravitational waves,
both in axisymmetry and the full 3D case, remains one of the most important 
unresolved issues in this field.

In $D+1$ dimensions, with $D$ even, application of a co-homogeneity two symmetry reduction
to the vacuum Einstein equations yields 
a set of wave equations dependent only on a single radial dimension.
In contrast to those resulting from a spherically symmetric reduction, these equations admit 
asymptotically-flat, radiative 
solutions~\cite{Bizon:2005cp,Bizon:2006qi,Bizon:2005af,Szybka:2007fx}.
For $D=4$, and adopting the so-called biaxial ansatz, Type II DSS behaviour was found, 
with $\Delta\approx0.47$ and $\gamma\approx0.33$~\cite{Bizon:2005cp}.  Analogous results 
were found for $D=8$ where $\Delta\approx0.78$ and $\gamma\approx1.64$~\cite{Bizon:2005af}.
The more general triaxial ansatz for $D=4$ was considered in~\cite{Bizon:2006qi}.  Here, 
the biaxial critical solution still appears at threshold, but due to a discrete symmetry 
in the model, the critical surface actually contains three copies of the configuration. As well, 
on the boundaries of the basins of attractions of these copies, a different DSS solution
with two unstable modes was predicted and computed using a two-parameter tuning process.
Additional numerical experiments have shown that the critical-surface 
boundaries have a fractal structure~\cite{Szybka:2007fx} .

\subsubsection{Fluids}
Studies of critical behaviour with perfect fluid sources have been extremely 
important in the development of the subject, not least since it was in this 
context that understanding of the phenomena in terms of unstable
perturbative modes was developed.  The first calculations focused on 
spherically symmetric simulations with a fluid equation of state (EOS), \index{equation of state} 
$P=k\rho$, where $P$ and $\rho$ are the fluid pressure and energy density, 
respectively, and with the specific choice $k=1/3$ (radiation fluid)~\citep{Evans:1994pj}.  
A continuously 
self-similar critical solution was found with a mass-scaling exponent $\gamma\approx0.36$.
In addition, the critical solution was computed independently by adopting 
a self-similar ansatz, and was shown to be in excellent agreement with the simulation
results, and it was suggested that a perturbation analysis could be used
to at least approximately compute $\gamma$. Such an analysis was carried out
in~\citep{Koike:1995jm}, where both the critical solution and its linear perturbations
were determined, and it was shown that there was a single unstable mode 
whose inverse Lyapunov exponent yielded the same value of $\gamma$ seen in the simulations.
Interestingly, at this time the values of $\gamma$ that had emerged from the 
three models for which threshold solutions had been identified were numerically the 
same to the estimated level of numerical accuracy, suggesting that the mass-scaling 
exponent might be universal across all matter models.
However, the results 
of~\cite{Maison:1995cc} (performed at the same time as~\citep{Koike:1995jm}), where critical 
solutions and their perturbative modes were determined via the self-similar ansatz for 
many values of the EOS parameter $k$ in the range $0.01$--$0.888$, showed definitively that 
$\gamma$ was in general model-dependent.  
A more extensive perturbation analysis~\cite{Gundlach:1997nb,Gundlach:1997vy} suggested that the spherical
solutions
will appear at threshold when spherically symmetry is relaxed only for values of $k$ in 
the range $1/9<k\lesssim 0.49$; for other values of $k$,  additional 
unstable modes were found.  These conclusions have yet to be verified through 
simulations, and it will be very interesting to do so.

The $P=k\rho$ EOS is scale-invariant (and in fact is the only EOS compatible 
with self-similarity~\citep{Cahill1971}) so Type II critical behaviour is expected.
For more general equations of state, including the commonly adopted 
ideal gas law, intrinsic length scales appear and, as anticipated from the 
massive scalar field studies~\citep{Brady:1997fj}, the 
critical phenomenology becomes richer.  In particular, the expectation that 
unstable stars can appear as Type I critical solutions was confirmed 
in~\citep{Noble:2003xx} using the ideal gas EOS and the same 
type of experiments performed in~\citep{Hawley:2000dt}.  Type II behaviour 
with this EOS also appears when the fluid internal density of the configuration
is much larger than the rest energy density~\cite{Neilsen:1998qc,Noble:2007vf,Novak:2001ck},
in which case the EOS limits to the scale-free equation, and the measured mass-scaling 
exponents,  agree with those computed from a scale-invariant ansatz.  

A possible cosmological application of Type II fluid collapse was posited 
in~\citep{Niemeyer:1997mt},  where it was argued that the mass-scaling relation
should apply to the formation of primordial black holes, since the exponential
decay of the scale of density fluctuations entering the horizon at 
any epoch provides an intrinsic
fine tuning mechanism.  This leads to a modification of the usual 
mass function for the primordial black holes, which incorporates the prediction
that holes of sub-horizon scale could form at all times.

Over the past few years, significant progress has been made in extending 
the investigations of Type I critical behaviour with fluids to the 
axisymmetric~\citep{Jin:2006gm,Kellermann:2010rt,Radice:2010rw,Wan:2011wg}
and 3D~\citep{Liebling:2010bn} arenas.  Almost all studies have adopted a stiff ($k=1$) 
ideal gas EOS (with static or stationary 
solutions interpreted as neutron stars), and the work reported 
in~\citep{Liebling:2010bn} also incorporated
rotational and magnetic effects. In~\citep{Jin:2006gm} a Type I transition was 
observed in the head-on collisions of two neutron stars where
several different tuning parameters, including the stellar mass and the index $k$, were 
employed.  Clear evidence of lifetime scaling for subcritical 
evolutions was seen. It was also suggested that the change 
in the EOS that occurs as a real post-collision remnant cools could provide a 
natural tuning mechanism, so that if the cooling was sufficiently slow,
the critical solution might have astrophysical relevance.  Further simulations
of head-on collisions~\citep{Kellermann:2010rt,Wan:2011wg} have corroborated 
these findings, and it was demonstrated in~\cite{Kellermann:2010rt}
that the end state of the marginally subcritical collision
was well-described by a perturbed star on the stable branch.  Intriguingly, the lifetime 
scaling measured in~\citep{Kellermann:2010rt}
exhibits a periodic modulation of $\sigma$---analogous 
to that seen in the mass-scaling exponent for DSS Type II 
transitions---that has yet to be explained.   The fact that stars on an unstable branch can be
identified as Type I solutions has also been demonstrated in a more direct fashion, 
through the use of initial data families where the tuning parameter 
perturbs (or effectively perturbs) a star
known or suspected to be one-mode unstable.  This strategy was employed 
in~\cite{Radice:2010rw} to demonstrate the criticality of an unstable spherical
configuration, with an accurate computation of $\sigma$.  Finally, 
in~\cite{Liebling:2010bn} evidence for the threshold nature of rotating 
unstable stars---both non-magnetized and magnetized---with preliminary evidence of 
lifetime scaling was reported.  This last study, along with~\citep{Healy:2013xia}, provides 
a tantalizing glimpse of what lies in store for this field as symmetry restrictions
are relaxed and the physical realism of models is enhanced.
\subsubsection{Other Types of Matter}
Spherical collapse of an $SU(2)$ Yang-Mills field within the magnetic ansatz was studied
in~\cite{Choptuik:1996yg}, and was the first model where both Type I and Type II behaviour was observed. 
Here the $n=1$ member of the Bartnik and McKinnon \index{Bartnik and McKinnon sequence} countable sequence of static 
configurations~\cite{Bartnik:1988am}, which had previously been shown to have one unstable
mode, is the attractor for the Type I transition, while a DSS solution with $\Delta=0.74$ and 
$\gamma=0.20$ was also found.  The model exhibits another transition, 
strictly in the black-hole sector of solution space, where colored black holes arise at 
the threshold and where $\mbh$ has a gap as one tunes across it~\citep{Choptuik:1999gh}.

Spherically symmetric self-gravitating $\sigma$-models (wave-maps), \index{nonlinear $\sigma$ models} \index{wave maps} which typically 
incorporate dimensionless tunable coupling constants, have been shown to display 
especially rich critical phenomenology.  Notably, transitions between CSS and DSS Type II behaviour
as the coupling is varied have been seen in both the 
2-dimensional nonlinear model~\citep{Liebling:1996dx} and 
the $SU(2)$ case~\citep{Lechner:2001ng}.  The transition in the latter instance is particularly
interesting, displaying behaviour where near-critical evolutions approach and depart from a 
CSS solution episodically.

Finally, Type I critical behaviour has been seen in the collapse of collisionless matter
in spherical symmetry---with or without a particle mass---where the threshold solutions are 
static~\citep{Andreasson:2006gx,Olabarrieta:2001wy,Rein:1998uf} and appear to 
exhibit the expected properties of Type I solutions, including lifetime scaling.   In the massless case 
it has been argued that there should be {\em no} one-mode unstable 
solutions~\citep{MartinGarcia:2001nh},
and this apparent contradiction with the numerical results remains another unsolved 
puzzle.

\subsection{Binary Black Hole Mergers}\label{sec_BBH}
The non-linear nature of general relativity has several interesting consequences for how it describes
particles and the gravitational interaction between them. First, technical caveats aside,
the simplest possible solution describing the geometry of an idealized point-like distribution
of chargeless, spinning matter is a Kerr black hole. From an external observer's perspective
there is thus {\em no} geometrical realization of a point-like structure, as the event horizon prevents
length scales smaller than the energy (in geometric units) of the black hole from being probed.
Second, there is no analogue of a Newtonian potential that can be superposed to come up with
a simple description of the interaction of two black holes.
In consequence, a detailed understanding of one of the most basic interactions in gravity, 
the two body problem, 
requires numerical solution of the field equations. 
On the other hand, thanks to the ``no-hair'' properties of black holes, the merger of two Kerr
black holes is expected to describe the merger of all astrophysical black holes essentially exactly,
the only idealization being that the presence of surrounding matter is ignored.

The discussion in the previous paragraph assumes many properties of solutions to the field
equations not yet proven with mathematical rigor. Chief among them are
that cosmic censorship holds in these scenarios, and that any black hole that forms in our universe
(specifically here via the merger of two black holes, but implicitly also by processes that led
to the initial black holes) evolves to a geometry locally describable by a unique
member of the Kerr family (again modulo perturbations from the exterior
universe). Other than intrinsic theoretical interest to understand
merger geometries, finding numerical solutions for specific examples can provide strong evidence
for these assumptions. However, the most pressing reason to study
the binary black hole problem in recent years has been to support the effort to observe the universe
in the gravitational wave spectrum. As discussed in more detail elsewhere in this volume (see Chapter 6), 
theoretical models of expected waveforms are necessary for successful detection and to decipher the properties
of sources. A host of tools have been developed to tackle this problem for black hole mergers,
including post Newtonian expansions, \index{post Newtonian expansions} black hole perturbation theory, \index{black hole perturbation theory} the effective one body (EOB) approach, \index{effective one body approach} and the geodesic self-force problem \index{geodesic self-force problem} applicable to extreme mass
ratio mergers. For comparable mass ratio mergers,
perturbative methods break down near coalescence, and this is where numerical relativity contributes
most to the problem. The rest of this section is devoted to an overview of what has
been learned about these final stages of the merger from numerical solutions, restricting to the
four spacetime-dimension case. For more detailed reviews see~\cite{Pretorius:2007nq,2010ARNPS..60...75C}.

\begin{figure}
\hspace{-.7cm}
\includegraphics[width=2.35in,clip]{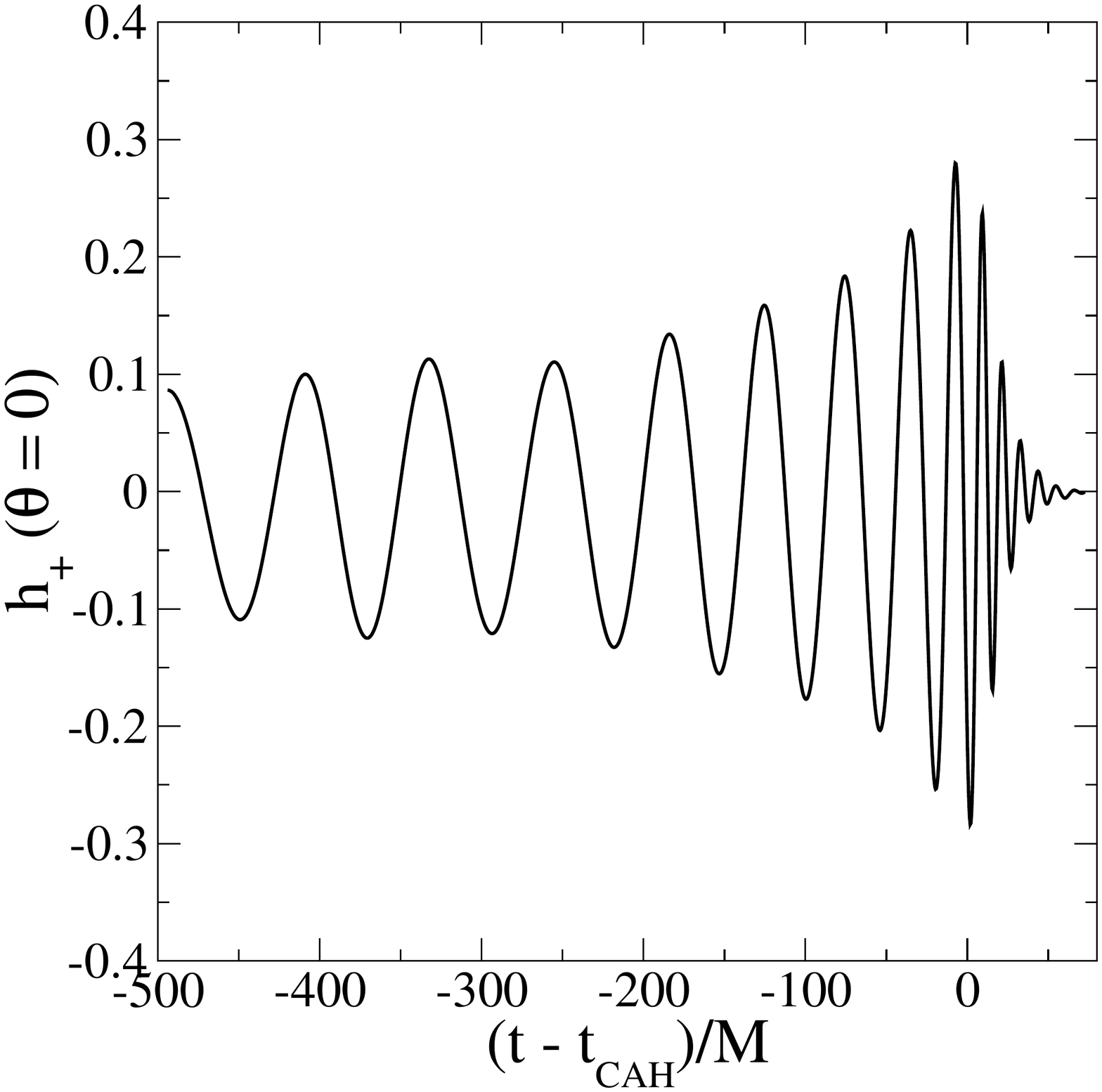} \hspace{0.2cm}
\includegraphics[width=2.35in,clip]{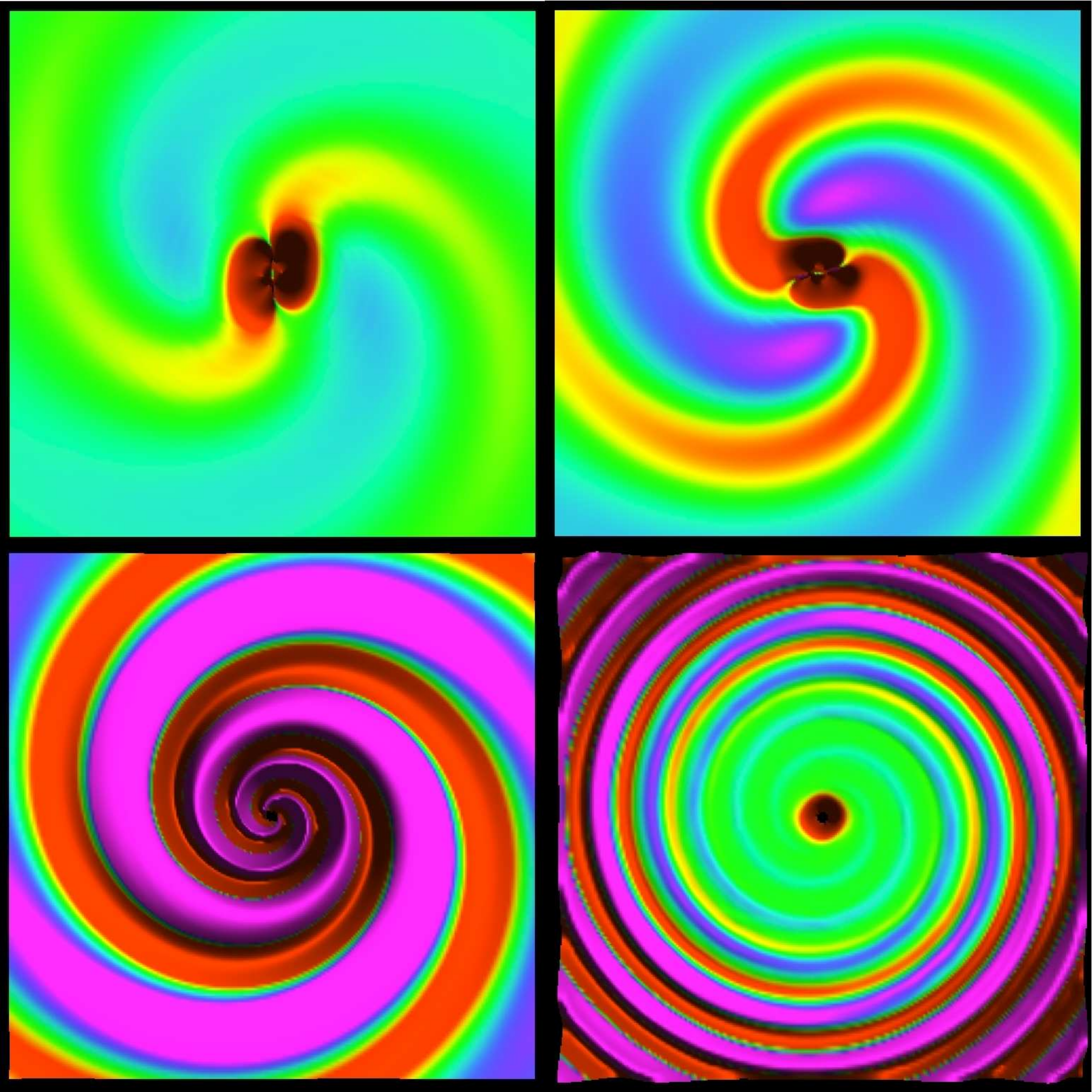}
\caption[Gravitational wave emission from the simulation of an equal mass binary black hole merger]
{Depictions of the gravitational waves emitted during
the merger of two equal mass (approximately) non-spinning black 
holes~\cite{Buonanno:2006ui}. Left: The plus-polarized component $h_+$
of the wave measured along the axis perpendicular to the orbital
plane. $t_{CAH}$ on the horizontal axis is the time
a common apparent horizon is first detected. Right: A color-map
of the real component of the Newman-Penrose scalar $\Psi_4$ (proportional
to the second time derivative of $h_+$ far from the BH) multiplied
by $r$ along a slice through the orbital plane
(grey is $0$, toward white (black) positive (negative)).
From top left to bottom right the time $(t-t_{CAH})/M$ of each panel
is approximately $-150,-75,0,75$.}
\label{fig:bhbh1}
\end{figure}

One of the results that was immediately obvious from the first full merger simulations
of equal mass, non-spinning black holes~\citep{Pretorius:2005gq,Campanelli:2005dd,Baker:2005vv}, and 
since then for the large swath of parameter space simulated (see for example~\cite{Hinder:2013oqa}), 
is the relative simplicity of the structure in the emitted waves
during the transition from inspiral to ringdown (see the left panel of Fig.~\ref{fig:bhbh1}). 
This is the regime of evolution where the strongest-field
dynamics is manifest, and the perturbative approaches applicable before and after should be least reliable.
Certainly the perturbative inspiral calculations do break down evolving forwards to merger, and similarly
for extending the quasi-normal ringdown backwards to this time. However, there is no significant intermediate regime
of dynamics between the two, and with guidance from the numerical simulations, the perturbative waveforms
can be stitched together with relatively simple matching conditions
(this is a rapidly advancing sub-field; see~\citep{2011PhRvL.106x1101A,2013arXiv1307.6232P,2013PhRvD..87h4035D} for a few
recent examples at the time this chapter was written).

With regards to the issues of theoretical interest discussed above,
no simulation has shown a violation of cosmic censorship, and the final state, to within
the accuracy of the simulations and the level that researchers have scrutinized the geometry, is a member of the
Kerr family. Moreover, though it is unlikely that the quasi-normal mode spectrum of Kerr \index{quasi-normal mode spectrum of Kerr} is 
able to describe all possible perturbations, in cases studied
to date the post-merger waveforms can indeed be well approximated as a sum of quasi-normal
modes. Of course, here we have a rather restrictive class of astrophysically minded ``initial conditions'' for the perturbed Kerr black hole formed by the merger of two black holes. We note that a couple of studies of single black holes perturbed by 
gravitational waves have also been studied numerically beyond
the linear regime, and similar conclusions hold~\citep{1997PhRvD..56.6298B,2011PhRvD..83j4018C}.

\begin{figure}
\includegraphics[width=2.5in]{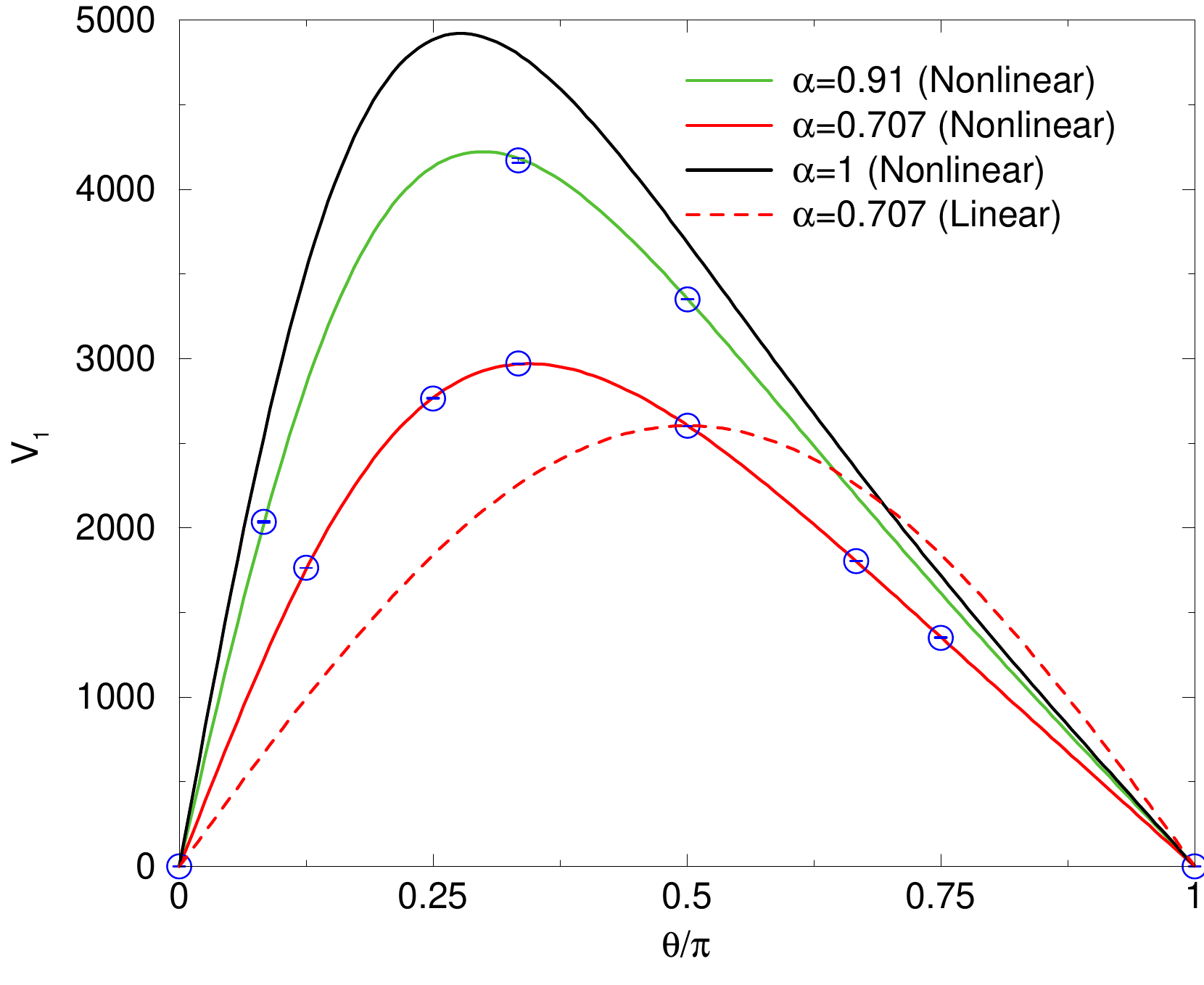} 
\caption[Recoil velocities from equal mass, spinning binary black hole merger
simulations]{Recoil velocities from equal mass, spinning binary black hole merger
simulations (circles) together with analytical fitting functions.
Each black hole has the same spin magnitude $\alpha$, equal but opposite
components of the spin vector within the orbital plane, and $\theta$ is
the initial angle between each spin vector and the orbital angular momentum. 
The dashed line corresponds
to a fitting formula that depends linearly on the spins, while solid lines add non-linear spin contributions
(from~\cite{Lousto:2011kp}).
\label{kick}}
\end{figure}

Some of the more important numbers that have been provided by numerical simulations
include the total energy and angular momentum radiated during merger (and consequently
the final mass and spin of the remnant black hole), the spectra of quasi-normal modes
excited, and the recoil, or ``kick'' velocity \index{recoil velocity} of the final black hole to balance 
net linear momentum radiated. It is beyond the scope of this chapter to list
all these numbers. However in brief, for a baseline reference, it has been found that two equal mass, non-spinning black holes
beginning on a zero eccentricity orbit at ``infinite'' separation radiate $\sim4.8\%$ of the
net gravitational energy during inspiral, merger and ringdown, ultimately becoming a Kerr black hole
with dimensionless spin parameter $a\sim0.69$ (due to the symmetry
of this system, there is zero recoil). The waveform spectrum is dominated by the quadrupole mode in a 
spin-weight 2 spheroidal harmonic mode decomposition; the next-to-leading order is the octupole mode,
which is strongly sub-dominant, though it briefly grows to an amplitude around $1/5$th that
of the quadrupole mode near merger~\citep{Buonanno:2006ui} (the energy of a mode scales as its amplitude squared).
Changing the mass ratio decreases the energy radiated by roughly the square of the symmetric mass ratio $\eta$,
the final black hole spin drops linearly with $\eta$, new multipole moments in the waveform are
introduced (reflecting the quadrupole moment of the effective energy distribution of the two particle source), and can 
produce recoil velocities as high as $\sim175{\rm km/s}$~\citep{Gonzalez:2006md,Baker:2006vn,2007CQGra..24S..33H,Berti:2007fi}.
Introducing spin for the initial black holes can alter the radiated energies by up to a factor of roughly 2 (higher for spins
aligned with the orbital angular momenta, lower otherwise)~\citep{Campanelli:2006uy},
increase (decrease) the final spin for initial spin aligned (anti-aligned) with the orbital angular
momentum (the largest aligned spin cased simulated to date begins with equal initial spins of $a\sim0.97$, merging
to a black hole with $a\sim0.95$~\citep{Hemberger:2013hsa}), 
introduces precession of the orbital plane which correspondingly modulates the multipole structure of the waveform observed
along a given line of sight~\citep{2011PhRvD..84b4046S,2011PhRvD..84l4011B,2011PhRvD..84l4002O}, 
and perhaps most remarkably can produce recoil velocities \index{recoil velocity} of several
thousand km/s for appropriately aligned high-magnitude spins~\citep{2007ApJ...659L...5C,2007PhRvL..98w1101G}.
Figure~\ref{kick} illustrates some of the results obtained for equal mass, fast spinning binary black holes.

\begin{figure}
\hspace{0.3in}
\includegraphics[width=2.3in,height=2in,clip]{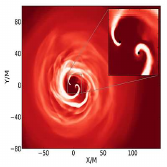} \hspace{0.25in}
\includegraphics[width=2.0in,clip]{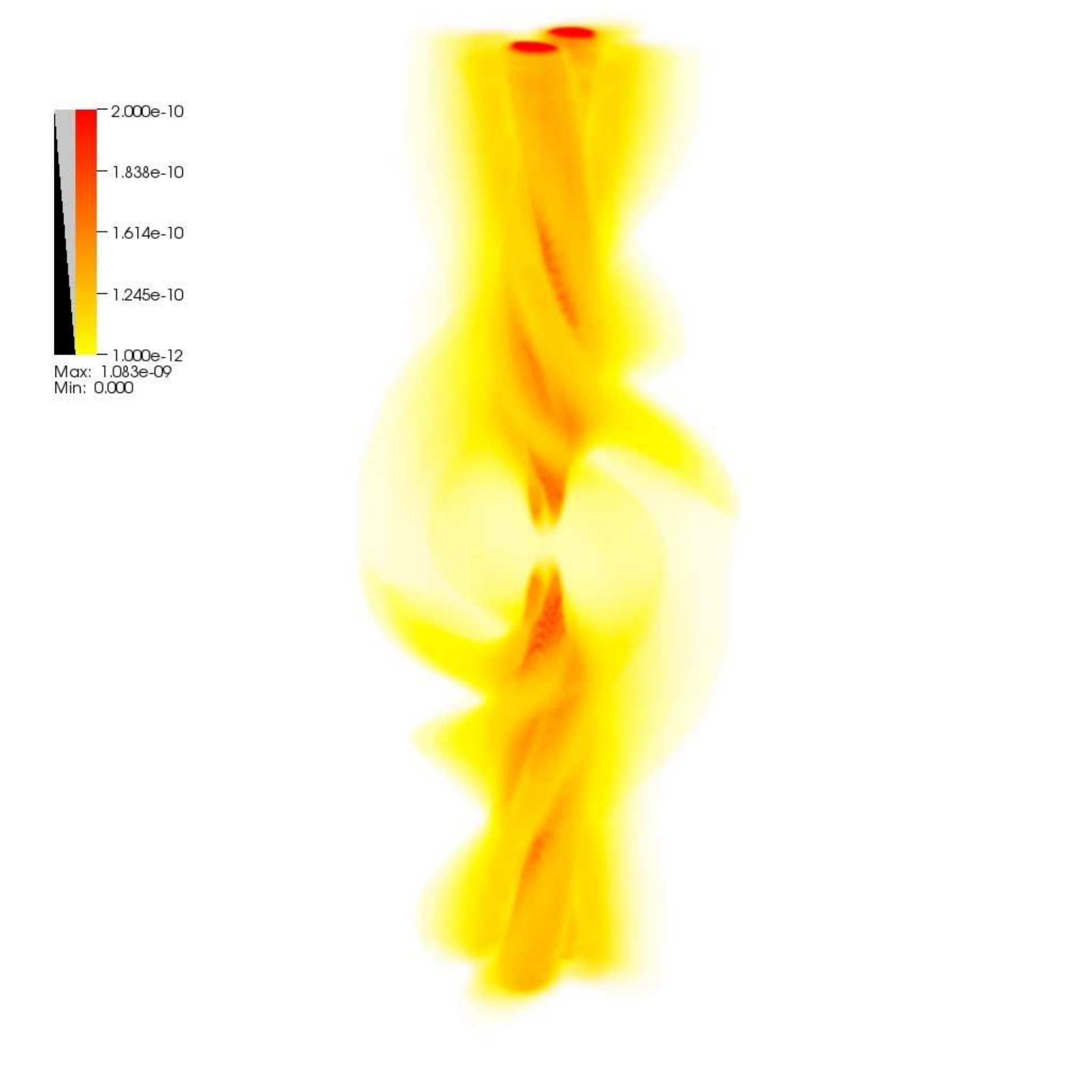} \hspace{0.0in}
\caption[Interaction of a binary black hole with a magnetized accretion disk]
{Left: Rest-mass density induced by a supermassive black hole binary 
interacting with a magnetized disk prior to when the binary ``decouples'' from the disk, namely
when the gravitational wave backreaction timescale becomes smaller than the viscous timescale
(from~\cite{2012PhRvL.109v1102F}).
Right: Poynting flux produced by the interaction of an orbiting binary black hole
interacting with a surrounding magnetosphere. The ``braided'' jet structure
is induced by the orbital motion of the black holes (from~\cite{Palenzuela:2010nf}). 
\label{gasandjets}}
\end{figure}

There are many astrophysical consequences of large recoil velocities, in particular for supermassive
black hole mergers; \index{super massive black mergers} we briefly mention a few here, together with some broader consequences
of mergers on surrounding matter (for recent more detailed reviews 
see~\cite{2012AdAst2012E..14K,2013arXiv1307.3542S}). 
First, the velocities for near equal mass, high spin mergers are large enough
to significantly displace the remnant from the galactic core, or for the highest velocities
even eject the black hole from the host galaxy altogether. This may be in some tension with observations
that seem to suggest that all sufficiently massive galaxies harbour supermassive black holes.
If the system has a circumbinary accretion disk, the recoil would carry the inner part of the 
disk with it, and this could be observable in Doppler-displaced emission lines relative to the galactic
rest frame~\citep{Komossa:2008qd}. The near-impulsive perturbation to the gravitational potential
in the outer parts of the accretion disk could lead to the formation of strong shocks,
producing observable electromagnetic emission on timescales of a month to a year afterwards~\citep{Lippai:2008fx}. 
(Note that regardless of the recoil, the entire accretion disk will experience an impulsive
change in potential due to the near instantaneous loss of energy from gravitational wave
emission at merger, also producing electromagnetic emission post-merger~\citep{Milosavljevic:2004cg}).
Earlier studies have suggested that prior to merger the accretion rate, and hence the luminosity of the nucleus, would be low as the relatively slow migration of the inner edge of the accretion disk decouples
from the rapidly shrinking orbit of the binary. Post merger then, AGN-like emission could be re-ignited once
the inner edge of the disk reaches the new innermost stable circular orbit (ISCO) \index{inner most stable circular orbit (ISCO)} of the 
remnant black hole. This will be displaced from the
galactic center if a large recoil occurred, and could be observable in nearby 
galaxies (see for example~\citep{2007PhRvL..99d1103L}).
However, more recent simulations of circumbinary disks using ideal magnetohydrodynamics for the matter
shows that complete decoupling does not occur, and relatively high accretion rates can be maintained
all the way to merger~\cite{2012ApJ...755...51N,2012PhRvL.109v1102F}. (The left panel of Fig.~\ref{gasandjets}
illustrates a binary black hole system accreting surrounding gas). The binary orbit
can cause a modulation in the induced luminosity of the system, which may be observable.
A displaced central black hole will also have its loss-cone refilled, increasing the frequency
of close encounters with stars and their subsequent tidal disruption by the black hole, with rates
as high as $0.1/{\rm yr}$; the disruption could produce observable electromagnetic emission~\citep{2011MNRAS.412...75S}.
Yet another exciting prospect for an electromagnetic counterpart
is an analog of the standard Blandford-Znajek mechanism (to extract rotational
energy from a spinning black hole) induced by a tightening binary within a circumbinary disk. In particular, numerical simulations 
have uncovered that binary black holes can extract both rotational and translational kinetic
energies when there is surrounding plasma~\cite{Palenzuela:2010nf}. This not only can
power strong dual Poynting jets (emanating from each black hole), but the jets will increase in strength until
merger, making them indirect ``spacetime tracers''---the right panel of Fig.~\ref{gasandjets}
depicts the resulting ``braided'' structure of the Poynting flux.

As a final comment we note that the majority of work, both numerical and analytic, has been devoted to studying
zero-eccentricity mergers, due to the prevailing view that these will dominate event rates. However, there
are binary formation mechanisms that can produce high-eccentricity mergers (see the discussion in~\citep{East:2012xq}
for an overview and references). One of the interesting results from the handful of studies including large 
eccentricity performed to date~\citep{Pretorius:2007jn,2009PhRvL.103m1101H,2012arXiv1209.4085G} is that zoom-whirl orbital \index{zoom-whirl orbital dynamics}
dynamics is possible for comparable mass binaries. In the test particle limit, zoom-whirl orbits are perturbations
of the class of unstable circular geodesics that exist within the ISCO; further, they exhibit 
extreme sensitivity to initial conditions where sufficiently fine-tuned data can exhibit an arbitrary number
of near-circular ``whirls'' at periapse for a fixed eccentricity geodesic.
Away from the test particle limit gravitational wave emission
adds dissipation to the system, though what the simulations show is that even in the comparable mass limit the
dissipation is not strong enough to eradicate zoom-whirl dynamics, but merely limits how long it can persist.

\subsection{Black Hole-Neutron Star/Binary Neutron Star Mergers}\label{sec_BHNS_BNS}
Non-vacuum compact binary systems--i.e., those involving at least one neutron star--are also the subject
of intense scrutiny. These systems produce powerful gravitational waves and likely
also lead to intense neutrino and electromagnetic emission that 
could be detected by transient surveys or by dedicated follow up by the 
astronomical community.
In particular they are posited to be the progenitors
of short gamma ray bursts (sGRBs) \index{short gamma ray bursts} and a host of other transient phenomena~\cite{Metzger:2011bv,2013MNRAS.430.2121P}.
Signals from these systems can thus carry a wealth of information about gravity,
the behavior of matter at nuclear densities, and binary populations and their environments.
The challenge for simulations is to obtain predictions to confront with observations. 

Relative to the two black hole case, the most obvious complication in the simulation 
of binaries with neutron stars \index{neutron stars} is the need to include non-gravitational physics.
The simplest relativistic model of a neutron star couples
relativistic hydrodynamics to the Einstein equations and, using a simplified
equation of state (EOS), \index{equation of state (EOS)} the first successful simulations of binary
neutron star mergers within this framework were presented in~\cite{Shibata:1999wm,Nakamura99a}.
Since the time of those studies, the community has made steady
progress in exploring the full parameter space relevant to astrophysical mergers, while
simultaneously 
increasing the fidelity of the matter modeling through inclusion of the 
electromagnetic interaction, neutrino and radiation transport, nuclear reactions, and other
physics.
A crucial unknown here is the EOS that describes matter at nuclear densities: it
plays a leading role in the phenomenology of the system as, for a given stellar mass, it regulates the star's radius, 
affects its response to tidal forces, and affects its ability to resist collapse to a black hole when it accretes matter (or collides with another star).
Given the difficulty of first-principles calculations or probing similar conditions in laboratories,
detailed knowledge of the nuclear density EOS is likely to come only through astronomical observations, 
and prospects
for doing this through gravitational waves are particularly exciting---see for example~\cite{Read:2013zra,Lackey:2013axa}.
 
\begin{figure}
\includegraphics[height=3.99cm,angle=0]{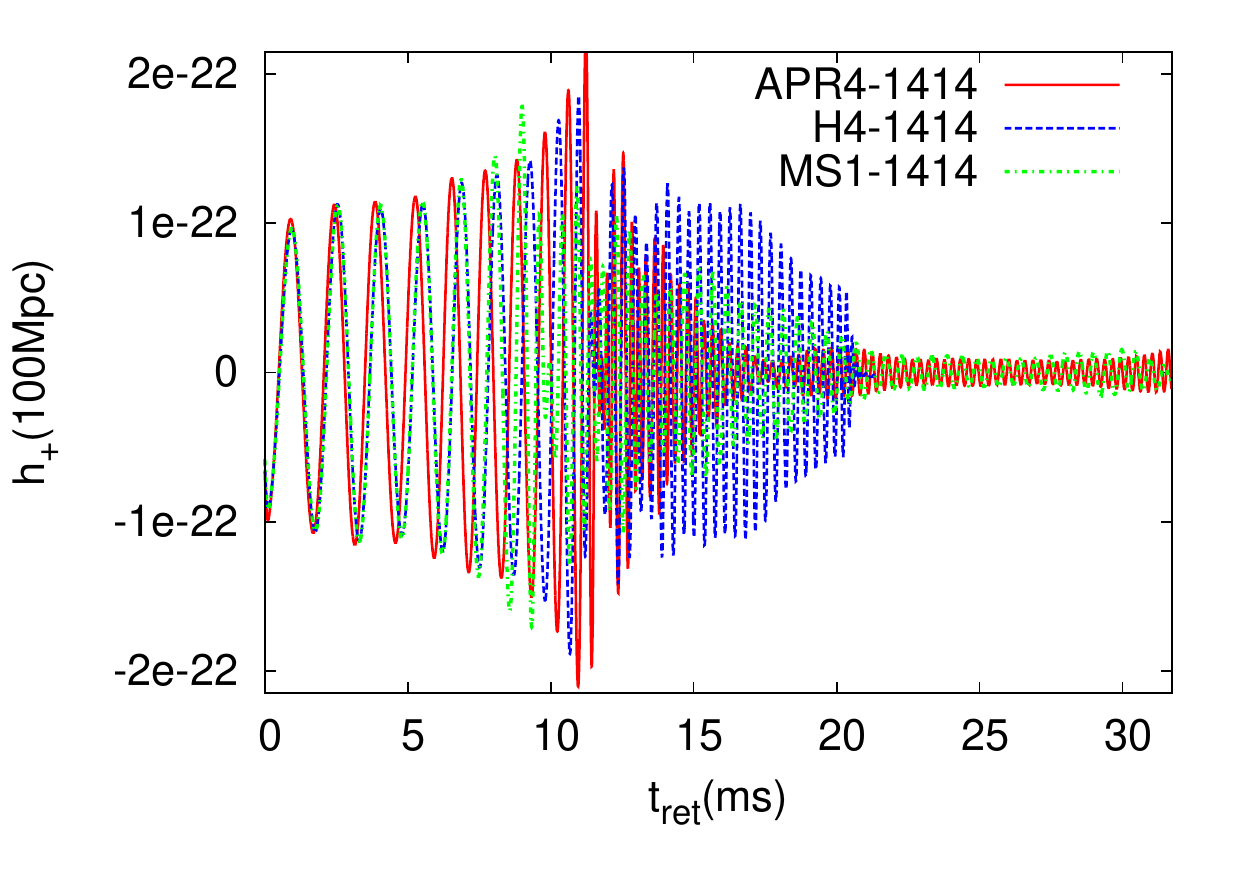}
\includegraphics[height=4.01cm,angle=0]{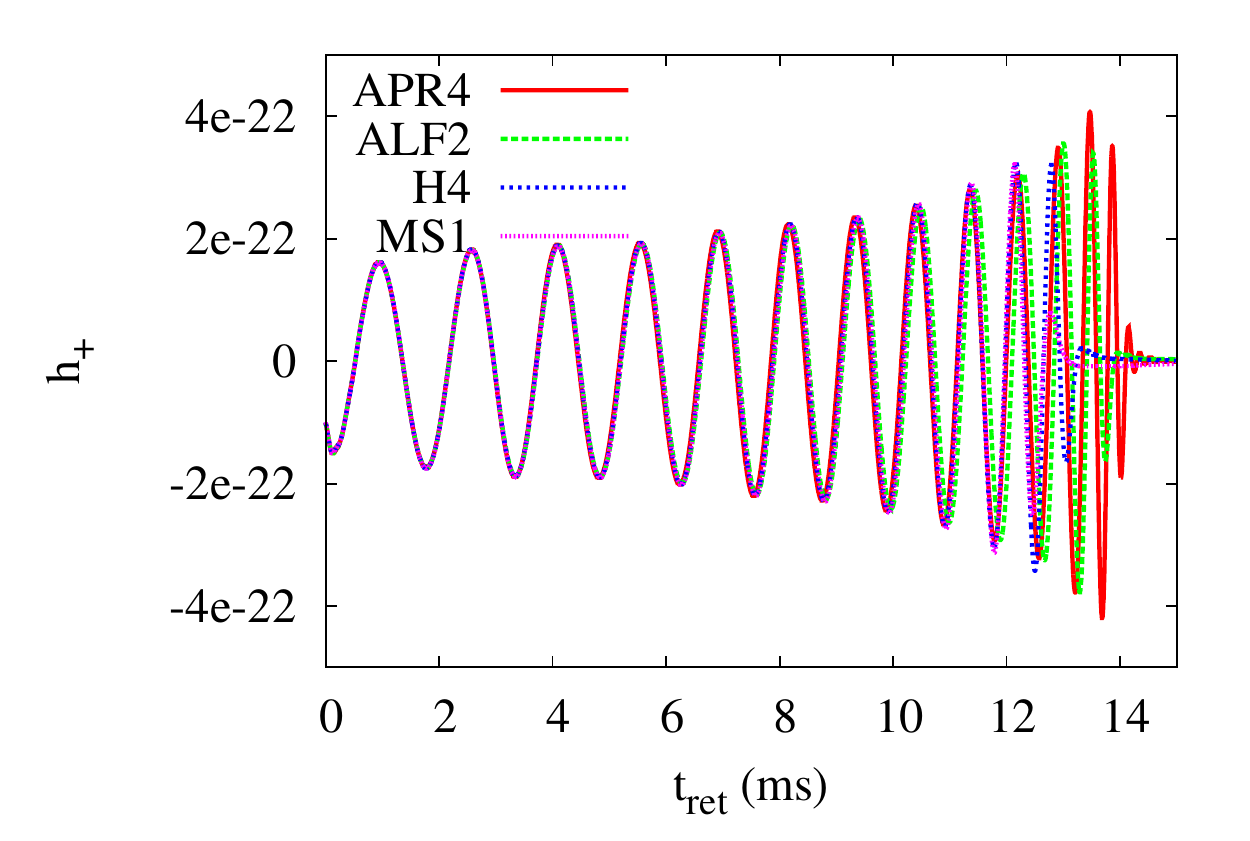}
\caption[Gravitational wave emission from binary neutron star merger simulations]
{Examples of the ``plus'' polarization component of gravitational waves from binary neutron star mergers,
measured $100$ Mpc from the source along the direction of the orbital angular momentum. 
The different curves correspond to different choices of the EOS of the neutron star matter,
labeled APR4, ALF2, H4 and MS1.
For a $1.4M_{\odot}$ neutron star, the APR4, ALF2, H4, MS1 EOS give radii of $11.1,12.4,13.6,14.4$km respectively.
Left: Mergers of an equal mass binary neutron star system (with $m_1=m_2=1.4M_{\odot}$). A hypermassive
neutron star (HMNS) is formed at merger, but how long it survives before collapse to a black hole
strongly depends on the EOS. The H4 case collapses to a black hole $\approx 10$ms after 
merger; the APR and MS1 cases have not yet collapsed $\simeq 35$ms after merger when the
simulations where stopped (the MS1 EOS allows a maximum total mass
of $2.8 M_{\odot}$, so this remnant may be stable). The striking difference in gravitational wave signatures
is self evident (from~\cite{2013PhRvD..88d4026H}). Right: Emission from black hole-neutron star mergers, with 
$m_{{\rm BH}} = 4.05 M_{\odot}, m_{{\rm NS}} = 1.35 M_{\odot}$. Variation with EOS is primarily due to coalescence taking
place earlier for larger radii neutron stars (from~\cite{Kyutoku:2013wxa}).}
\label{nonvacuumwaves}
\end{figure}

While the pericenter is large these systems evolve much like black hole binaries. The orbit shrinks due to
the emission of gravitational radiation with internal details of the stars playing essentially no role. However,
finite body effects become important as the orbit tightens. In the remainder of this
paragraph we focus on binary neutron stars, \index{binary neutron stars} returning to black hole-neutron star systems in the following paragraph.
Tidal forces deform both stars (which
can even induce crust-shattering~\cite{Tsang:2011ad}), leaving subtle imprints in the ensuing gravitational waves. This
behavior intensifies until the point of merger, when the local velocities reach a sizable
fraction of the speed of light, ending in a violent collision that ejects neutron rich matter due to shock heating
and extreme tidal forces. Figure~\ref{nonvacuumwaves} (left panel) illustrates waveforms obtained 
in an equal mass binary neutron star system for different EOS models, demonstrating how significantly this can
affect the behavior. In general terms, 
for the typically expected neutron star masses of $1.2 - 1.8 M_{\odot}$ the merger yields a hot, 
differentially rotating, hyper-massive neutron star (HMNS). \index{hyper-massive neutron star (HMNS)} Such an object will promptly collapse to a black hole
if the total binary mass is above $2.6-2.8 M_{\odot}$, depending on the stiffness of the EOS. Otherwise, a delayed collapse
takes places as the star is initially supported by differential rotation and thermal pressure. During this stage, the HMNS
rotates and emits gravitational waves with frequencies in the range $2 \lesssim f \lesssim 4$Khz, with a characteristic 
frequency proportional (and relatively close) to the Keplerian velocity $(M_{{\rm HMNS}}/R^3_{{\rm HMNS}})^{1/2}$ (e.g.,~\cite{Hotokezaka:2011dh}). 
On a scale of tens of milliseconds however, such support diminishes due to gravitational radiation, angular momentum transport via 
hydrodynamical and electromagnetic effects, and cooling due to neutrino emission 
(these effects have just begun to be studied, e.g.~\citep{Anderson:2008zp,Sekiguchi:2012uc,Kaplan:2013wra}).
The black holes that form in both prompt or delayed cases are (reasonably) well-described by a Kerr solution with a spin 
parameter $J/M^2 \lesssim 0.8$, surrounded by left-over material, much of which is bound (e.g.,~\cite{Rezzolla:2010fd}) and can
form an accretion disk with a mass on the order of $\simeq 0 - 0.3 M_{\odot}$. The amount depends on the EOS, mass ratio, and electromagnetic
fields (though this latter effect is still largely unexplored) and is intuitively larger for longer lived HMNS as more angular momentum
is transferred outwards to the envelope.  Importantly, this is enough
material to form a sufficiently massive disk as called for in models of sGRBs.
Some material will be ejected (again, the amount depending upon various parameters) and will decompress to form heavy elements through
the r-process; if these merger events are frequent this could account for a significant fraction,
if not the majority of such elements in the Universe. Subsequent decay of the more radioactive isotopes could lead
to a so-called kilo- or macronova (reports of the afterglow of the recent sGRB 130603B are
consistent with this~\cite{2013Natur.500..547T,2013ApJ...774L..23B}).
Observation of these signatures together with gravitational wave observations will allow us
to make contact between simulations and the birth of a black hole. However, gravitational waves emitted during the HMNS and
collapse stages have a higher frequency than those reachable by LIGO/VIRGO/KAGRA, \index{LIGO/VIRGO/KAGRA} and will take third-generation 
facilities to detect.
Nevertheless, up to the frequencies that existing (and near future) detectors can probe, subtle differences in the gravitational waveforms 
should allow for constraining the radius of the neutron stars to within $10\%$~\citep{Hinderer:2009ca,Markakis:2011vd}.
Simulations are further probing possible counterpart signals 
from neutrino production~\cite{Sekiguchi:2011zd} and
electromagnetic interactions~\cite{Lehner:2011aa,Kyutoku:2012fv,Palenzuela:2013hu}.

\begin{figure}
\includegraphics[width=4cm]{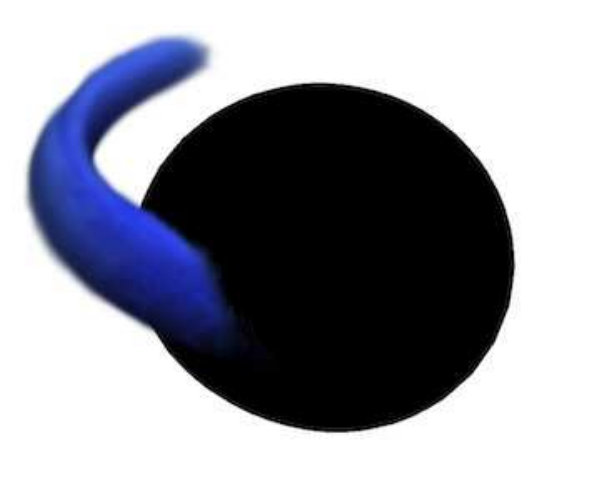}
\includegraphics[width=4cm]{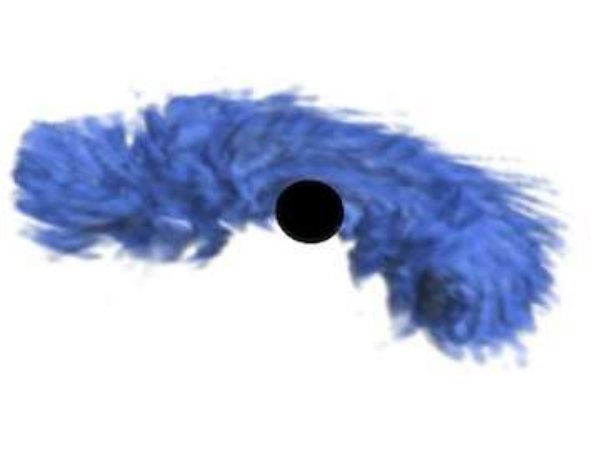}\\
\includegraphics[width=4cm]{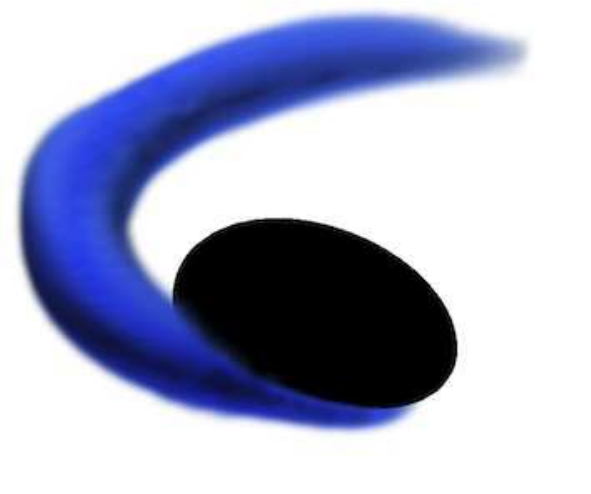}
\includegraphics[width=4cm]{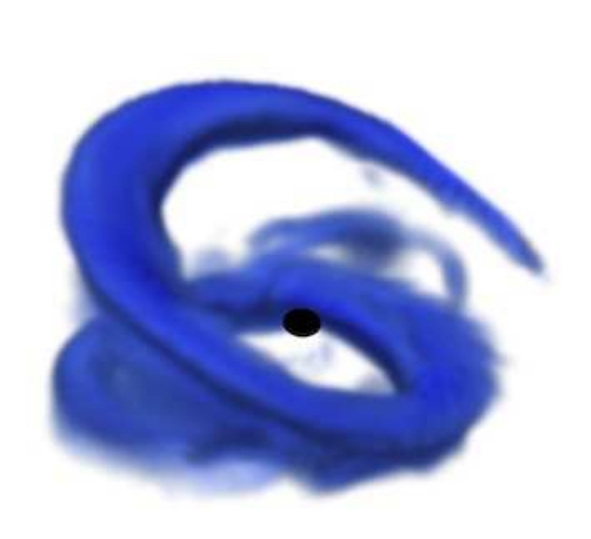}
\caption[Tidal disruption of a neutron star merging with a black hole]
{General relativistic hydrodynamic simulations of the merger of a $9.8 M_\odot, a=0.9$ black hole
with a $1.4 M_\odot$ neutron star, from~\cite{Foucart:2012vn}. The top two panels
are from a case where the spin and orbital angular momentum vectors are aligned; the bottom two where the initial ($\sim9$
orbits before merger) misalignment is $40^\circ$. The left two panels are at a time when half the material
has been absorbed by the black hole, showing matter densities above $\sim 6\times 10^{10}{\rm g/cm^3}$; the right two are $5$ms later,
showing densities above $\sim 6\times 10^{9}{\rm g/cm^3}$, and the facing quadrant has been cut from the top-right rendering.
These results illustrate the profound affect spin-induced precession can have on the matter disruption
and subsequent accretion.}
\label{fig:bhns_focaurt}
\end{figure}

Black hole-neutron star binaries \index{black hole-neutron star binaries} display even more complex merger dynamics. Indeed, at an intuitive level
one expects significant differences to arise depending on whether the tidal radius \index{tidal radius $R_T$}
$R_T$ ($\propto R_{{\rm NS}} \left(3M_{{\rm BH}}/M_{{\rm NS}}\right)^{1/3}$) 
lies inside or outside the black hole's inner most  stable circular orbit radius ($R_{{\rm  ISCO}}$), which ranges from $M_{{\rm BH}}$ to $9 M_{{\rm BH}}$ for a 
prograde to retrograde orbit about a maximally spinning black hole.  This is clearly borne out in simulations 
exploring a range of mass ratios and black hole spins, showing markedly different behavior in the ensuing dynamics
and gravitational waves produced. Qualitatively,
for sufficiently high spins and/or sufficiently low mass ratios, the star significantly disrupts instead of plunging
into the black hole. As a result, gravitational waves promptly ``shut-off'' at a frequency related to the star's EOS. 
Figure~\ref{nonvacuumwaves} (right panel) illustrates this for different EOS models
in a  $3:1$ mass ratio black hole-neutron star system. When disruption occurs during the merger, 
a significant amount of material, in the range $0.01-0.3 M_{\odot}$, can remain outside $R_{{\rm ISCO}}$. 
This material will be on trajectories having a range of eccentricities,
with the fraction that is bound falling back to accrete onto the black hole
at a rate governed by the familiar law $\dot M \propto t^{-5/3}$~\cite{Chawla:2010sw,Foucart:2012vn}. The details however depend on 
many factors, including spin-orbit
precession \index{spin-orbit precession} as illustrated in Fig.~\ref{fig:bhns_focaurt}. The matter that is ejected ($\lesssim 0.05 M_{\odot}$) can be have speeds up to
$\simeq 0.2c$~\cite{Foucart:2012nc,Kyutoku:2013wxa}. This, together with the amount of likely accretion,
is in the range assumed by models predicting that black hole-neutron stars mergers can power sGRBs, kilonovae and related 
electromagnetic counterparts. Consequently, 
a similar array of electromagnetic signatures and r-process \index{r-process} elements could result as with binary neutron star mergers,
and the gravitational wave signals could be ideal to differentiate between them.
For the subset of black hole neutron star mergers where $R_T \lesssim R_{{\rm ISCO}}$, the star plunges into the black hole 
with little or no material left behind, and the resulting 
gravitational wave signal will be much like that of a binary black hole system with the same binary parameters. Counterparts
such as sGRBs or kilonova requiring significant accretion disks or unbound matter are therefore not
favored for this sub-class of binary.  Nevertheless, interesting electromagnetic
precursors could be induced by magnetosphere-black hole interactions prior to merger (e.g.~\cite{Hansen:2000am,McWilliams:2011zi,Paschalidis:2013jsa}). 

An important observation is that black hole-neutron star systems are, in all likelihood, more massive than binary neutron star
systems. Therefore, the wave frequency peaks at lower frequencies than binary neutron star mergers,
offering better prospects for observing non-linear effects by near-future
 detectors. Indeed, since the characteristics of gravitational waves
depend on masses, spins and the EOS, black hole-neutron star systems provide perhaps the best prospects for extracting key physical information about
neutron stars~\cite{Lackey:2013axa,Lackey:2011vz}. To date the majority of simulations have focused on a black hole spin aligned with the 
orbital angular momentum, with the exception of~\citep{Foucart:2012vn}, which showed that the above conclusions hold qualitatively
even with inclinations $\lesssim 30^o$ of the spin axis away from alignment. For larger inclinations, of the disrupted material
a smaller fraction forms a disk on a
short timescale following merger, while a larger fraction follows an eccentric trajectory and returns to interact with the black hole
on longer timescales.

As in the binary black hole case, incipient efforts are examining encounters with high initial orbital
eccentricity in non-vacuum binaries (e.g.~\cite{bhns_astro_letter,Gold2011}). Qualitatively, much
of the same phenomenology of outcomes can occur as with quasi-circular inspirals (except 
that now zoom-whirl orbital dynamics is also possible), though the details can be drastically different.
For example, in high eccentricity encounters of a neutron star with a black hole, tidal disruption
can occur for higher mass ratio systems and smaller black hole spin, as the effective inner most stable
orbit is closer to the black hole for eccentric orbits. There can also be multiple, partial disruptions
on each of the last several periapse passages, ejecting larger amounts of material and leaving behind
more massive accretion disks than otherwise possible. On close periapse passages (even without disruption)
f-modes \index{f-modes} can be impulsively excited in the star, or both stars in a binary neutron star encounter.
These modes are too low amplitude/high-frequency to be directly observed with
the current generation of ground-based gravitational wave detectors, though they may indirectly
be measured in the leading order part of the waveform, since from the perspective of the binary the
f-modes are a new channel of energy dissipation. The impulsive tidal interaction may also
cause crust-shattering~\cite{Tsang:2013mca}, leading to electromagnetic emission similar to the resonant excitation
induced shattering in quasi-circular inspiral~\cite{Tsang:2011ad}.

For merger simulations involving neutron stars, the current frontier is to add more matter physics to the 
models (resistive magnetohydrodynamics, radiation and neutrino physics, multi-component fluids, ``realistic''
high temperature equations of state, etc.).
Given the many orders of magnitude of spatial and temporal scales involved, as well as the complexity of the microphysics, 
it will likely be several years before both realistic models and the computational power necessary to simulate 
them accurately are available. Due to space constraints we will not list all the directions
currently being pursued, referring the reader to recent reviews in~\cite{Duez:2009yz,Pfeiffer:2012pc,Faber:2012rw}.

\subsection{Gravitational Collapse to a Neutron Star or Black Hole}\label{sec_collapse}
\begin{figure}
\includegraphics[width=4.5in]{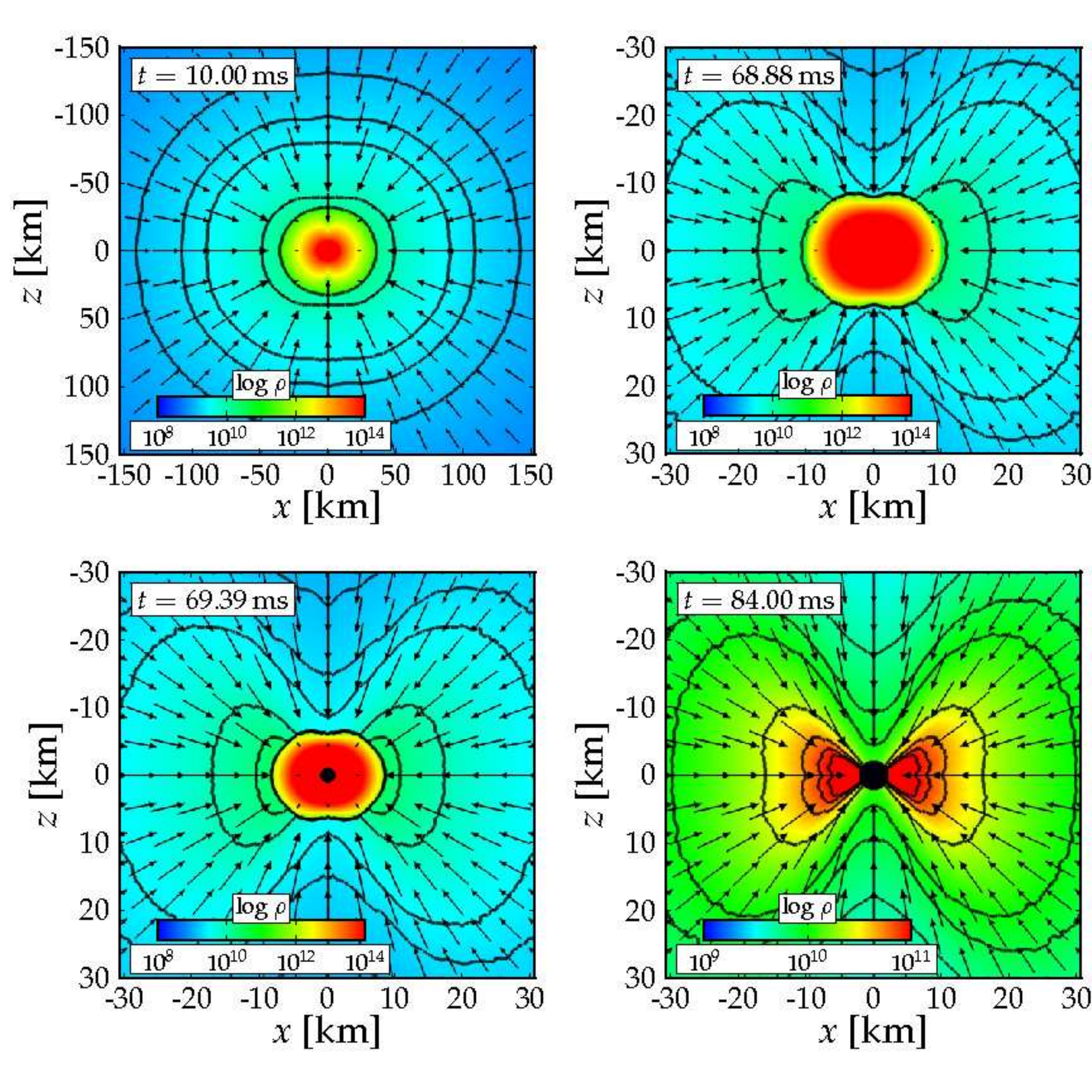}
\caption[Simulation of stellar collapse to a black hole]
{Density colormaps of the meridional plane of a collapsing
$75 M_{\odot}$ star superposed with velocity vectors at various
times after bounce (and with different spatial ranges to zoom-in on particularly relevant behavior). 
The collapse first forms a proto-neutron star which
later collapses to a BH (shown in the bottom panels). (From~\cite{2011PhRvL.106p1103O}).
}
\label{fig:core-collase}
\end{figure}

Considerable efforts have been undertaken to study gravitational collapse to a neutron star or a black hole,
in particular within the context of core-collapse supernovae. \index{core-collapse supernovae}
Here, stars with masses in the range $10 M_{\odot} \lesssim M \lesssim 100 M_{\odot}$ at zero-age main sequence
form cores which can exceed the Chandrasekhar mass \index{Chandrasekhar mass} and become gravitationally unstable. 
This leads to collapse which compresses the inner core to nuclear densities, at which point
the full consequences of general relativity must be accounted for. Depending upon the mass of the core, it
can ``bounce'' or collapse to a black hole. 
Figure~\ref{fig:core-collase} displays representative snapshots of the behavior of a collapsing $75 M_{\odot}$ star at different
times. The collapse forms a proto-neutron star which later collapses to a black hole.
In the case of a bounce, an outward propagating shock wave is launched which collides with still infalling material 
and stalls. Observations of core-collapse supernovae imply that  some mechanism is capable of reviving the shock, 
which is then able to plow through  the stellar envelope and blow up the star. This process is extremely energetic, 
releasing energies on the order of $10^{53}$erg,  the majority of which is emitted in 
neutrinos. For several decades now, the primary motivation driving
theoretical and numerical studies has been to understand what process (or combination of processes)
mediates such revival, and how (for a recent review see~\cite{2009CQGra..26f3001O}).
Several suspects have been identified: heating by neutrinos, (multidimensional) hydrodynamical instabilities, magnetic
fields and nuclear burning (see e.g.~\cite{2007PhR...442...38J,2007PhR...442...23B}). With the very disparate
time and space scales involved, a multitude of
physically relevant effects to consider, and the intrinsic cost to accurately model them
(e.g., radiation transport is a $7$-dimensional problem) progress
has been slow. Moreover, electromagnetic observations
do not provide much guidance to constrain possible mechanisms, as they can not peer deep into the central
engine. On the other hand, observations of gravitational waves and neutrinos have the potential to do so, provided
the explosion is sufficiently close to us. Thus, in addition to exploring mechanisms capable of reviving the stalled
shock, simulations have also concentrated on predicting specific gravitational wave and neutrino signatures.

Modeling gravity using full general relativity has only recently been undertaken~\citep{Ott:2012mr}, though prior
to this some of the more relevant relativistic effects were incorporated
(e.g.~\cite{Dimmelmeier:2002bk,Obergaulinger:2006qr,2010ApJS..189..104M,Wongwathanarat:2012zp}).
While the full resolution of the problem is still likely years ahead, 
interesting insights into fundamental questions and observational prospects have been garnered.
For example, simulations have shown that in rotating core collapse scenarios, gravitational waves
can be produced and their characteristics are strongly dependent on properties of the collapse:
the precollapse central angular velocity, the development of non-axisymmetric rotational instabilities, postbounce
convective overturn, the standing accretion shock instability (SASI), protoneutron star pulsations, etc. 
If a black hole forms, gravitational wave
emission is mainly determined by the quasi-normal modes \index{quasi-normal modes} of the newly formed black hole.
The typical frequencies of gravitational radiation can lie in the range $\simeq 100-1500$Hz, and so are potential sources 
for advanced earth-based gravitational wave detectors (though the amplitudes are sufficiently
small that it would need to be a galactic event). As mentioned, the characteristics of these waveforms
depend on the details of the collapse, and hence could allow 
us to distinguish the mechanism inducing the explosion. 
Neutrino signals have also been calculated,
revealing possible correlations between oscillations of gravitational waves
and variations in neutrino luminosities. However, current estimates suggest neutrino detections
would be difficult for events taking place farther than kpc distances~\citep{Ott:2012mr}.

\subsection{Ultra-relativistic Collisions}\label{sec_URC}
Some of the early interest in the ultra-relativistic collision problem stemmed
from investigations by Penrose~\cite{Penrose} into its relevance to questions
of cosmic censorship. \index{cosmic censorship conjecture}  It was known that collisions of gravitational
waves with planar symmetry in 4d lead to the formation of naked singularities
regardless of how ``weak'' the initial curvature. This is not considered
a serious counter-example to cosmic censorship as the spacetime is not
asymptotically flat, nor for that matter are there black hole solutions
with planar symmetry in 4d vacuum Einstein
gravity (with zero cosmological constant), so in a sense the question of censorship is not particularly
meaningful here. However, taking the infinite boost limit of the Schwarzschild
metric (scaling the rest mass $m$ to zero as the boost 
$\gamma\rightarrow\infty$ while the energy $E=m\gamma$ remains finite
~\cite{Aichelburg:1970dh}) results in the Lorentz contraction of  the curvature to a plane-fronted gravitational shock wave, 
with Minkowski spacetime on either side. One can then consider
what happens when two such shock waves, traveling in opposite directions, collide.
Given the resemblance between the two scenarios, the infinitely
boosted black hole collision is a natural place to test cosmic censorship, especially
since the geometry approaches Minkowski spacetime transverse to the center
of each shock sufficiently rapidly to remove the trivial objections to the plane-symmetric gravitational
wave collisions. \index{plane-symmetric gravitational wave collisions} 
Penrose found a trapped surface \index{trapped surface} in a zero-impact parameter, infinite $\gamma$ black hole 
collision, and though the metric to the causal future of the collision is unknown,
this is a good indication that cosmic censorship does holds here.

More recently two additional lines of research have come to the fore motivating
the study of ultra-relativistic collision geometries. The first is a consequence
of the observation that if extra spatial dimensions exist, then the true Planck scale
could be much different from the effective 4-dimensional scale one would
otherwise expect~\cite{1998PhLB..429..263A,1999PhRvL..83.3370R}. In particular, a ``natural'' solution to the
hierarchy problem results if the Planck energy is on the order of a TeV.
If that is the case it was conjectured that particle collisions at the
Large Hadron Collider (LHC) \index{large Hadron Collider (LHC)} and in cosmic ray collisions with the earth with center of mass energies
above this could result in black hole formation~\cite{Giddings:2001bu,2002PhRvL..88b1303F}~\footnote{To date, 
the LHC has not seen evidence for black hole formation
in searches of collisions with center of mass energies up to 
8 TeV~\cite{Chatrchyan:2013xva,Aad:2013lna};  likewise, no signs of black hole formation have
yet been observed in cosmic ray collisions~\cite{2007astro.ph..1333D}, the most energetic of which can
have much larger center of mass energies.}.
The conjecture is essentially based on two premises: that Thorne's hoop conjecture~\cite{thorne_hoop} \index{hoop conjecture} can be
applied to the collision to deduce whether
the purely classical gravitational interaction between particles will
cause a black hole to form, and if so, that the quantum interactions are sufficiently
``local'' to not alter this conclusion (until Hawking evaporation becomes significant).
The second motivation comes from applications of the AdS/CFT correspondence \index{ADS/CFT correspondence}
of string theory to attempt to explain the formation and early time dynamics
(before hadronization) of the quark-gluon plasma formed in relativistic heavy
ion collisions (RHIC) \index{relativistic heavy ion collisions (RHIC)} (see Sec.~\ref{ADSCFT} for more on this). Here, the gravitational dual
to a heavy ion collision is conjectured to be an ultrarelativistic black hole collision 
in the bulk asymptotically AdS spacetime.

The first 4d ultra-relativistic head-on black hole collision simulations (up to $\gamma\approx 3$)
were carried out in~\cite{Sperhake:2008ga}, followed by several studies with general 
impact parameters~\cite{Shibata:2008rq,Sperhake:2009jz},
including the effects of black hole spin~\cite{Sperhake:2010uv,Sperhake:2012me}, and collisions in 
higher dimensions~\cite{Okawa:2011fv}.
A wealth of interesting 
results have emerged, a select few of which we briefly summarize here.
Outcomes of most interest to LHC searches
include the critical impact parameters for black hole formation, and the energy and angular
momentum lost to gravitational waves as a function of impact parameter. This determines the 
formation cross-section and initial spectrum of black hole masses that will subsequently Hawking evaporate. 
Extrapolated results from 4d head-on collisions give $14\pm3\%$ energy emitted in gravitational
waves, roughly 1/2 the Penrose trapped surface calculation, though consistent with the $16\%$ obtained
using perturbative analytic methods~\cite{1992PhRvD..46..694D}. As the impact
parameter increases, the radiated energy and now angular momentum increases, though the former
is still less than trapped surface estimates~\cite{Eardley:2002re}. Qualitative features of the spectrum
of emitted waves can be understood appealing to the analytic zero-frequency and point particle limit
calculations~\cite{Berti:2010ce}. The largest gravitational wave
fluxes arise near the threshold impact parameter. Here, in the 4d case, the binary exhibits
behavior akin to zoom-whirl dynamics of black hole geodesics, though not in 5d
(presumably due to the stronger effective gravitational potential, related to the fact
that there are no stable circular orbits about Myers-Perry black holes in dimensions 
greater than 4)~\cite{Sperhake:2009jz,Okawa:2011fv}.

Because of the zoom-whirl behavior in 4d, it was argued in~\cite{Pretorius:2007jn} that turning to the threshold
in the large $\gamma$ limit essentially all the initial kinetic energy of
the black holes would be converted and radiated out as gravitational waves. However, the results presented in~\cite{Sperhake:2012me}
show this is likely not true, due to what {\em appears} to be strong self-absorption of the
emitted gravitational energy by the black holes. These simulations only went
to $\gamma\approx2.5$; however if they in fact provide a decent approximation of the large 
$\gamma$ limit,then one concludes that 
as much as $1/2$ the kinetic energy
could be converted to rest-mass energy in the black holes (the rest to gravitational waves), 
{\em even} in close scattering encounters (which is in fact consistent with a perturbative 
calculation in the extreme mass ratio limit~\cite{Gundlach:2012aj}). The surprising
consequence of this is that two ``microscopic'' black holes each of rest-mass $m$ scattered
off one another with $\gamma\gg1$ and finely tuned impact parameter could grow
to two ``macroscopic'' black holes moving apart sub-relativistically, each with rest-mass $\sim m \gamma/2$.

A further intriguing result for the 5d case presented in~\cite{Okawa:2011fv} is that for a small range of 
impact parameters near threshold, curvature invariants grow rapidly at the center of mass shortly
after what appears to be a scattering event, though the code crashed before the final outcome
could be determined. No encompassing apparent horizon is detected then, which could simply
be because of the nature of the time coordinate employed, or could be due to a naked
singularity that is forming. If the former, this would be a new outcome to the black hole scattering
problem in 5d, \index{black hole scattering problem in 5d} namely three  black holes; if the latter, this would be 
another example (in additional
to the Gregory-Laflamme instability of black strings, and possible prolate dust collapse~\cite{Shapiro:1991zza}) 
showing  a  violation of cosmic censorship. 

The high-speed limit has also shed some light on the mechanism responsible for large recoil velocities seen
in merger simulations of astrophysically relevant inspirals with certain spin
configurations (see the discussion in Sec.~\ref{sec_BBH}). In particular, there have
been suggestions that the large recoils require the formation of a common horizon
to effect the gravitational wave emission of ``field momentum'' 
associated with what is otherwise purely kinematical properties
of the orbit; in reaction the merger remnant
receives a kick in a direction that conserves linear momentum~\cite{2010PhRvD..81j4012G}.
However, in high speed merger simulations with similar black hole spin setups,
even in scattering cases where a common horizon does {\em not} form, large recoils are observed~\cite{Sperhake:2010uv}. 
This is consistent with the heuristic explanation of the superkicks presented in~\cite{Pretorius:2007nq} 
as arising from frame-dragging \index{frame-dragging} induced Doppler boosting of the radiation emitted by the
binary motion.

The first ultra-relativistic collision simulations of ``solitons'' \index{solitons} (non-singular compact
distributions of matter) were carried out in~\cite{Choptuik:2009ww}, 
consisting of the head-on collision of two boson stars each with compactness $2M/R\approx1/20$,
and center of mass boosts up to $\gamma=4$ ($v\approx 0.968$). The main goal
of the study was to test the hoop conjecture \index{hoop conjecture} arguments for black hole formation;
hence the use of boson stars as model particles given that their self-interaction
is weak compared to gravity in this limit.
Black hole formation was observed above
a critical boost $\gamma_c \approx 3$, roughly one third the value predicted
by the hoop conjecture. Similar results were later obtained using an ideal fluid
(fermion) star as the model particle~\cite{East:2012mb,Rezzolla:2012nr}. The study in ~\cite{East:2012mb} used less
compact stars that pushed the critical boost to $\gamma_c\approx 8.5$, but again found
this to be a similar factor less than the hoop conjecture estimate. It was argued
that the lower thresholds are due to the compression of one particle by gravitational focusing
of the near-shock geometry of the other particle, and vice-versa. This conclusion was
anticipated by a geodesic model of black hole formation presented in~\cite{Kaloper:2007pb}. It is remarkable
that such a simple model, and for that matter the trapped surface calculations as well,
predict the qualitative properties of what is ostensibly the regime where the most dynamical,
non-linear aspects of the Einstein equations are manifest. On the other hand, a recent
calculation of the gravitational self-force using effective field theory techniques 
in the large $\gamma$ limit show that many simplifications arise here; in particular
the non-linear interactions coming from gravitational bulk vertices 
are suppressed by factors of $1/\gamma^4$~\cite{Galley:2013eba} (see also~\cite{tsov:2012sn}).
Aside from giving strong
evidence that the hoop conjecture is applicable to the classical collision problem, 
these studies also support the expectation that the outcome of sufficiently 
supercritical $\gamma>\gamma_c$ collisions are insensitive to the details of the particle self-interactions.
This is essential for black hole formation in super-Planck particle collisions to be a robust conclusion,
despite the lack of detailed calculations in a full quantum (gravity) theory.
This also justifies the use of black holes as model particles, which from a 
classical gravity perspective is (in theory) a simpler problem to simulate, due to the absence of matter.

The motivation and applications of the AdS/CFT correspondence in string theory
are discussed below in \ref{ADSCFT}; here we briefly comment on what RHIC-motivated \index{RHIC} 
studies have taught us about ultra-relativistic collisions.
The relevant spacetime for this problem is 5d AdS, and in particular the
Poincare wedge, as its boundary is conformal to 4d Minkowski spacetime. Solving
the full Einstein equations in 5 spacetime dimensions without symmetries and 
resolving the geometry dual to highly boosted concentrations of energy would be an extremely
challenging computation to perform. To date then, existing studies (see~\cite{Chesler:2013lia}
for a review) have made 
simplifying approximations:
each particle is modeled as a finite-width gravitational wave with planar
symmetry transverse to the collision axis\footnote{Note added in proof : a 
first simulation without symmetry assumptions in 5d AdS was presented in 
\cite{Chesler:2015wra}}. 
This effectively reduces the numerical evolution to $2+1$ dimensions,
and characteristic approaches \index{characteristic formulation} have proven highly successful for this problem.
Though the topology and asymptotics are quite different from the 4d asymptotically flat
case, there is some similarity. Most relevant to this discussion is that the
infinite boost limit 
is similar to the Penrose/Aichelburg-Sexl superposed
shock-wave construction; in both cases trapped surfaces can be found~\cite{Grumiller:2008va}, 
yet the full solution to the causal future of the collision is unknown. The numerics have solved the finite
width planar collision problem, showing that a black hole (with planar topology) does form in this 
case, and resolving the spacetime to the future of the shock. In particular, post-collision 
along the future lightcone of the collision, the amplitude of the shock, as projected onto the Minkowski boundary, 
decays as a power law in time; within
the lightcone, after a time roughly consistent with inferred thermalization times
in RHIC experiments, the near boundary metric fluctuations transform to a state that
can be characterized as an expanding, cooling hydrodynamic flow~\cite{Chesler:2013lia,Casalderrey-Solana:2013aba}.
For further details on numerical relativity applications in the realm of high energy see~\cite{Cardoso:2012qm}.

\subsection{Gravity in $d\ne4$}\label{sec_HDG}

Beyond ultrarelativistic collisions,
numerical relativity has also been crucial in exploring the
behavior of gravity in both stationary and time dependent scenarios beyond $d=4$. 
There are several motivations for doing so.
On one end there is the desire to understand gravity at a fundamental level by contrasting known behavior
in $d=4$ to what arises in different dimensions. Higher dimensions are required by string theory,
and this has inspired many speculative theories: for example TeV scale 
gravity/braneworlds~\cite{1998PhLB..436..257A,1999PhRvL..83.3370R}, some models of 
inflation~\cite{Baumann:2009ds} and modern cyclic models~\cite{Lehners:2008vx} of the universe.
Lower dimensions have also been used to provide a simpler setting to gain intuition about quantum 
gravity (e.g., in $2+1$~\cite{lrr-2005-1} and in $1+1$ dimensional dilaton gravity~\cite{Gegenberg:2009ny}).
At the other end, compelling
practical reasons are provided by the role gravity may play to understand phenomena
described by field theories through holography~\citep{Maldacena:1997re,Aharony:1999ti}. 

For more information, readers are directed to the recent book~\cite{Horowitz:2012nnc}. Here,
for brevity we mainly focus on time-dependent problems, though
we briefly review stationary solutions;  in particular, those that are relevant to existing
or future dynamical studies.

\subsubsection{Black holes in dimensions $d>4$}
Understanding the landscape of stationary solutions with event horizons has
been the focus of considerable effort~\cite{Emparan:2008eg,Reall:2012it}. 
This work has illustrated how much richer the space of stationary black object solutions in higher
dimensions is compared to the $d=4$ case. A case in point is the broader class of topological 
structures allowed, which
includes hyper-spherical black holes, black rings \index{black rings} \index{black strings} \index{black Saturns} and a combination of these latter two giving
``black Saturns'', black strings, black branes, etc. Interestingly, 
several topologically distinct solutions can have the same
asymptotic charges, showing some degree of non-uniqueness of black hole solutions in higher dimensions. 
However, a particularly intriguing conjecture is that uniqueness can be restored
by the additional requirement of {\em stability}. This possibility is implied by the
fact that linear perturbations of many of these solutions are unstable.
As a further contrast with $d=4$ stationary black holes, there is
no  ``Kerr-like'' bound for spinning black holes in $d\ge5$ as they can have
arbitrarily large angular momenta. Again, related to the uniqueness issue, these ultra-spinning
black holes are unstable~\cite{Emparan:2003sy}.
Numerical solutions are required to understand the non-linear dynamics of unstable black objects,
and to-date this has only been achieved for black strings in $d=5$~\cite{Lehner:2010pn} and 
ultra-spinning black holes in $d\in5..8$~\cite{Shibata:2009ad,Shibata:2010wz}.

\begin{figure}
\includegraphics[width=3.5in,clip]{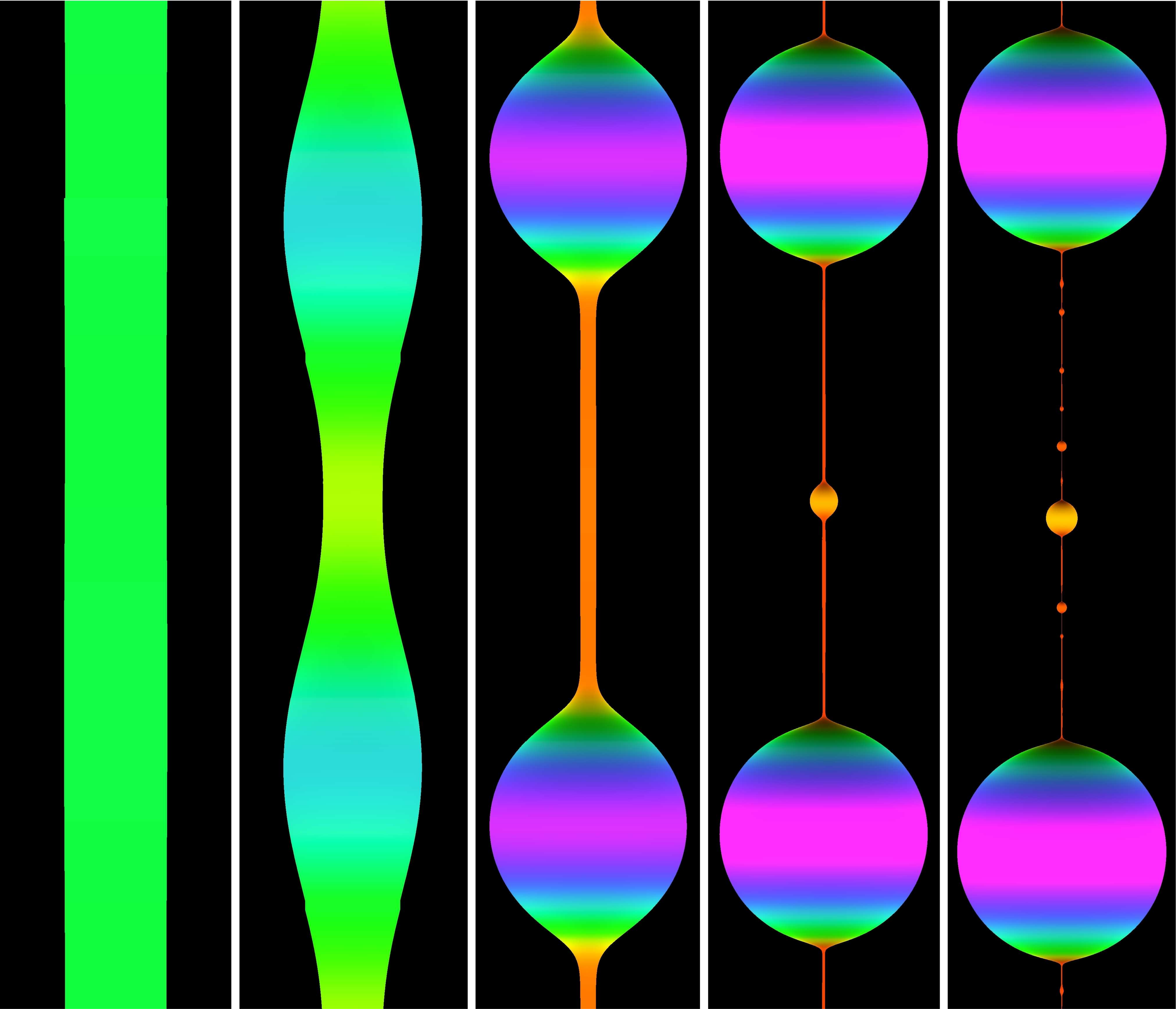}
\caption[Evolution of an unstable black string]{A sequence of snapshots showing the evolution (left to right) of
an unstable black string; see~\cite{Lehner:2010pn} for details.}
\label{fig:bs}
\end{figure}

Black strings are black hole solutions extended along a trivial (optionally) compactified extra-dimension. For simplicity,
and because it is the one studied numerically, we restrict to $d=5$, and so the static black
string is given by the $d=4$ Schwarzschild solution cross a circle with (asymptotic) length $L$. Gregory and Laflamme 
showed that linearized perturbations of such a black string admits exponentially growing modes above
some critical $L/M$ (with $M$ the mass per unit length of the black string)~\cite{Gregory:1993vy}. Further,
thermodynamical arguments suggested that above this ratio the entropically preferred solution would
be a $d=5$ Schwarzschild-Tangherlini black hole. \index{Schwarzschild-Tangherlini black hole} Thus, it appeared possible that the effect of these growing
modes would be to eventually cause the black string to pinch-off and give rise to a spherical configuration.
Naturally, if that happened, cosmic censorship would be violated, indicating yet again that gravity in $d=4$ is
rather special.

To understand the dynamical behavior of the solution, a full non-linear---and so necessarily numerical---analysis is required.
Such a study was presented in~\cite{Lehner:2010pn}, and revealed that the instability unfolds in a self-similar
fashion, where the black string horizon at any given time could be seen as thin strings connected
by hyperspherical black holes of different radii (see Fig~\ref{fig:bs}). As the evolution
proceeds, pieces of the string shrink further while others give rise to spherical black holes bulges, and the horizon
develops a fractal structure. Interestingly, such behavior is reminiscent of the one displayed
by a thin column of fluid through the Rayleigh-Plateau instability (see ~\cite{eggersprl}). 
In the case of the black string, extrapolating the numerical results shows that the ever-thinning string regions 
eventually reach zero size, revealing a
massless naked singularity in {\em finite time}. Thus, perturbed black strings {\em do} provide a 
counter-example 
to the cosmic censorship conjecture, though in $d=5$. In still higher dimensions, the outcome is expected to be qualitatively
similar up to  a critical dimension beyond which stable, non-uniform black strings states are entropically favored.
Perturbative analysis indicates that the critical dimension is $d=13$~\cite{Sorkin:2004qq}, though recent
work making use of a local Penrose inequality suggests it may be as low as $d=11$~\cite{Figueras:2012xj}.

This result has application beyond black string spacetimes, as many of the higher dimensional black
hole solutions have a near horizon geometry that can be mapped, in appropriate regions, to black strings.
For instance, ultra-spinning black holes satisfy the
Gregory-Laflamme instability \index{Gregory-Laflamme instability} condition around the polar region~\cite{Emparan:2003sy}. Such black holes are thus expected to develop growing 
deformations about the poles of the horizon when perturbed. These can be both axisymmetric modes
that would evolve toward axially ``pinched'' or ring-like configurations~\cite{Dias:2009iu}, and also non-axisymmetric modes.
The latter however would induce a time-varying quadrupole moment that would radiate angular momentum,
allowing for the possibility that gravitational wave emission could regulate the instability, in particular
if the non-axisymmetric modes are the dominant unstable ones.
Numerical simulations for systems  with angular momentum mildly higher than the
critical value show this is precisely the case~\cite{Shibata:2009ad,Shibata:2010wz},
where a ``bar-mode'' develops that radiates angular momentum until the black hole settles down to a sub-critical, stable state.
For larger initial spins it has been speculated that non-axisymmetric modes can grow more rapidly
than gravitational wave emission can reduce the spin to sub-critical, and the horizon might then
fragment into multiple pieces~\cite{Emparan:2003sy,Shibata:2010wz}. These cases have yet to be explored
beyond the linear level.

As a last example we mention that solutions describing large black holes
in Randall-Sundrum models \index{Randall-Sundrum models} were numerically constructed 
in~\cite{Figueras:2011gd}, disproving a conjecture that such solutions
could not exist~\cite{Tanaka:2002rb,Emparan:2002px}. Moreover, the particular Ricci flow method employed to 
obtain the solutions argues implicitly in favour of their stability.

\subsubsection{AdS/CFT duality applications}\label{ADSCFT}
The AdS/CFT correspondence~\cite{Maldacena:1997re,Aharony:1999ti} \index{ADS-CFT correspondence} provides a remarkable framework to
study  certain strongly coupled gauge theories in $d$ dimensions by mapping to weakly coupled 
gravitational systems in $d+1$ dimensions. A large body of work has been built since the 
introduction of this correspondence; we will not review it here. Rather, we concentrate on a handful of
applications where numerical simulations have been crucial to the understanding of  the gravitational
aspects of the problems. The relevant spacetimes typically involve black holes, 
and are asymptotically AdS, the latter property creating delicate issues on both analytical and numerical fronts.
This is in part due to the timelike nature of the AdS  boundary, with the consequence that  the 
correct specification 
of boundary data (in addition to the initial configuration) is crucial for 
a well-defined evolution that can be mapped to the CFT description of the problem.
The boundary conditions can be derived from the limiting behavior of the Einstein equations
approaching the boundary, together with constraints from imposing that the spacetime approaches
AdS at the appropriate rate in a suitable gauge.

\begin{figure}
\includegraphics[trim=0.1in 1.75in 2in 0.75in,width=4.35in,angle=0,clip]{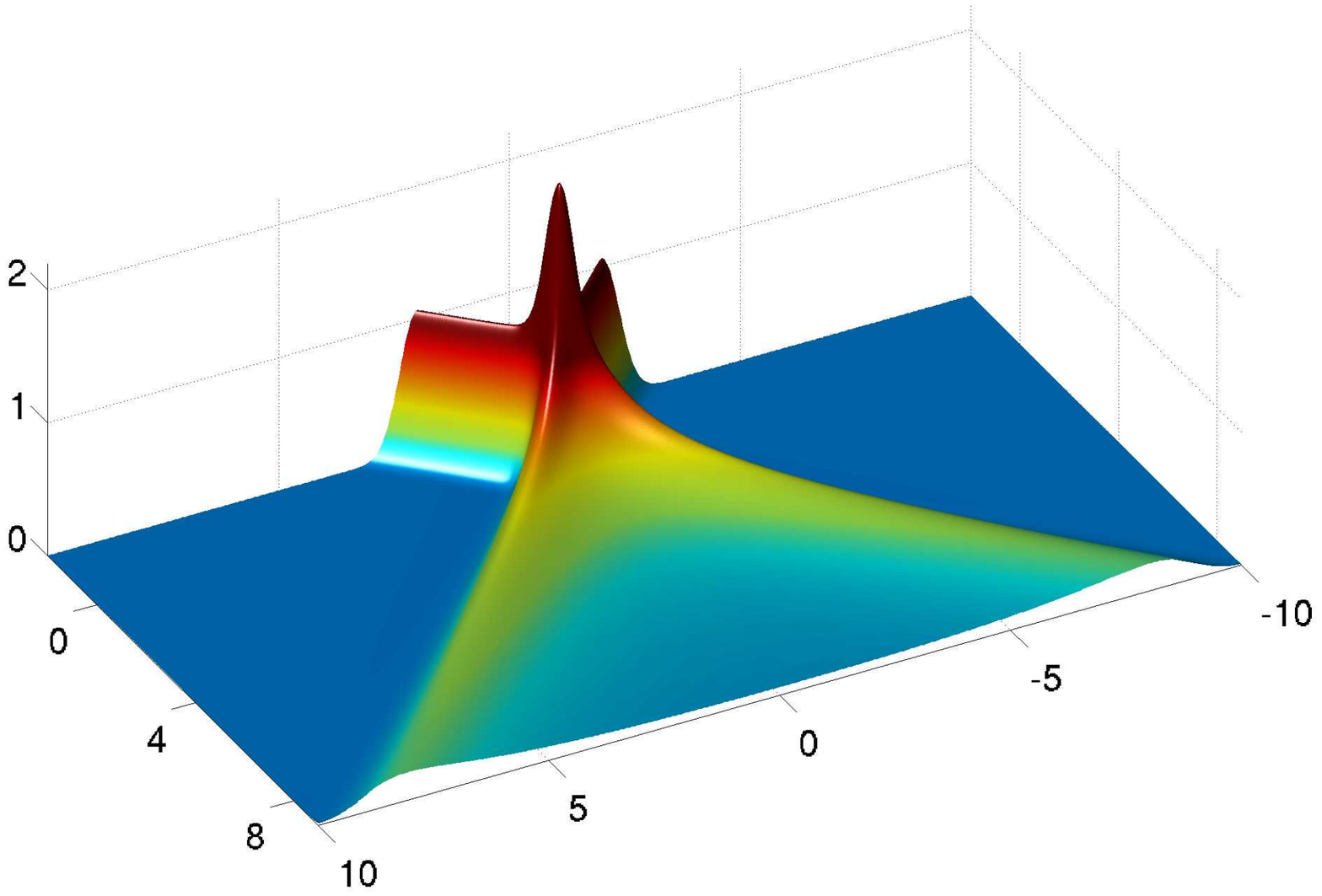}
\includegraphics[width=2.35in,clip]{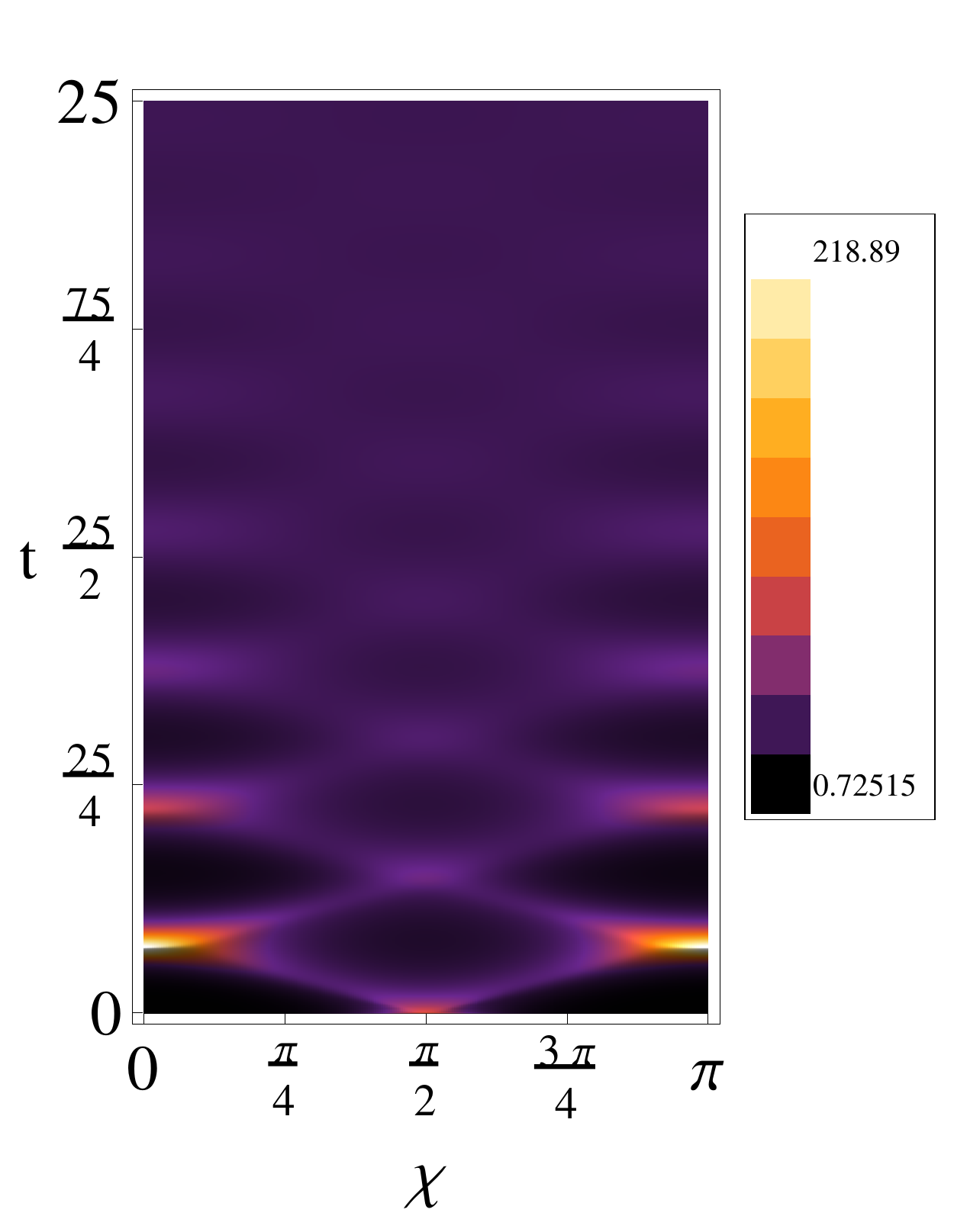}
\caption[Shock wave collision and perturbed black hole in AdS]{ 
{Top: Energy density in planar shock collisions as a function
of time $t$ and longitudinal position $z$. The shocks approach each
other along the $z$ axis and collide at $t=0,z=0$. The collision produces
``debris'' that fills the forward light cone (from~\cite{Chesler:2013lia}.)}
Bottom: Depiction of the energy density of a 4d boundary flow
dual to the evolution of a highly perturbed 5d black hole 
in asymptotically global AdS spacetime (the radius of the black hole settles
to $5$ in geometric units, where the AdS length scale is 
$L=1$) (from~\cite{Bantilan:2012vu}). The boundary has topology $\mathbb{R} {\rm x} S^3$,
and $\chi$ is an angular coordinate; hence
the image represents an initial high density (hence pressure)
enhancement on the equator ($\chi=\pi/2$) that propagates
back and forth between the equator and the poles ($\chi=0,\pi$). 
This result is from a pure 5d vacuum gravity simulation, yet 
the projected boundary dynamics matches that of a relativistic conformal
fluid to within better than $1\%$, even in the early stages
when the perturbation is highly non-linear.
\label{theramlization}}
\end{figure}

Many interesting applications have been pursued using holography, \index{holography} spurred by
work beginning soon after the formulation of the correspondence indicating  the rather
spectacular breadth of possible applications to finite-temperature field theory (see~\citep{Son:2007vk} for a review).
Some highlights include that the hydrodynamic
behavior of field theory is captured by correlation functions in the low-momentum limit, that
hydrodynamics modes in relevant field theory states correspond to low-lying quasinormal modes of an AdS black brane solution,
and for such states there is a universal viscosity to entropy density ratio $\eta/s = 1/4\pi$ for a broad class of theories 
with gravitational duals.
The value of $\eta/s$ is remarkably close to that inferred from hydrodynamic models of the quark-gluon plasma (QGP) \index{quark-gluon plasma (QGP)}
formed in relativistic heavy ion collisions, and this observation has led to the new approach
of using AdS/CFT to try to understand the QGP (for a review see~\cite{DeWolfe:2013cua}). 
Though the $N=4$ SYM theory of the duality is not deconfined QCD,
there are sufficient similarities that one might hope the former can give insights into aspects of the problem
not easily calculable via traditional techniques (perturbative Feynman diagrams and lattice QCD).
For example, using AdS/CFT purely gravitational studies can be used to estimate the thermalization timescales 
post-collision, and the subsequent evolution of the expanding plasma to the point of hadronization.
As mentioned above in Sec.~\ref{sec_URC}, a series of groundbreaking 
works~\cite{Chesler:2008hg,Chesler:2011ds,Chesler:2010bi} studied
the behavior of the spacetime when two gravitational shock waves collide.
This is expected to offer a decent approximation to the dynamics along the beam axis
in central (head-on) collisions. The results for the thermalization time (one definition of which
is the time after collision when the boundary stress tensor of the CFT is well approximated
by the leading order terms in the hydrodynamic expansion) are broadly consistent with the
times inferred from experiment. Moreover, the subsequent hydrodynamic flow exhibits a form
of boost-invariance similar to the predicted Bjorken flow~\cite{Bjorken:1982qr,Luzum:2008cw},  
the difference being characterizable as a modest dependence of the energy scale of the flow
on rapidity~\cite{Chesler:2013lia,Casalderrey-Solana:2013aba}. Figure~\ref{theramlization} illustrates
the energy density measured at the AdS boundary from simulations on the gravitational side; the
left image is from a shock collision simulation, and the right is the relaxation of a highly perturbed black hole.
Soon after the collision (left) and from the beginning of ringdown (right),  a hydrodynamical
description on the field theory side matches the observed near-boundary metric behavior to an excellent degree.

Another front where the duality is being exploited is to understand the behavior of a system in the
ground state of a given Hamiltonian when a ``quenched interaction'' is introduced.
Here the response of an initial thermal equilibrium state of the theory under rapid variations of suitable 
operators can be studied using the correspondence. As the behavior on the gravitational side is governed by the 
dynamics of an appropriately perturbed black hole, a universal response is uncovered which, on the CFT side, 
means the system responds in a way only dependent on the conformal dimension of the quench operator in the
vicinity of the ultraviolet fixed point of the theory~\cite{Buchel:2012gw,Buchel:2013lla,Buchel:2013gba}.

\begin{figure}
\includegraphics[trim=1cm 9cm 14cm 1cm, height=5.cm,clip]{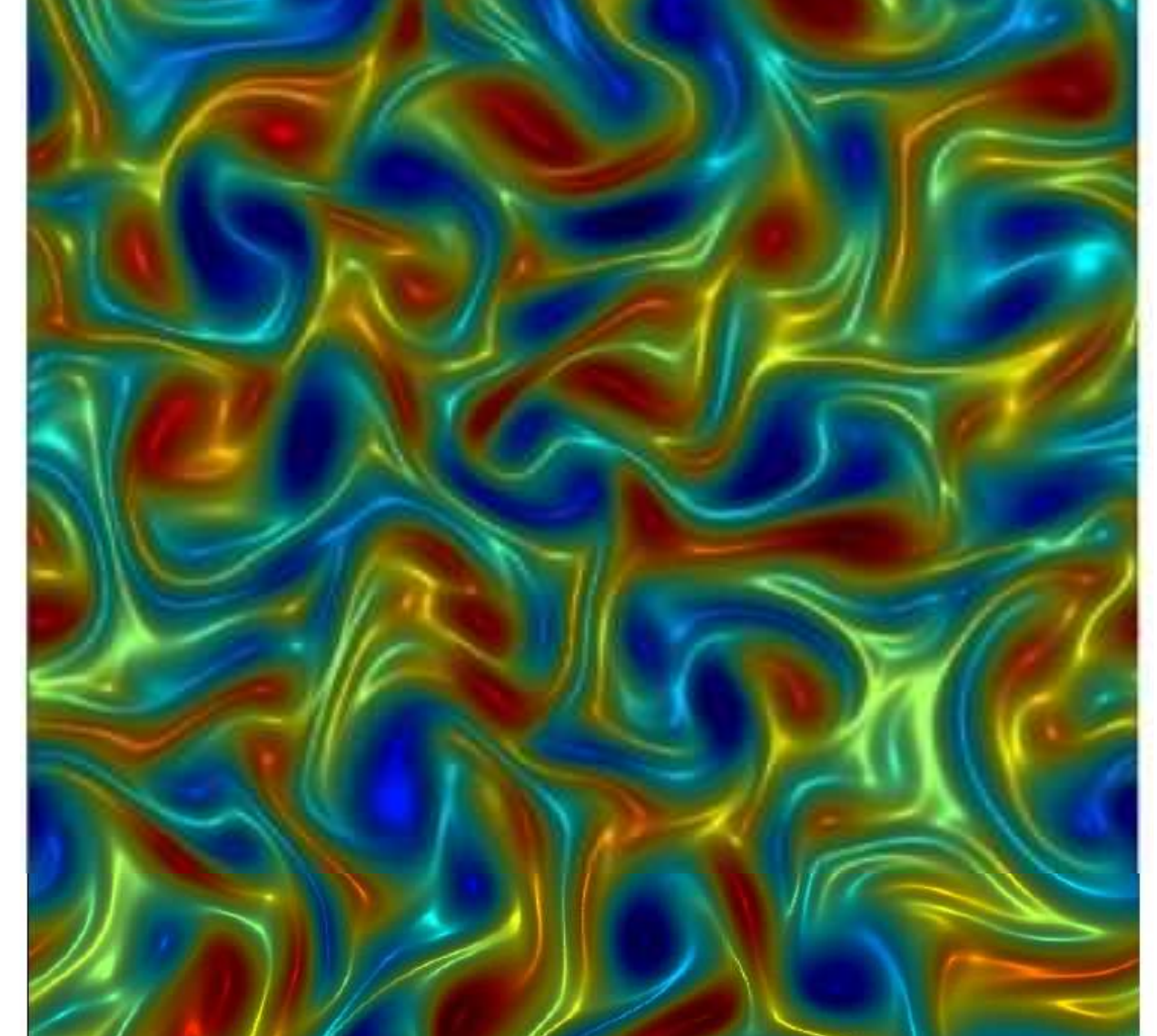}
\includegraphics[height=5.0cm]{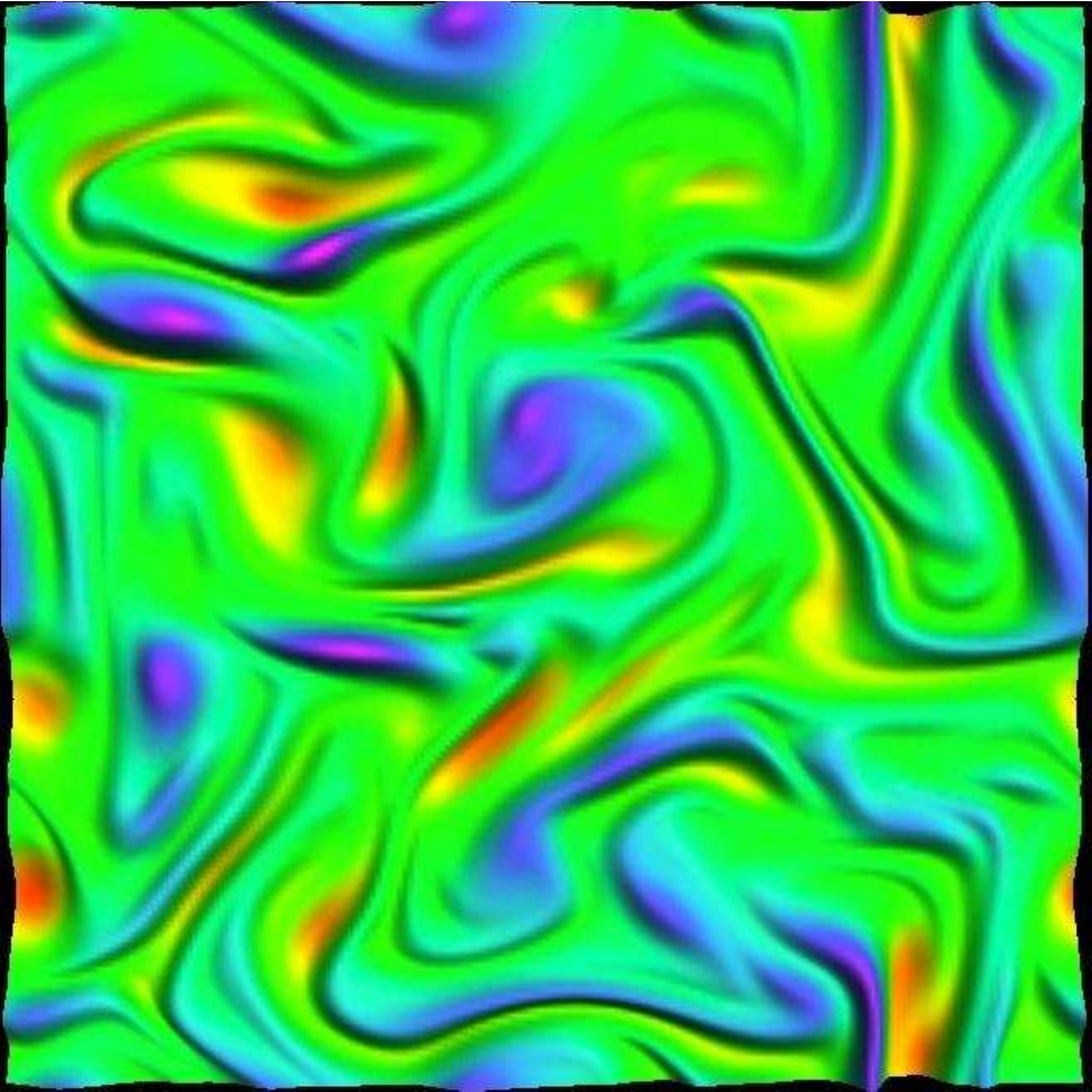}
\caption[Turbulence in black branes and relativistic fluids]
{Left: Vorticity of gravitational perturbations of a planar black hole as obtained through a $3+1$ simulation 
of Einstein equations in AdS (from~\cite{Chesler:2013lia}). Right: Vorticity of a hydrodynamical
field obtained in a $2+1$ viscous hydrodynamic simulation with a background fluid configuration dual to a planar
black hole (from~\cite{Green:2013zba}). Exploiting the fluid/gravity duality allows for constructing the
full metric of the dual $3+1$ spacetime to excellent accuracy.}
\label{turbulence} 
\end{figure}

As a last example we mention an application of the duality in the opposite direction: 
using knowledge of the behavior on the field theory side to discover and analyze  novel features on the
gravitational side. It is well known that field theories at sufficiently high energies admit
a hydrodynamical description; this motivated works that established a duality between gravity and
hydrodynamics for relativistic, conformal fluids. \index{gravity-hydrodynamics duality} Specifically, it was shown that 
in the limit of long wavelength perturbations of black holes the
Einstein equations projected onto the AdS boundary reduce to the familiar relativistic
hydrodynamics for a viscous fluid (e.g.~\cite{Hubeny:2011hd}). Numerical work has demonstrated
that the hydrodynamic description
matches the behavior of full, non-linear solutions of the Einstein
equations surprisingly well in may situations as mentioned above with RHIC applications (and see Figure~\ref{theramlization}).
Taken the other direction then,
this duality implies that phenomena familiar in hydrodynamics should arise in gravity.
In particular, motivated by this, arguments were presented that gravity could exhibit
turbulent dynamics, with a direct energy cascade in $4+1$ dimensions and the opposite in $3+1$~\cite{VanRaamsdonk:2008fp}. 
Furthermore, in $3+1$ dimensional gravity a quasi-conserved quantity should arise that is related to the 
conservation of entrophy in hydrodynamics~\cite{Carrasco:2012nf}. These observations have recently
been demonstrated in ground-breaking numerical simulations of perturbed black branes~\cite{Adams:2013vsa} (see Fig.~\ref{turbulence}),
showing that the horizon geometry reflects the turbulent behavior
of the boundary projection, and develops a fractal-like structure over the corresponding range of lengthscales.

\subsection{Singularities}\label{sec_CS}

Numerical simulations have played a significant role in analyzing  the nature of singularities,
in particular those which are often called ``cosmological'' due to the spacetimes having compact spatial topology (for a
thorough review, see ~\cite{Berger:2002st}). One of the longstanding questions has been the generic nature of 
singularities; i.e., what is the geometry of a spacetime approaching a singularity if no symmetries
are presumed? \index{singularities in cosmological spacetimes} For a vacuum spacetime, much of the research was inspired by the early work of 
Belinski, Lifschitz and Khalatnikov (BKL,\cite{Belinskii:1972sg}), \index{BKL conjecture} who conjectured that the generic singularity is spacelike,
local, and oscillatory. The ``local'' part of the conjecture is that, in an appropriate gauge, the spatial
gradients in the field equations become irrelevant compared to the temporal gradients, and hence the dynamics
at any spatial point reduces to a set of ordinary differential equations in time. The oscillatory (or ``mixmaster'') aspect \index{mixmaster} then
describes the dynamics of one of these points, claiming that the solution consists of an infinite, chaotic sequence
of transitions between epochs, and in each epoch the geometry is well-described by one member of the Kasner family of geometries. A Kasner geometry \index{Kasner space times} is a homogeneous but anisotropic solution to the field equations consisting of 
two contracting and one expanding spatial direction (in the approach to the singularity). 
Several objections were raised to the BKL conjecture, in particular that the assumptions they employed
restricted their conclusions to local aspects of homogeneous cosmologies, and hence had little bearing
on the generic, global properties of the spacetimes~\citep{1979PhR....56..371B}. Numerical simulations have been key in resolving these disputes, gathering evidence in favour of the BKL 
conjecture~\citep{Berger:1993ff,Berger:1998vxa,Garfinkle:2003bb}, though discovering a surprising caveat
in the process. This discovery was of so-called ``spikes'' that \index{spikes} develop at isolated regions in the geometry~\citep{Berger:1993ff}.
A spike is a small lengthscale feature where the spatial gradients are {\em not} small, and hence are important
in governing the local dynamics of the geometry. Spikes seem to undergo oscillatory transitions similar
to the mixmaster behavior of non-spike worldlines~\citep{Lim:2009dg}. However since (in the approach to the singularity) they shrink rapidly with time, even numerical simulations imposing planar symmetry (so $1+1$ dimensional evolution) have not been able to follow their dynamics for long enough to conclusively demonstrate this. Due to these resolution challenges spikes have not been studied in scenarios with less symmetry, and so whether spike-like features beyond co-dimension 1 exist is also not known.

An important point to make with regards to the above discussion of genericity and singularities is that it strictly
applies only to these so-called cosmological singularities, and not {\em necessarily} to those formed in gravitational collapse to black holes. There is some expectation that local properties of the singularities should
be the same whether in a cosmological or black hole setting (indeed, the interior geometry of Schwarzschild
is locally Kasner). However, the interior (Cauchy) horizon of rotating and charged black holes develop
into null singularities when perturbed, and have very different structure from the spacelike singularities in
cosmology~\citep{1990PhRvD..41.1796P}. There are also arguments that null singularities are ``as generic'' as spacelike
singularities~\citep{1996PhRvD..53.1754O}, and may also be relevant in a cosmological setting~\citep{Dafermos:2012np}. 

\subsection{Miscellaneous}\label{sec_misc}

Here we briefly discuss two miscellaneous topics where numerics
have played an important role, and do not naturally fit into the
main topic sections above.

\subsubsection{Stability of AdS}\label{ads_stab}
As opposed to Minkowski and deSitter spacetimes where global non-linear stability with respect to small perturbations
has been established~\citep{Christodoulou93,friedrichds}, the related question has yet
to be resolved in AdS. A key difference between AdS and the other two spacetimes is
the fact that infinity is timelike, and acts like a confining boundary; namely the future light cone from any event
on an interior timelike observer's worldline will reach the boundary and return to intersect the worldline a finite proper time
later.
Ground-breaking numerical and analytical work~\citep{Bizon:2011gg} studied the spherically symmetric Einstein-Klein-Gordon 
system in asymptotically AdS spacetime, and uncovered that a black hole eventually forms from an arbitrarily 
small initial perturbation. This can heuristically be understood as a direct result of the confining property
of AdS---energy cannot dissipate away, and due to non-linear interaction eventually a configuration
will be explored where the central energy density becomes sufficiently large to cause gravitational collapse.
A more quantitative explanation was given in~\citep{Bizon:2011gg}, where through a resonant mechanism
there is a secular transfer of energy from large to small scales, ending when a black hole forms.
Furthermore, at the threshold of black hole formation, the
spacetime behaves self-similarly, and the solution corresponds to the one seen in the 
asymptotically flat case~\citep{Choptuik:1992jv} (as expected
since the AdS scale is irrelevant for a small black hole). Related work has argued that this behavior should still
be present in the absence of symmetries, and also when only gravitational perturbations are considered~\citep{Dias:2011ss}. 
While these studies suggested AdS is unstable to arbitrarily small, generic perturbations, more recent follow up work has
demonstrated the existence of large classes of initial data that are stable~\citep{Buchel:2013uba,Dias:2012tq,Maliborski:2013jca}.
Applying these results to the AdS/CFT correspondence, given that black hole formation is synonymous with thermalization,
this implies (perhaps unsurprisingly) that there are large classes of states in the dual CFT that do not thermalize.

\subsubsection{Formation and evaporation of CGHS Black Holes}\label{cghs}
Two dimensional dilaton models of black hole evaporation were a popular subject
of research a couple of decades ago, and though much was learned about the
quantum nature of black holes from them, one could argue that no consensus results
were obtained regarding the final near-Planck stages of evaporation
(whether the black hole evaporates completely, if there is a remnant, a naked
singularity, baby universe, etc.), or whether information is lost.
One such popular model is that of Callan, Giddings, Harvey and Strominger (CGHS)~\cite{Callan:1992rs}.
\index{Callan, Giddings, Harvey, Strominger (CGHS) two dimensional spacetimes}
Though extensively studied before, many interesting quantitative and qualitative
features of solutions to the semi-classical CGHS equations of motion were missed until a
recent numerical study~\cite{Ashtekar:2010hx,Ashtekar:2010qz,Ramazanoglu:2010aj}.
One of the more interesting results revealed here is that
there are two distinct classes of solutions:  those that can
be identified as microscopic (with initial masses of order the Planck mass
or less), and those that are macroscopic (with initial masses a few times the
Planck mass or larger).
Remarkably, for macroscopic cases, after a brief transient,
the evaporating spacetime and Hawking flux asymptote
to a universal solution,
irrespective of details of the matter distribution that
formed the black hole.  Evaporation continues until the dynamical
horizon shrinks to zero area, whence it encounters a singularity
of the semi-classical equations (though this singularity is
weaker than that arising in the classical solution). The future
Cauchy horizon of this singularity is regular, in contrast
with earlier suggestions that it would propagate
to infinity in a ``thunderbolt''. An improvement to the
Bondi mass of the spacetime proposed in~\cite{Ashtekar:2008jd} shows
that there is still on the order of a Planck mass ``remnant'' in the
singularity, though this would presumably be resolved with higher order
quantum corrections. 

This behavior is very different from 
that of black holes initially formed with only of the
order of the Planck mass (the microscopic branch). Earlier studies
had missed this distinction, and focused all attention
on the physically less relevant microscopic solutions. The macroscopic branch also 
turns out to be quite challenging to solve numerically,
where scales of the order $M_{{\rm Planck}}$ in the initial vacuum state
are exponentially ``inflated'' to scales of order $e^{M/M_{{\rm Planck}}}$ in the
outgoing Hawking flux (a manifestation of the red-shift of outgoing
radiation in black hole spacetimes, though in the evaporating case
this red-shift remains finite, since what was a null event horizon becomes instead a time-like dynamical horizon). 

One unusual aspect of 2d dilaton gravity
that prevents straight-forward application of 2d results
to the more relevant 4d case, is that in the former case there are two distinct,
causally disconnected null infinities (``left'' and ``right'').
This effectively disassociates the quantum state of the ``ingoing'' (right to left
moving quanta, say) matter that forms the black hole from the ``outgoing'' (left to
right) vacuum that becomes the Hawking flux. The semi-classical results together
with the arguments presented in~\cite{Ashtekar:2008jd} suggest
that the evolution of this vacuum sector is unitary, though little
information about the infalling matter is retrievable from the Hawking
flux. There is also no sign of any ``firewall''~\cite{Almheiri:2012rt} \index{firewall} along the dynamical
horizon at the semi-classical level.

\subsubsection{Cosmic bubble collisions}
Within the eternal inflation paradigm, our observable universe is contained in one of many bubbles \index{bubbles} formed from an
inflating metastable vacuum~\cite{Guth:2007ng}. Collisions between bubbles can potentially leave a detectable imprint on the cosmic 
microwave background radiation (see reviews~\cite{Aguirre:2009ug,Kleban:2011pg}). \index{cosmic microwave background (CMB) radiation} While this scenario was initially studied through phenomenological
models, recent works have concentrated on providing a quantitative connection between particular
scalar field models giving rise to eternal inflation and the detailed signatures imprinted on the CMB.
To this end, the intrinsically non-linear nature of the bubbles and their
collisions have been studied numerically within full general 
relativity~\cite{2012PhRvD..85h3516J,2013arXiv1312.1357W}. Simulations have 
revealed, in particular, the following: i) the energy released in the collision of identical vacuum 
bubbles goes mostly into the formation of localized field configurations
such as oscillons; ii) the structure of the potential considered is the dominant factor
determining the immediate outcome of a collision; and iii) slow-roll inflation can occur to the
future of a collision. Interestingly, these studies indicate that the signature in the CMB is well-described
by a set of four phenomenological parameters whose values can be only probabilistically determined. 

\subsubsection{Inhomogeneity in cosmology}
The majority of applications of general relativity to cosmology over the past decades
have utilized analytical methods. For observational cosmology this is because the observed homogeneity and isotropy
of the universe implies that its large scale structure is well-described by known exact solutions (the
Friedmann-Robertson-Walker-Lemaitre metrics), \index{Friedmann-Robertson-Walker-Lemaitre (FRWL) metrics } with deviations from  the FRWL solutions  small and hence amenable to treatment
by perturbation theory. There has however been some concern, in particular in light of the 
discovery of the present day accelerated
expansion of the universe, that large scale inhomogeneities such as filaments and voids,
or small scale non-linear inhomogeneities such as stars,  can alter
the assumptions made to study the largest scale dynamics of the cosmos (see ~\citep{2012ARNPS..62...57B}
for a recent review). Some of these questions can be addressed by numerical solutions within
full general relativity, in particular whether local, non-linear inhomogeneities can affect
the overall expansion compared to a homogeneous universe with the same average stress-energy
context (though ``averaging'' is itself an issue of some delicacy here). Only recently have
simulations of such scenarios been considered~\citep{Zhao:2009yp,2013PhRvL.111p1102Y,2014arXiv1404.1435Y}. 
In the latter two studies,  
universes with a positive cosmological constant and filled with a periodic lattice of black holes 
(thus the most extreme example of non-linearity possible in general relativity) were evolved. The results 
were that the effective expansion rate was consistent with that of an equivalent 
homogeneous dust-filled universe. 

\section{Unsolved Problems}\label{sec_unsolved}
The aforementioned list of studies, while impressive in its own right,
is only a portion of interesting phenomena where numerical relativity 
can shed light on important questions, as well as open up new research directions
from them. The following is a (necessarily incomplete) list of such questions.
\begin{itemize}
\item {\em Strongly gravitating/highly dynamical scenarios and astrophysics.}
While it is clear that simulations have played a key role in uncovering the behavior
and characteristics of gravitational wave emission from compact binaries in astrophysical settings,
much work, and many opportunities, remain.
Indeed, even in the case of binary black holes where ``only'' the Einstein equations in vacuum are required,
higher mass ratios and/or nearly-maximal spinning
configurations have proven difficult and costly. The possible existence of intermediate mass black holes
strongly motivates understanding the former class of binary. Non-vacuum systems require a more
complex description due to the additional, and often involved, matter physics. The rewards for this complexity 
are that now in addition to gravitational waves, electromagnetic and/or neutrino emission become possible, 
with the consequence that the opportunity for simulations to make contact with observation is extremely rich. 
The overarching goals of such simulations are to obtain first-principles 
descriptions of the detailed observational signatures across the range of emission channels
the binaries might produce. The challenge to do so comes largely from the 
disparate time/length scales introduced by a plethora of physical processes, and by
the complexity of the microphysics. Unlike the Einstein equations, to make
simulations of realistic matter tractable invariably requires simplified models of the 
fundamental equations. There is much opportunity here for synergy 
among the relevant communities: numerical relativity, gravitational wave observation,
theoretical and observational astronomy, and nuclear physics.
\item {\em Fundamental questions.} Numerical simulations will continue to play a key role in exploring
questions about the fundamental nature of Einstein gravity. There is no shortage of tantalizing questions
remaining to be explored, including the non-linear development of superradiant and
other black hole instabilities in four and higher dimensional spacetimes (see~\citep{2014PhRvD..89f1503E}
for a first such study in 4d), 
the nature of generic singularities inside rotating black holes (c.f. the ``mass inflation''
phenomenon~\cite{1990PhRvD..41.1796P}), the dynamics of near-extremal black holes, 
cosmological domain wall and gravitational shock-wave collisions without symmetry assumptions, 
critical collapse without symmetries, black hole collisions and other dynamical non-linear 
interactions in asymptotically AdS spacetimes (in particular with AdS/CFT applications in mind),
testing the limits of the hoop and cosmic censorship conjectures, the possible development --and consequences-- 
of turbulence in gravity
and fractal horizon 
structures,
etc.(some of these
are discussed further below).
If past discoveries, such as critical phenomena and the ``turbulent'' instability in AdS spacetimes
are any indication, many surprises await. Furthermore, these might have counterparts in other physical
systems, and important insights might be gained in both directions by analogy and their mathematical similarity or
equivalence.
\item {\em Critical collapse.} As reviewed above, the vast majority of 
studies of the threshold of gravitational collapse have been carried out in 
spherical symmetry. That the original study of
the axisymmetric gravitational wave critical solution has defied
attempts at a detailed solution for almost twenty years now hints
at a very interesting and rich geometric structure awaiting discovery.
For axisymmetric scalar field collapse, the inconsistency between
perturbative results suggesting that all non-spherical perturbations
decay and a numerical study that hinted at a second, ``focusing''
instability remains to be resolved. Collapse without any symmetry
assumptions is essentially uncharted territory.
\item {\em Black hole instabilities in higher dimensional, asymptotically flat spacetimes}.
A number of black holes in higher dimensions have been argued to be unstable, in particular by making
connection to Gregory-Laflamme type instabilities. These arguments stem from the realization that the near-horizon
geometry in (at least portions of) these black hole spacetimes can be mapped to unstable black string solutions, and so
should display related phenomenology. This is the case for ultra-spinning black holes, black rings, black
Saturns, etc. (e.g.~\cite{Emparan:2003sy}). Whether in all cases these black holes yield the rich behavior observed in perturbed, unstable
black strings is yet unknown. For instance, in rapidly spinning black holes the instability
induces a non-trivial, time-dependent quadrupole that radiates angular momentum that could shut off
the instability. This has already been observed in numerical simulations of Myers-Perry black 
holes in 6-8 dimensions, though only for cases with 
relatively mild angular momenta~\cite{Shibata:2010wz}. For sufficiently large
angular momentum (recall that there is no upper bound in higher dimensions)
the time scale of the Gregory-Laflamme instability is shorter than the expected
gravitational wave emission time required to reduce the spin by enough
to stabilize the system.
A related problem is to consider
highly prolate,``cigar-shaped'' black holes  in higher dimensions.
Certainly, barring small scale length-wise perturbations, such a black hole would 
tend to becoming spherical on a time-scale of order equal to 
the light crossing time $\tau_{{\rm CT}}$ of the black hole. However, such horizons
that are sufficiently thin should locally be Gregory Laflamme unstable
on a time scale much quicker than $\tau_{{\rm CT}}$. 
Both of  the aforementioned problems appear tractable in the near future. 
\item {\em Gravitational behavior in $d\ne4$ and holography.} As discussed above, holography has opened
the door for numerical relativity to be exploited in problems outside the gravitational arena.
Indeed, applications relevant to quark-gluon plasmas, condensed matter physics and quantum quenches 
(processes in which the physical couplings of a quantum system are abruptly changed), have recently been
undertaken.
While there is already an impressive body of work
in this context, it is important to point out that most studies in this field have been in non-dynamical settings,
and existing dynamical studies have assumed symmetries to yield a tractable
computational problem. As a consequence, current results have certain limitations to the applicability
and generality of the physics than can be drawn from them. This leaves much room for novel
future work.
\item {\em High speed/soliton collisions.} Many questions remain in this topic. 
For soliton collisions, the nature of the black hole formation threshold solution
is unknown; possibilities include a ``universal'' gravitational critical solution
irrespective of the nature of the matter, or alternatively the critical solution
of the matter field that the soliton is composed of. 
In the infinite boost limit,
the geometry to the causal future of the shock wave collision is unknown.
Very few studies of finite boost, higher dimensional collisions relevant to super-Planck scale
particle collisions have been conducted. In particular only trivial topologies without
brane tension have been considered, and charge has been ignored, which could be
important at LHC energies. The intriguing suggestion of naked singularity formation
in grazing 5d collisions shown in~\cite{Okawa:2011fv} needs further investigation.
A detailed study of the radiation emitted in large impact parameter encounters in 4d
would allow comparison with the effective field theory calculations that
suggest the problem simplifies in this limit~\cite{Galley:2013eba,tsov:2012sn} 
(the difficulty with such a study is that the black holes loose little energy
in the encounter, and hence a very long time numerical evolution will be required
to allow the gravitational waves to get sufficiently far ahead of the black holes,
unless novel gravitational wave extraction methods are developed).
For high speed collision applications to heavy ion collisions via AdS/CFT,
future work includes relaxing symmetries to model non-central collisions,
and introducing refinements to allow the dual CFT to better approximate QCD (for example,
trying to model effects of confinement with additional matter fields, or via
dynamics in the $S^5$ manifold of $AdS_5 \times S^5$ that are usually assumed
to be trivial).
\item {\em Alternative theories of gravity.} Numerical relativity has also recently ventured into studying
astrophysical binary systems within alternative gravity theories. Incipient investigations within Scalar-Tensor
theories have uncovered an unexpected dynamical scalarization phenomena driven by the dynamics of binary 
neutron stars~\cite{Barausse:2012da}. This phenomena has a significant impact on the orbiting behavior with
clear consequences for  the gravitational wave signals from the system. In all likelihood
this behavior is only a token of the rich phenomenology awaiting to be discovered upon closer inspection
of relevant theories, and can have astrophysical/observational consequences.
\item {\em Supernova}. Most core-collapse supernova simulations to date have not incorporated
full general relativity. Given that the problem of the explosion mechanism(s) is still
unsolved and likely to be highly sensitive to the underlying physics, making the codes
fully relativistic is another crucial step in the direction of more realistic modeling of the physics of

this highly complex problem.
\end{itemize}

As is clear from this list, there is no shortage of interesting applications for
numerical studies. With the rapid development of numerical relativity over the past
decades and its expansion to fields outside of pure classical general relativity,
it is impossible to tell what a future review might
have in store. At the same time however, it is safe to predict that many exciting results will fill its pages!

\section*{Acknowledgements}
This work was supported in part by CIFAR; NSERC Discovery Grants (MWC and LL);
NSF grants PHY-1065710, PHY1305682 and the Simons Foundation (FP).
Research at Perimeter Institute
is supported by the Government of Canada through Industry Canada and by the Province of Ontario through the
Ministry of Research and Innovation.

\bibliographystyle{cambridgeauthordate}
\bibliography{SFG_NR}\label{refs}

\begin{thebibliography}{315}
\expandafter\ifx\csname natexlab\endcsname\relax\def\natexlab#1{#1}\fi
\expandafter\ifx\csname selectlanguage\endcsname\relax
  \def\selectlanguage#1{\relax}\fi

\bibitem[1]{300yrs}
Israel, Werner, and Hawking, Stephen~William. 1987.
\newblock {\em {Three hundred years of gravitation}}.
\newblock Cambridge: Cambridge Univ. Press.

\bibitem[2]{Hahn64}
Hahn, Susan~G., and Lindquist, Richard~W. 1964.
\newblock {\em Ann. Phys.}, {\bf 29}, 304--331.

\bibitem[3]{1960PhRv..118.1110M}
{Misner}, C.~W. 1960.
\newblock {\em Physical Review}, {\bf 118}(May), 1110--1111.

\bibitem[4]{Smarr75}
Smarr, Larry~L. 1975.
\newblock {\em The Structure of General Relativity with a Numerical
  Illustration: The Collision of Two Black Holes}.
\newblock Ph.D. thesis, University of Texas, Austin, Austin, Texas.

\bibitem[5]{Smarr79a}
Smarr, Larry~L. 1979.
\newblock Basic Concepts in Finite Differencing of Partial Differential
  Equations.
\newblock {Page  139 of:} Smarr, Larry~L. (ed), {\em Sources of Gravitational
  Radiation}.
\newblock Cambridge, U.K.: Cambridge University Press.

\bibitem[6]{Eppley75}
Eppley, Kenneth. 1975.
\newblock {\em The Numerical Evolution of the Collision of Two Black Holes}.
\newblock Ph.D. thesis, Princeton University, Princeton, New Jersey.

\bibitem[7]{1976PhRvD..14.2443S}
{Smarr}, L., {et~al.} 1976.
\newblock {\em \prd}, {\bf 14}, 2443--2452.

\bibitem[8]{Anninos:1993zj}
Anninos, Peter, {et~al.} 1993.
\newblock {\em Phys. Rev. Lett.}, {\bf 71}(18), 2851--2854.

\bibitem[9]{Bruegmann:1997uc}
Br\"{u}gmann, Bernd. 1999.
\newblock {\em Int. J. Mod. Phys. D}, {\bf 8}, 85--100.

\bibitem[10]{Pretorius:2005gq}
Pretorius, Frans. 2005.
\newblock {\em Phys. Rev. Lett.}, {\bf 95}, 121101.

\bibitem[11]{Campanelli:2005dd}
Campanelli, Manuela, {et~al.} 2006.
\newblock {\em Phys. Rev. Lett.}, {\bf 96}, 111101.

\bibitem[12]{Baker:2005vv}
Baker, John~G., {et~al.} 2006.
\newblock {\em Phys. Rev. Lett.}, {\bf 96}, 111102.

\bibitem[13]{Pretorius:2007nq}
Pretorius, Frans. 2009.
\newblock Binary Black Hole Coalescence.
\newblock {Pages  305--369 of:} Colpi, Monica, Casella, P., Gorini, V.,
  Moschella, U., and Possenti, A. (eds), {\em Physics of Relativistic Objects
  in Compact Binaries: from Birth to Coalescence}.
\newblock Heidelberg, Germany: Springer.

\bibitem[14]{1966PhRv..141.1232M}
{May}, M.~M., and {White}, R.~H. 1966.
\newblock {\em Physical Review}, {\bf 141}, 1232--1241.

\bibitem[15]{1971ApJ...163..209W}
{Wilson}, J.~R. 1971.
\newblock {\em \apj}, {\bf 163}, 209.

\bibitem[16]{Wilson79a}
Wilson, James~R. 1979.
\newblock A Numerical Method for Relativistic Hydrodynamics.
\newblock {Pages  423--445 of:} Smarr, Larry~L. (ed), {\em Sources of
  Gravitational Radiation}.
\newblock Cambridge, U.K.: Cambridge University Press.

\bibitem[17]{Shapiro80}
Shapiro, Stuart~L., and Teukolsky, Saul~A. 1980.
\newblock {\em Astrophys. J.}, {\bf 235}, 199--215.

\bibitem[18]{Stark:1985da}
Stark, R.~F., and Piran, Tsvi. 1985.
\newblock {\em Phys. Rev. Lett.}, {\bf 55}, 891--894.
\newblock {Erratum}: ibid. {\bf 56}, 97 (1986).

\bibitem[19]{Nakamura81}
Nakamura, Takashi. 1981.
\newblock {\em Prog. Theor. Phys.}, {\bf 65}, 1876--1890.

\bibitem[20]{Nakamura83}
Nakamura, Takashi. 1983.
\newblock {\em Prog. Theor. Phys.}, {\bf 70}, 1144--1147.

\bibitem[21]{Evans86}
Evans, Charles~R. 1986.
\newblock An Approach for Calculating Axisymmetric Gravitational Collapse.
\newblock {Pages  3--39 of:} Centrella, Joan~M. (ed), {\em Dynamical Spacetimes
  and Numerical Relativity}.
\newblock Cambridge, U.K.: Cambridge University Press.

\bibitem[22]{Shapiro:1991zza}
Shapiro, Stuart~L., and Teukolsky, Saul~A. 1991.
\newblock {\em Phys. Rev. Lett.}, {\bf 66}, 994--997.

\bibitem[23]{Thorne72a}
Thorne, Kip~S. 1972.
\newblock Nonspherical Gravitational Collapse: A Short Review.
\newblock {Page  231 of:} Klauder, J. (ed), {\em Magic Without Magic: John
  Archibald Wheeler}.
\newblock San Francisco: Freeman.

\bibitem[24]{Piran80}
Piran, Tsvi. 1980.
\newblock {\em J. Comp. Phys.}, {\bf 35}, 254--283.

\bibitem[25]{Cook:2000LR}
Cook, Gregory~B. 2000.
\newblock {\em Living Rev. Relativity}, {\bf 3}(5).

\bibitem[26]{Gourgoulhon:2007tn}
Gourgoulhon, Eric. 2007.
\newblock {\em J. Phys. Conf. Ser.}, {\bf 91}, 012001.

\bibitem[27]{Pfeiffer:2005zm}
Pfeiffer, Harald~Paul. 2005.
\newblock {\em Initial data for black hole evolutions}.
\newblock Ph.D. thesis, Cornell University, Ithaca, New York.

\bibitem[28]{York79}
York~Jr., James~W. 1979.
\newblock Kinematics and Dynamics of General Relativity.
\newblock {Pages  83--126 of:} Smarr, Larry~L. (ed), {\em Sources of
  Gravitational Radiation}.
\newblock Cambridge, U.K.: Cambridge University Press.

\bibitem[29]{York-Piran-1982-in-Schild-lectures}
York~Jr., James~W., and Piran, Tsvi. 1982.
\newblock The Initial Value Problem and Beyond.
\newblock {Pages  147--176 of:} Matzner, Richard~A., and Shepley, Lawrence~C.
  (eds), {\em Spacetime and Geometry: The {A}lfred {S}child Lectures}.
\newblock Austin (Texas): University of Texas Press.

\bibitem[30]{Arnowitt:1962hi}
Arnowitt, R., Deser, S., and Misner, Charles~W. 1962.
\newblock The dynamics of general relativity.
\newblock {Pages  227--265 of:} Witten, Louis (ed), {\em Gravitation: An
  Introduction to Current Research}.
\newblock New York: Wiley.

\bibitem[31]{Bowen:1980yu}
Bowen, Jeffrey~M., and York~Jr., James~W. 1980.
\newblock {\em Phys. Rev. D}, {\bf 21}, 2047--2056.

\bibitem[32]{Brandt:1997tf}
Brandt, Steve~R., and Br\"{u}gmann, Bernd. 1997.
\newblock {\em Phys. Rev. Lett.}, {\bf 78}(19), 3606--3609.

\bibitem[33]{Thornburg87}
Thornburg, Jonathan. 1987.
\newblock {\em Class. Quantum Grav.}, {\bf 4}, 1119--1131.

\bibitem[34]{Centrella:1980np}
Centrella, Joan~M. 1980.
\newblock {\em Phys. Rev. D}, {\bf 21}, 2776--2784.

\bibitem[35]{Centrella:1983cj}
Centrella, Joan~M., and Wilson, James~R. 1984.
\newblock {\em Astrophys. J. Supp. Ser.}, {\bf 54}, 229--249.

\bibitem[36]{Anninos91a}
Anninos, Peter, Centrella, Joan~M., and Matzner, Richard~A. 1991.
\newblock {\em Phys. Rev. D}, {\bf 43}, 1808.

\bibitem[37]{Kurki-Suonio93}
Kurki-Suonio, Hannu, Laguna, Pablo, and Matzner, Richard~A. 1993.
\newblock {\em Phys. Rev. D}, {\bf 48}, 3611--3624.

\bibitem[38]{Berger:1993ff}
Berger, Beverly~K., and Moncrief, Vincent. 1993.
\newblock {\em Phys.Rev.}, {\bf D48}, 4676--4687.

\bibitem[39]{PhysRev.172.1331}
Kaup, David~J. 1968.
\newblock {\em Phys. Rev.}, {\bf 172}(Aug), 1331--1342.

\bibitem[40]{PhysRev.187.1767}
RUFFINI, REMO, and BONAZZOLA, SILVANO. 1969.
\newblock {\em Phys. Rev.}, {\bf 187}(Nov), 1767--1783.

\bibitem[41]{PhysRevLett.57.2485}
Colpi, Monica, Shapiro, Stuart~L., and Wasserman, Ira. 1986.
\newblock {\em Phys. Rev. Lett.}, {\bf 57}(Nov), 2485--2488.

\bibitem[42]{Liebling:2012fv}
Liebling, Steven~L., and Palenzuela, Carlos. 2012.
\newblock {\em Living Rev.Rel.}, {\bf 15}, 6.

\bibitem[43]{Seidel:1992vd}
Seidel, Edward, and Suen, Wai-Mo. 1992.
\newblock {\em Phys. Rev. Lett.}, {\bf 69}(13), 1845--1848.

\bibitem[44]{Bona92b}
Bona, Carles, and Mass\'{o}, Joan. 1993.
\newblock A vacuum fully relativistic 3D numerical code.
\newblock {Pages  258--264 of:} d'Inverno, Ray~A. (ed), {\em Approaches to
  Numerical Relativity}.
\newblock Cambridge, U.K.: Cambridge University Press.

\bibitem[45]{Cook:1997na}
Cook, Gregory~B., {et~al.} 1998.
\newblock {\em Phys. Rev. Lett.}, {\bf 80}, 2512--2516.

\bibitem[46]{Abrahams:1997ut-etal}
Abrahams, Andrew~M., {et~al.} 1998.
\newblock {\em Phys. Rev. Lett.}, {\bf 80}, 1812--1815.

\bibitem[47]{Gomez:1998uj}
G\'{o}mez, Roberto, {et~al.} 1998.
\newblock {\em Phys. Rev. Lett.}, {\bf 80}, 3915--3918.

\bibitem[48]{Lehner:2001wq}
Lehner, Luis. 2001.
\newblock {\em Class. Quantum Grav.}, {\bf 18}, R25--R86.

\bibitem[49]{Bona05}
Bona, Carles, and Palenzuela, Carlos (eds). 2005.
\newblock {\em Elements of Numerical Relativity}.
\newblock Lecture Notes in Physics, vol. 673.
\newblock Berlin/Heidelberg: Springer.

\bibitem[50]{Alcubierre08}
Alcubierre, Miguel. 2008.
\newblock {\em Introduction to 3+1 Numerical Relativity}.
\newblock Oxford, U.K.: Oxford University Press.

\bibitem[51]{2010nure.book.....B}
{Baumgarte}, T.~W., and {Shapiro}, S.~L. 2010.
\newblock {\em {Numerical Relativity: Solving Einstein's Equations on the
  Computer}}.

\bibitem[52]{2012LNP...846.....G}
{Gourgoulhon}, E. (ed). 2012.
\newblock {\em {3+1 Formalism in General Relativity}}.
\newblock Lecture Notes in Physics, Berlin Springer Verlag, vol. 846.

\bibitem[53]{Kreiss:2001cu}
Kreiss, Heinz~O., and Ortiz, Omar~E. 2002.
\newblock {\em Lect.Notes Phys.}, {\bf 604}, 359.

\bibitem[54]{Gustafsson95}
Gustafsson, Bertil, Kreiss, Heinz-Otto, and Oliger, Joseph. 1995.
\newblock {\em Time dependent problems and difference methods}.
\newblock New York: Wiley.

\bibitem[55]{Sarbach:2012pr}
Sarbach, Olivier, and Tiglio, Manuel. 2012.
\newblock {\em Living Rev.Rel.}, {\bf 15}, 9.

\bibitem[56]{Friedrich:2000qv}
Friedrich, Helmut, and Rendall, Alan~D. 2000.
\newblock {\em Lect. Notes Phys.}, {\bf 540}, 127--224.

\bibitem[57]{Reula:1998LR}
Reula, Oscar~A. 1998.
\newblock {\em Living Rev. Relativity}, {\bf 1}(3).

\bibitem[58]{Alic:2011gg}
Alic, Daniela, {et~al.} 2012.
\newblock {\em Phys. Rev. D}, {\bf 85}, 064040.

\bibitem[59]{Nakamura:1987zz}
Nakamura, Takashi, Oohara, Ken-Ichi, and Kojima, Y. 1987.
\newblock {\em Prog. Theor. Phys. Suppl.}, {\bf 90}, 1--218.

\bibitem[60]{Shibata:1995we}
Shibata, Masaru, and Nakamura, Takashi. 1995.
\newblock {\em Phys. Rev. D}, {\bf 52}, 5428--5444.

\bibitem[61]{Baumgarte:1998te}
Baumgarte, Thomas~W., and Shapiro, Stuart~L. 1998.
\newblock {\em Phys. Rev. D}, {\bf 59}, 024007.

\bibitem[62]{1999ewgr.book...87R}
{Renn}, J., and {Sauer}, T. 1999.
\newblock {Heuristics and Mathematical Representation in Einstein's Search for
  a Gravitational Field Equation}.
\newblock {Page ~87 of:} {Goenner}, H., {Renn}, J., {Ritter}, J., and {Sauer},
  T. (eds), {\em The Expanding Worlds of General Relativity}.

\bibitem[63]{Lindblom:2005qh}
Lindblom, Lee, {et~al.} 2006.
\newblock {\em Class. Quantum Grav.}, {\bf 23}, S447--S462.

\bibitem[64]{Garfinkle:2001ni}
Garfinkle, David. 2002.
\newblock {\em Phys. Rev. D}, {\bf 65}, 044029.

\bibitem[65]{Gundlach:2005eh}
Gundlach, Carsten, {et~al.} 2005.
\newblock {\em Class. Quantum Grav.}, {\bf 22}, 3767--3774.

\bibitem[66]{Brodbeck:1998az}
Brodbeck, O, {et~al.} 1999.
\newblock {\em J. Math. Phys.}, {\bf 40}, 909--923.

\bibitem[67]{Palenzuela:2009hx}
Palenzuela, Carlos, Lehner, Luis, and Yoshida, Shin'ichirou. 2010.
\newblock {\em Phys. Rev. D}, {\bf 81}, 084007.

\bibitem[68]{Headrick:2009pv}
Headrick, Matthew, Kitchen, Sam, and Wiseman, Toby. 2010.
\newblock {\em Class.Quant.Grav.}, {\bf 27}, 035002.

\bibitem[69]{Hannam:2006vv}
Hannam, Mark~D., Husa, Sascha, Pollney, Denis, Br\"{u}gmann, Bernd, and
  \'{O}~Murchadha, Niall. 2007.
\newblock {\em Phys. Rev. Lett.}, {\bf 99}, 241102.

\bibitem[70]{Hannam:2008sg}
Hannam, Mark~D., Husa, Sascha, Ohme, Frank, Br\"{u}gmann, Bernd, and
  \'{O}~Murchadha, Niall. 2008.
\newblock {\em Phys. Rev. D}, {\bf 78}, 064020.

\bibitem[71]{Winicour:1998LR}
Winicour, Jeffrey. 1998.
\newblock {\em Living Rev. Relativity}, {\bf 1}(5).

\bibitem[72]{Gomez:1997pd}
G\'{o}mez, Roberto, {et~al.} 1998.
\newblock {\em Phys. Rev. D}, {\bf 57}, 4778--4788.

\bibitem[73]{Bishop:1996gt}
Bishop, Nigel~T., {et~al.} 1997.
\newblock {\em Phys. Rev. D}, {\bf 54}(10), 6153--6165.

\bibitem[74]{Reisswig:2009rx}
Reisswig, Christian, {et~al.} 2010.
\newblock {\em Class. Quantum Grav.}, {\bf 27}, 075014.

\bibitem[75]{1995PhRvL..75.1256B}
{Brady}, P.~R., and {Smith}, J.~D. 1995.
\newblock {\em Physical Review Letters}, {\bf 75}, 1256--1259.

\bibitem[76]{Chesler:2013lia}
Chesler, Paul~M., and Yaffe, Laurence~G. 2013.
\newblock [arXiv:1309.1439].

\bibitem[77]{Friedrich:2002xz}
Friedrich, Helmut. 2002.
\newblock {\em Lect.Notes Phys.}, {\bf 604}, 1--50.

\bibitem[78]{Frauendiener:1997zc}
Frauendiener, J\"{o}rg. 1998.
\newblock {\em Phys. Rev. D}, {\bf 58}, 064002.

\bibitem[79]{Husa:2002kk}
Husa, Sascha. 2002.
\newblock Problems and Successes in the Numerical Approach to the Conformal
  Field Equations.
\newblock {Pages  239--260 of:} Frauendiener, J\"{o}rg, and Friedrich, Helmut
  (eds), {\em The Conformal Structure of Spacetimes: Geometry, Analysis,
  Numerics}.
\newblock Lecture Notes in Physics, vol. 604.
\newblock Berlin/Heidelberg: Springer.

\bibitem[80]{Berger:1984zza}
Berger, Marsha~J., and Oliger, Joseph. 1984.
\newblock {\em J. Comp. Phys.}, {\bf 53}, 484.

\bibitem[81]{Choptuik89}
Choptuik, Matthew~W. 1989.
\newblock Experiences with an Adaptive Mesh Refinement Algorithm in Numerical
  Relativity.
\newblock {In:} Evans, Charles~R., Finn, Lee~S., and Hobill, David~W. (eds),
  {\em Frontiers in Numerical Relativity}.
\newblock Cambridge, U.K.: Cambridge University Press.

\bibitem[82]{Lehner:2005vc}
Lehner, Luis, Liebling, Steven~L., and Reula, Oscar~A. 2006.
\newblock {\em Class. Quantum Grav.}, {\bf 23}, S421--S446.

\bibitem[83]{Boyd89a}
Boyd, J.~P. 1989.
\newblock {\em {C}hebyshev and {F}ourier {S}pectral {M}ethods}.
\newblock New York: Springer-Verlag.

\bibitem[84]{Grandclement:2009LR}
Grandclement, Philippe, and Novak, Jerome. 2009.
\newblock {\em Living Rev. Relativity}, {\bf 12}(1).

\bibitem[85]{Szilagyi:2009qz}
Szil\'{a}gyi, B\'{e}la, Lindblom, Lee, and Scheel, Mark~A. 2009.
\newblock {\em Phys. Rev. D}, {\bf 80}, 124010.

\bibitem[86]{2008LRR....11....7F}
{Font}, J.~A. 2008.
\newblock {\em Living Reviews in Relativity}, {\bf 11}(Sept.), 7.

\bibitem[87]{Leveque92}
LeVeque, Randall~J. 1992.
\newblock {\em Numerical Methods for Conservation Laws}.
\newblock Basel: Birkhauser Verlag.

\bibitem[88]{upcomingNRLR}
Cardoso, Vitor, Gualtieri, Leonardo, Herdeiro, Carlos, and Sperhake, Ulrich.
  2014.
\newblock {\em Exploring New Physics Frontiers Through Numerical Relativity}.
\newblock To appear in Living Reviews in Relativity.

\bibitem[89]{Evans:1994pj}
Evans, Charles~R., and Coleman, Jason~S. 1994.
\newblock {\em Phys.Rev.Lett.}, {\bf 72}, 1782--1785.

\bibitem[90]{Koike:1995jm}
Koike, Tatsuhiko, Hara, Takashi, and Adachi, Satoshi. 1995.
\newblock {\em Phys.Rev.Lett.}, {\bf 74}, 5170--5173.

\bibitem[91]{Maison:1995cc}
Maison, Dieter. 1996.
\newblock {\em Phys.Lett.}, {\bf B366}, 82--84.

\bibitem[92]{Gundlach:2007gc}
Gundlach, Carsten, and Martin-Garcia, Jose~M. 2007.
\newblock {\em Living Rev.Rel.}, {\bf 10}, 5.

\bibitem[93]{Gundlach:1997wm}
Gundlach, Carsten. 1998.
\newblock {\em Adv. Theor. Math. Phys}, {\bf 2}, 1--49.

\bibitem[94]{Choptuik:1992jv}
Choptuik, Matthew~W. 1993.
\newblock {\em Phys. Rev. Lett.}, {\bf 70}, 9--12.

\bibitem[95]{Gundlach:1996eg}
Gundlach, Carsten. 1997.
\newblock {\em Phys. Rev. D}, {\bf 55}, 695--713.

\bibitem[96]{Hod:1996ar}
Hod, Shahar, and Piran, Tsvi. 1997.
\newblock {\em Phys.Rev.}, {\bf D55}, 3485--3496.

\bibitem[97]{MartinGarcia:1998sk}
Martin-Garcia, Jose~M., and Gundlach, Carsten. 1999.
\newblock {\em Phys. Rev. D}, {\bf 59}, 064031.

\bibitem[98]{Choptuik:2003ac}
Choptuik, Matthew~W., {et~al.} 2003.
\newblock {\em Phys. Rev. D}, {\bf 68}, 044007.

\bibitem[99]{Healy:2013xia}
Healy, James, and Laguna, Pablo. 2013.
\newblock [arXiv:1310.1955].

\bibitem[100]{Brady:1997fj}
Brady, Patrick~R., Chambers, Chris~M., and Goncalves, Sergio~M.C.V. 1997.
\newblock {\em Phys.Rev.}, {\bf D56}, 6057--6061.

\bibitem[101]{Seidel:1991zh}
Seidel, Edward, and Suen, Wai-Mo. 1991.
\newblock {\em Phys. Rev. Lett.}, {\bf 66}, 1659--1662.

\bibitem[102]{Hawley:2000dt}
Hawley, Scott~H., and Choptuik, Matthew~W. 2000.
\newblock {\em Phys. Rev. D}, {\bf 62}, 104024.

\bibitem[103]{Husain:2000vm}
Husain, Viqar, and Olivier, Michel. 2001.
\newblock {\em Class.Quant.Grav.}, {\bf 18}, L1--L10.

\bibitem[104]{Pretorius:2000yu}
Pretorius, Frans, and Choptuik, Matthew~W. 2000.
\newblock {\em Phys. Rev. D}, {\bf 62}, 124012.

\bibitem[105]{Bizon:2011gg}
Bizon, Piotr, and Rostworowski, Andrzej. 2011.
\newblock {\em Phys.Rev.Lett.}, {\bf 107}, 031102.

\bibitem[106]{Garfinkle:2000br}
Garfinkle, David. 2001.
\newblock {\em Phys.Rev.}, {\bf D63}, 044007.

\bibitem[107]{Abrahams:1993wa}
Abrahams, Andrew~M., and Evans, Charles~R. 1993.
\newblock {\em Phys. Rev. Lett.}, {\bf 70}, 2980--2983.

\bibitem[108]{Sorkin:2010tm}
Sorkin, Evgeny. 2011.
\newblock {\em Class. Quantum Grav.}, {\bf 28}, 025001.

\bibitem[109]{Garfinkle:1998va}
Garfinkle, David, and Duncan, G.~Comer. 1998.
\newblock {\em Phys.Rev.}, {\bf D58}, 064024.

\bibitem[110]{Bizon:2005cp}
Bizon, Piotr, Chmaj, Tadeusz, and Schmidt, Bernd~G. 2005.
\newblock {\em Phys.Rev.Lett.}, {\bf 95}, 071102.

\bibitem[111]{Bizon:2006qi}
Bizon, Piotr, Chmaj, Tadeusz, and Schmidt, Bernd~G. 2006.
\newblock {\em Phys.Rev.Lett.}, {\bf 97}, 131101.

\bibitem[112]{Bizon:2005af}
Bizon, P., {et~al.} 2005.
\newblock {\em Phys.Rev.}, {\bf D72}, 121502.

\bibitem[113]{Szybka:2007fx}
Szybka, Sebastian~J., and Chmaj, Tadeusz. 2008.
\newblock {\em Phys.Rev.Lett.}, {\bf 100}, 101102.

\bibitem[114]{Gundlach:1997nb}
Gundlach, Carsten. 1998.
\newblock {\em Phys. Rev. D}, {\bf 57}, R7080.

\bibitem[115]{Gundlach:1997vy}
Gundlach, Carsten. 1998.
\newblock {\em Phys. Rev. D}, {\bf 57}, R7075--R7079.

\bibitem[116]{Cahill1971}
Cahill, M.~E., and Taub, A.~H. 1971.
\newblock {\em Commun.Math.Phys.}, {\bf 21}, 1--40.

\bibitem[117]{Noble:2003xx}
Noble, Scott~Charles. 2003.
\newblock {\em {A Numerical study of relativistic fluid collapse}}.
\newblock Ph.D. thesis, The University of British Columbia, Vancouver, British
  Columbia.
\newblock [arXiv:gr-qc/0310116].

\bibitem[118]{Neilsen:1998qc}
Neilsen, David~W., and Choptuik, Matthew~W. 2000.
\newblock {\em Class.Quant.Grav.}, {\bf 17}, 761--782.

\bibitem[119]{Noble:2007vf}
Noble, Scott~C., and Choptuik, Matthew~W. 2008.
\newblock {\em Phys.Rev.}, {\bf D78}, 064059.

\bibitem[120]{Novak:2001ck}
Novak, Jerome. 2001.
\newblock {\em Astron.Astrophys.}, {\bf 376}, 606--613.

\bibitem[121]{Niemeyer:1997mt}
Niemeyer, Jens~C., and Jedamzik, K. 1998.
\newblock {\em Phys.Rev.Lett.}, {\bf 80}, 5481--5484.

\bibitem[122]{Jin:2006gm}
Jin, Ke-Jian, and Suen, Wai-Mo. 2007.
\newblock {\em Phys.Rev.Lett.}, {\bf 98}, 131101.

\bibitem[123]{Kellermann:2010rt}
Kellermann, Thorsten, Rezzolla, Luciano, and Radice, David. 2010.
\newblock {\em Class.Quant.Grav.}, {\bf 27}, 235016.

\bibitem[124]{Radice:2010rw}
Radice, David, Rezzolla, Luciano, and Kellermann, Thorsten. 2010.
\newblock {\em Class.Quant.Grav.}, {\bf 27}, 235015.

\bibitem[125]{Wan:2011wg}
Wan, Mew-Bing. 2011.
\newblock {\em Class.Quant.Grav.}, {\bf 28}, 155002.

\bibitem[126]{Liebling:2010bn}
Liebling, Steven~L., {et~al.} 2010.
\newblock {\em Phys. Rev. D}, {\bf 81}, 124023.

\bibitem[127]{Choptuik:1996yg}
Choptuik, Matthew~W., Chmaj, Tadeusz, and Bizon, Piotr. 1996.
\newblock {\em Phys.Rev.Lett.}, {\bf 77}, 424--427.

\bibitem[128]{Bartnik:1988am}
Bartnik, R., and Mckinnon, J. 1988.
\newblock {\em Phys.Rev.Lett.}, {\bf 61}, 141--144.

\bibitem[129]{Choptuik:1999gh}
Choptuik, Matthew~W., Hirschmann, Eric~W., and Marsa, Robert~L. 1999.
\newblock {\em Phys.Rev.}, {\bf D60}, 124011.

\bibitem[130]{Liebling:1996dx}
Liebling, Steven~L., and Choptuik, Matthew~W. 1996.
\newblock {\em Phys.Rev.Lett.}, {\bf 77}, 1424--1427.

\bibitem[131]{Lechner:2001ng}
Lechner, Christiane, Thornburg, Jonathan, Husa, Sascha, and Aichelburg,
  Peter~C. 2002.
\newblock {\em Phys. Rev. D}, {\bf 65}(8), 081501(R).

\bibitem[132]{Andreasson:2006gx}
Andreasson, Hakan, and Rein, Gerhard. 2006.
\newblock {\em Class.Quant.Grav.}, {\bf 23}, 3659--3678.

\bibitem[133]{Olabarrieta:2001wy}
Olabarrieta, Ignacio, and Choptuik, Matthew~W. 2002.
\newblock {\em Phys.Rev.}, {\bf D65}, 024007.

\bibitem[134]{Rein:1998uf}
Rein, Gerhard, Rendall, Alan~D., and Schaeffer, Jack. 1998.
\newblock {\em Phys.Rev.}, {\bf D58}, 044007.

\bibitem[135]{MartinGarcia:2001nh}
Martin-Garcia, Jose~M., and Gundlach, Carsten. 2002.
\newblock {\em Phys.Rev.}, {\bf D65}, 084026.

\bibitem[136]{2010ARNPS..60...75C}
{Centrella}, J., {et~al.} 2010.
\newblock {\em Annual Review of Nuclear and Particle Science}, {\bf 60}(Nov.),
  75--100.

\bibitem[137]{Hinder:2013oqa}
Hinder, Ian, {et~al.} 2013.
\newblock {\em Class.Quant.Grav.}, {\bf 31}, 025012.

\bibitem[138]{2011PhRvL.106x1101A}
{Ajith}, P., {et~al.} 2011.
\newblock {\em Physical Review Letters}, {\bf 106}(24), 241101.

\bibitem[139]{2013arXiv1307.6232P}
{Pan}, Y., {et~al.} 2013.
\newblock [arXiv:1307.6232].

\bibitem[140]{2013PhRvD..87h4035D}
{Damour}, T., {Nagar}, A., and {Bernuzzi}, S. 2013.
\newblock {\em \prd}, {\bf 87}(8), 084035.

\bibitem[141]{Buonanno:2006ui}
Buonanno, Alessandra, Cook, Gregory~B., and Pretorius, Frans. 2007.
\newblock {\em Phys. Rev. D}, {\bf 75}, 124018.

\bibitem[142]{1997PhRvD..56.6298B}
{Bishop}, N.~T., {et~al.} 1997.
\newblock {\em \prd}, {\bf 56}(Nov.), 6298--6309.

\bibitem[143]{2011PhRvD..83j4018C}
{Chu}, T., {Pfeiffer}, H.~P., and {Cohen}, M.~I. 2011.
\newblock {\em \prd}, {\bf 83}(10), 104018.

\bibitem[144]{Gonzalez:2006md}
Gonz\'{a}lez, Jos\'{e}~A., {et~al.} 2007.
\newblock {\em Phys. Rev. Lett.}, {\bf 98}, 091101.

\bibitem[145]{Baker:2006vn}
Baker, John~G., {et~al.} 2006.
\newblock {\em Astrophys. J.}, {\bf 653}, L93--L96.

\bibitem[146]{2007CQGra..24S..33H}
{Herrmann}, F., {et~al.} 2007.
\newblock {\em Classical and Quantum Gravity}, {\bf 24}(June), 33.

\bibitem[147]{Berti:2007fi}
Berti, Emanuele, {et~al.} 2007.
\newblock {\em Phys. Rev. D}, {\bf 76}, 064034.

\bibitem[148]{Campanelli:2006uy}
Campanelli, Manuela, Lousto, Carlos~O., and Zlochower, Yosef. 2006.
\newblock {\em Phys. Rev. D}, {\bf 74}, 041501(R).

\bibitem[149]{Hemberger:2013hsa}
Hemberger, Daniel~A., {et~al.} 2013.
\newblock {\em Phys.Rev.}, {\bf D88}, 064014.

\bibitem[150]{2011PhRvD..84b4046S}
{Schmidt}, P., {et~al.} 2011.
\newblock {\em \prd}, {\bf 84}(2), 024046.

\bibitem[151]{2011PhRvD..84l4011B}
{Boyle}, M., {Owen}, R., and {Pfeiffer}, H.~P. 2011.
\newblock {\em \prd}, {\bf 84}(12), 124011.

\bibitem[152]{2011PhRvD..84l4002O}
{O'Shaughnessy}, R., {et~al.} 2011.
\newblock {\em \prd}, {\bf 84}(12), 124002.

\bibitem[153]{2007ApJ...659L...5C}
{Campanelli}, M., {et~al.} 2007.
\newblock {\em \apjl}, {\bf 659}(Apr.), L5--L8.

\bibitem[154]{2007PhRvL..98w1101G}
{Gonz{\'a}lez}, J.~A., {et~al.} 2007.
\newblock {\em Physical Review Letters}, {\bf 98}(23), 231101.

\bibitem[155]{Lousto:2011kp}
Lousto, Carlos~O., and Zlochower, Yosef. 2011.
\newblock {\em Phys. Rev. Lett.}, {\bf 107}, 231102.

\bibitem[156]{2012AdAst2012E..14K}
{Komossa}, S. 2012.
\newblock {\em Advances in Astronomy}, {\bf 2012}.

\bibitem[157]{2013arXiv1307.3542S}
Schnittman, Jeremy~D. 2013.
\newblock {\em Class.Quant.Grav.}, {\bf 30}, 244007.

\bibitem[158]{Komossa:2008qd}
Komossa, Stefanie, Zhou, H., and Lu, H. 2008.
\newblock {\em Astrophys. J.}, {\bf 678}, L81--L84.

\bibitem[159]{Lippai:2008fx}
Lippai, Zolt\'{a}n, Frei, Zsolt, and Haiman, Zolt\'{a}n. 2008.
\newblock {\em Astrophys. J.}, {\bf 676}, L5--L8.

\bibitem[160]{Milosavljevic:2004cg}
Milosavljevi\'{c}, Milo\v{s}, and Phinney, E.~Sterl. 2005.
\newblock {\em Astrophys. J.}, {\bf 622}, L93--L96.

\bibitem[161]{2007PhRvL..99d1103L}
{Loeb}, A. 2007.
\newblock {\em Physical Review Letters}, {\bf 99}(4), 041103.

\bibitem[162]{2012ApJ...755...51N}
{Noble}, S.~C., {et~al.} 2012.
\newblock {\em \apj}, {\bf 755}, 51.

\bibitem[163]{2012PhRvL.109v1102F}
{Farris}, B.~D., {et~al.} 2012.
\newblock {\em Physical Review Letters}, {\bf 109}(22), 221102.

\bibitem[164]{2011MNRAS.412...75S}
{Stone}, N., and {Loeb}, A. 2011.
\newblock {\em \mnras}, {\bf 412}, 75--80.

\bibitem[165]{Palenzuela:2010nf}
Palenzuela, Carlos, Lehner, Luis, and Leibling, Steven~L. 2010.
\newblock {\em Science}, {\bf 329}, 927--930.

\bibitem[166]{East:2012xq}
East, William~E., {et~al.} 2013.
\newblock {\em Phys.Rev.}, {\bf D87}(4), 043004.

\bibitem[167]{Pretorius:2007jn}
Pretorius, Frans, and Khurana, Deepak. 2007.
\newblock {\em Class. Quantum Grav.}, {\bf 24}, S83--S108.

\bibitem[168]{2009PhRvL.103m1101H}
{Healy}, J., {Levin}, J., and {Shoemaker}, D. 2009.
\newblock {\em Physical Review Letters}, {\bf 103}(13), 131101.

\bibitem[169]{2012arXiv1209.4085G}
Gold, Roman, and Bruegmann, Bernd. 2013.
\newblock {\em Phys.Rev.}, {\bf D88}, 064051.

\bibitem[170]{Metzger:2011bv}
Metzger, B.D., and Berger, E. 2012.
\newblock {\em Astrophys.J.}, {\bf 746}, 48.

\bibitem[171]{2013MNRAS.430.2121P}
{Piran}, T., {Nakar}, E., and {Rosswog}, S. 2013.
\newblock {\em Mon. Not. R. Astron. Soc.}, {\bf 430}, 2121--2136.

\bibitem[172]{Shibata:1999wm}
Shibata, Masaru, and Uryu, Koji. 2000.
\newblock {\em Phys. Rev. D}, {\bf 61}, 064001.

\bibitem[173]{Nakamura99a}
Nakamura, Takashi, and Oohara, Ken-Ichi. 1999.
\newblock {\em {A Way to 3D Numerical Relativity --- Coalescing Binary Neutron
  Stars}}.
\newblock {arXiv:gr-qc/9812054}.

\bibitem[174]{Read:2013zra}
Read, Jocelyn~S., {et~al.} 2013.
\newblock {\em Phys.Rev.}, {\bf D88}, 044042.

\bibitem[175]{Lackey:2013axa}
Lackey, Benjamin~D., {et~al.} 2013.
\newblock [arXiv:1303.6298.

\bibitem[176]{Tsang:2011ad}
Tsang, David, {et~al.} 2012.
\newblock {\em Phys.Rev.Lett.}, {\bf 108}, 011102.

\bibitem[177]{Hotokezaka:2011dh}
Hotokezaka, Kenta, {et~al.} 2011.
\newblock {\em Phys.Rev.}, {\bf D83}, 124008.

\bibitem[178]{Anderson:2008zp}
Anderson, Matthew, {et~al.} 2008.
\newblock {\em Phys.Rev.Lett.}, {\bf 100}, 191101.

\bibitem[179]{Sekiguchi:2012uc}
Sekiguchi, Yuichiro, {et~al.} 2012.
\newblock [arXiv:1206.5927].

\bibitem[180]{Kaplan:2013wra}
Kaplan, J.D., {et~al.} 2013.
\newblock {\em Phys.Rev.}, {\bf D88}, 064009.

\bibitem[181]{Rezzolla:2010fd}
Rezzolla, Luciano, {et~al.} 2010.
\newblock {\em Class. Quantum Grav.}, {\bf 27}, 114105.

\bibitem[182]{2013Natur.500..547T}
{Tanvir}, N.~R., {et~al.} 2013.
\newblock {\em \nat}, {\bf 500}, 547--549.

\bibitem[183]{2013ApJ...774L..23B}
{Berger}, E., {Fong}, W., and {Chornock}, R. 2013.
\newblock {\em \apjl}, {\bf 774}(Sept.), L23.

\bibitem[184]{Hinderer:2009ca}
Hinderer, Tanja, {et~al.} 2010.
\newblock {\em Phys. Rev. D}, {\bf 81}, 123016.

\bibitem[185]{Markakis:2011vd}
Markakis, Charalampos, {et~al.} 2009.
\newblock {\em J. Phys. Conf. Ser.}, {\bf 189}, 012024.

\bibitem[186]{Sekiguchi:2011zd}
Sekiguchi, Yuichiro, {et~al.} 2011.
\newblock {\em Phys.Rev.Lett.}, {\bf 107}, 051102.

\bibitem[187]{Lehner:2011aa}
Lehner, Luis, {et~al.} 2012.
\newblock {\em Phys.Rev.}, {\bf D86}, 104035.

\bibitem[188]{Kyutoku:2012fv}
Kyutoku, Koutarou, Ioka, Kunihito, and Shibata, Masaru. 2012.

\bibitem[189]{Palenzuela:2013hu}
Palenzuela, Carlos, {et~al.} 2013.
\newblock {\em Phys.Rev.Lett.}, {\bf 111}, 061105.

\bibitem[190]{2013PhRvD..88d4026H}
{Hotokezaka}, K., {et~al.} 2013.
\newblock {\em \prd}, {\bf 88}(4), 044026.

\bibitem[191]{Kyutoku:2013wxa}
Kyutoku, Koutarou, Ioka, Kunihito, and Shibata, Masaru. 2013.
\newblock {\em Phys.Rev.}, {\bf D88}, 041503.

\bibitem[192]{Chawla:2010sw}
Chawla, Sarvnipun, {et~al.} 2010.
\newblock {\em Phys. Rev. Lett.}, {\bf 105}, 111101.

\bibitem[193]{Foucart:2012vn}
Foucart, Francois, {et~al.} 2013.
\newblock {\em Phys.Rev.}, {\bf D87}, 084006.

\bibitem[194]{Foucart:2012nc}
Foucart, Francois. 2012.
\newblock {\em Phys.Rev.}, {\bf D86}, 124007.

\bibitem[195]{Hansen:2000am}
Hansen, Brad~M.S., and Lyutikov, Maxim. 2001.
\newblock {\em Mon.Not.Roy.Astron.Soc.}, {\bf 322}, 695.

\bibitem[196]{McWilliams:2011zi}
McWilliams, Sean~T., and Levin, Janna~J. 2011.
\newblock {\em Astrophys. J.}, {\bf 742}, 90.

\bibitem[197]{Paschalidis:2013jsa}
Paschalidis, Vasileios, Etienne, Zachariah~B., and Shapiro, Stuart~L. 2013.
\newblock {\em Phys.Rev.}, {\bf D88}, 021504.

\bibitem[198]{Lackey:2011vz}
Lackey, Benjamin~D., {et~al.} 2012.
\newblock {\em Phys.Rev.}, {\bf D85}, 044061.

\bibitem[199]{bhns_astro_letter}
Stephens, Branson~C., East, William~E., and Pretorius, Frans. 2011.
\newblock {\em Astrophys. J. Lett.}, {\bf 737}(1), L5.

\bibitem[200]{Gold2011}
Gold, Roman, {et~al.} 2012.
\newblock {\em Phys.Rev.}, {\bf D86}, 121501.

\bibitem[201]{Tsang:2013mca}
Tsang, David. 2013.
\newblock {\em Astrophys.J.}, {\bf 777}, 103.

\bibitem[202]{Duez:2009yz}
Duez, Matthew~D. 2010.
\newblock {\em Class.Quant.Grav.}, {\bf 27}, 114002.

\bibitem[203]{Pfeiffer:2012pc}
Pfeiffer, Harald~P. 2012.
\newblock {\em Class.Quant.Grav.}, {\bf 29}, 124004.

\bibitem[204]{Faber:2012rw}
Faber, Joshua~A., and Rasio, Frederic~A. 2012.
\newblock {\em Living Rev.Rel.}, {\bf 15}, 8.

\bibitem[205]{2009CQGra..26f3001O}
{Ott}, C.~D. 2009.
\newblock {\em Classical and Quantum Gravity}, {\bf 26}(6), 063001.

\bibitem[206]{2007PhR...442...38J}
{Janka}, H.-T., {et~al.} 2007.
\newblock {\em \physrep}, {\bf 442}(Apr.), 38--74.

\bibitem[207]{2007PhR...442...23B}
{Burrows}, A., {et~al.} 2007.
\newblock {\em \physrep}, {\bf 442}, 23--37.

\bibitem[208]{Ott:2012mr}
Ott, Christian~D., {et~al.} 2013.
\newblock {\em Astrophys.J.}, {\bf 768}, 115.

\bibitem[209]{Dimmelmeier:2002bk}
Dimmelmeier, H., Font, Jos\'{e}~A., and M\"{u}ller, E. 2002.
\newblock {\em Astron. Astrophys.}, {\bf 388}, 917--935.

\bibitem[210]{Obergaulinger:2006qr}
Obergaulinger, Martin, {et~al.} 2006.
\newblock {\em Astron.Astrophys.}, {\bf 457}, 209--222.

\bibitem[211]{2010ApJS..189..104M}
{M{\"u}ller}, B., {Janka}, H.-T., and {Dimmelmeier}, H. 2010.
\newblock {\em Astrophys. J. Supp. Ser.}, {\bf 189}(July), 104--133.

\bibitem[212]{Wongwathanarat:2012zp}
Wongwathanarat, A., Janka, H.-Th., and Mueller, E. 2013.
\newblock {\em AA 552,}, {\bf A126}.

\bibitem[213]{2011PhRvL.106p1103O}
{Ott}, C.~D., {et~al.} 2011.
\newblock {\em Physical Review Letters}, {\bf 106}(16), 161103.

\bibitem[214]{Penrose}
Penrose, Roger. 1966.
\newblock General Relativistic Energy Flux and Elementary Optics.
\newblock {In:} Hoffman, Banesh (ed), {\em Perspectives in Geometry and
  Relativity}.
\newblock Indiana University Press.

\bibitem[215]{Aichelburg:1970dh}
Aichelburg, Peter~C., and Sexl, R.~U. 1971.
\newblock {\em Gen. Rel. Grav.}, {\bf 2}, 303--312.

\bibitem[216]{1998PhLB..429..263A}
{Arkani-Hamed}, N., {Dimopoulos}, S., and {Dvali}, G. 1998.
\newblock {\em Physics Letters B}, {\bf 429}(June), 263--272.

\bibitem[217]{1999PhRvL..83.3370R}
{Randall}, L., and {Sundrum}, R. 1999.
\newblock {\em Physical Review Letters}, {\bf 83}(Oct.), 3370--3373.

\bibitem[218]{Giddings:2001bu}
Giddings, Steven~B., and Thomas, Scott~D. 2002.
\newblock {\em Phys. Rev. D}, {\bf 65}, 056010.

\bibitem[219]{2002PhRvL..88b1303F}
{Feng}, J.~L., and {Shapere}, A.~D. 2002.
\newblock {\em Physical Review Letters}, {\bf 88}(2), 021303.

\bibitem[220]{Chatrchyan:2013xva}
Chatrchyan, Serguei, {et~al.} 2013.
\newblock {\em JHEP}, {\bf 1307}, 178.

\bibitem[221]{Aad:2013lna}
Aad, Georges, {et~al.} 2013.
\newblock {\em Phys.Rev.}, {\bf D88}, 072001.

\bibitem[222]{2007astro.ph..1333D}
{de los Heros}, C. 2007.
\newblock {\em ArXiv Astrophysics e-prints}, Jan.

\bibitem[223]{thorne_hoop}
Thorne, K.~S. 1972.
\newblock {Page  231 of:} Klauder, J. (ed), {\em {Magic Without Magic: John
  Archibald Wheeler}}.
\newblock San Francisco: Freeman.

\bibitem[224]{Sperhake:2008ga}
Sperhake, Ulrich, {et~al.} 2008.
\newblock {\em Phys. Rev. Lett.}, {\bf 101}, 161101.

\bibitem[225]{Shibata:2008rq}
Shibata, Masaru, Okawa, Hirotada, and Yamamoto, Tetsuro. 2008.
\newblock {\em Phys. Rev. D}, {\bf 78}, 101501(R).

\bibitem[226]{Sperhake:2009jz}
Sperhake, Ulrich, {et~al.} 2009.
\newblock {\em Phys.Rev.Lett.}, {\bf 103}, 131102.

\bibitem[227]{Sperhake:2010uv}
Sperhake, Ulrich, {et~al.} 2011.
\newblock {\em Phys.Rev.}, {\bf D83}, 024037.

\bibitem[228]{Sperhake:2012me}
Sperhake, Ulrich, Berti, Emanuele, Cardoso, Vitor, and Pretorius, Frans. 2013.
\newblock {\em Phys.Rev.Lett.}, {\bf 111}, 041101.

\bibitem[229]{Okawa:2011fv}
Okawa, Hirotada, Nakao, Ken-ichi, and Shibata, Masaru. 2011.
\newblock {\em Phys.Rev.}, {\bf D83}, 121501.

\bibitem[230]{1992PhRvD..46..694D}
{D'eath}, P.~D., and {Payne}, P.~N. 1992.
\newblock {\em \prd}, {\bf 46}(July), 694--701.

\bibitem[231]{Eardley:2002re}
Eardley, Douglas~M., and Giddings, Steven~B. 2002.
\newblock {\em Phys. Rev. D}, {\bf 66}, 044011.

\bibitem[232]{Berti:2010ce}
Berti, Emanuele, {et~al.} 2010.
\newblock {\em Phys. Rev. D}, {\bf 81}, 104048.

\bibitem[233]{Gundlach:2012aj}
Gundlach, Carsten, {et~al.} 2012.
\newblock {\em Phys.Rev.}, {\bf D86}, 084022.

\bibitem[234]{2010PhRvD..81j4012G}
{Gralla}, S.~E., {Harte}, A.~I., and {Wald}, R.~M. 2010.
\newblock {\em \prd}, {\bf 81}(10), 104012.

\bibitem[235]{Choptuik:2009ww}
Choptuik, Matthew~W., and Pretorius, Frans. 2010.
\newblock {\em Phys. Rev. Lett.}, {\bf 104}, 111101.

\bibitem[236]{East:2012mb}
East, William~E., and Pretorius, Frans. 2013.
\newblock {\em Phys.Rev.Lett.}, {\bf 110}(10), 101101.

\bibitem[237]{Rezzolla:2012nr}
Rezzolla, Luciano, and Takami, Kentaro. 2013.
\newblock {\em Class.Quant.Grav.}, {\bf 30}, 012001.

\bibitem[238]{Kaloper:2007pb}
Kaloper, Nemanja, and Terning, John. 2008.
\newblock {\em Int.J.Mod.Phys.}, {\bf D17}, 665--672.

\bibitem[239]{Galley:2013eba}
Galley, Chad~R., and Porto, Rafael~A. 2013.
\newblock {\em JHEP}, {\bf 1311}, 096.

\bibitem[240]{tsov:2012sn}
Gal'tsov, Dmitry, Spirin, Pavel, and Tomaras, Theodore~N. 2013.
\newblock {\em JHEP}, {\bf 1301}, 087.

\bibitem[241]{Chesler:2015wra}
Chesler, Paul~M., and Yaffe, Laurence~G. 2015.

\bibitem[242]{Grumiller:2008va}
Grumiller, Daniel, and Romatschke, Paul. 2008.
\newblock {\em JHEP}, {\bf 0808}, 027.

\bibitem[243]{Casalderrey-Solana:2013aba}
Casalderrey-Solana, Jorge, Heller, Michal~P., Mateos, David, and van~der Schee,
  Wilke. 2013.
\newblock {\em Phys. Rev. Lett. 111,}, {\bf 181601}.

\bibitem[244]{Cardoso:2012qm}
Cardoso, Vitor, Gualtieri, Leonardo, Herdeiro, Carlos, Sperhake, Ulrich,
  Chesler, Paul~M., {et~al.} 2012.

\bibitem[245]{1998PhLB..436..257A}
{Antoniadis}, I., {et~al.} 1998.
\newblock {\em Physics Letters B}, {\bf 436}(Sept.), 257--263.

\bibitem[246]{Baumann:2009ds}
Baumann, Daniel. 2009.

\bibitem[247]{Lehners:2008vx}
Lehners, Jean-Luc. 2008.
\newblock {\em Phys.Rept.}, {\bf 465}, 223--263.

\bibitem[248]{lrr-2005-1}
Carlip, Steven. 2005.
\newblock {\em Living Reviews in Relativity}, {\bf 8}(1).

\bibitem[249]{Gegenberg:2009ny}
Gegenberg, J., and Kunstatter, G. 2009.
\newblock [arXiv:0902.0292].

\bibitem[250]{Maldacena:1997re}
Maldacena, Juan~Martin. 1998.
\newblock {\em Adv.Theor.Math.Phys.}, {\bf 2}, 231--252.

\bibitem[251]{Aharony:1999ti}
Aharony, Ofer, {et~al.} 2000.
\newblock {\em Phys.Rept.}, {\bf 323}, 183--386.

\bibitem[252]{Horowitz:2012nnc}
Horowitz, Gary~T. 2012.

\bibitem[253]{Emparan:2008eg}
Emparan, Roberto, and Reall, Harvey~S. 2008.
\newblock {\em Living Rev.Rel.}, {\bf 11}, 6.

\bibitem[254]{Reall:2012it}
Reall, Harvey~S. 2012.
\newblock [arXiv:1210.1402].

\bibitem[255]{Emparan:2003sy}
Emparan, Roberto, and Myers, Robert~C. 2003.
\newblock {\em JHEP}, {\bf 0309}, 025.

\bibitem[256]{Lehner:2010pn}
Lehner, Luis, and Pretorius, Frans. 2010.
\newblock {\em Phys.Rev.Lett.}, {\bf 105}, 101102.

\bibitem[257]{Shibata:2009ad}
Shibata, Masaru, and Yoshino, Hirotaka. 2010.
\newblock {\em Phys.Rev.}, {\bf D81}, 021501.

\bibitem[258]{Shibata:2010wz}
Shibata, Masaru, and Yoshino, Hirotaka. 2010.
\newblock {\em Phys.Rev.}, {\bf D81}, 104035.

\bibitem[259]{Gregory:1993vy}
Gregory, R., and Laflamme, R. 1993.
\newblock {\em Phys.Rev.Lett.}, {\bf 70}, 2837--2840.

\bibitem[260]{eggersprl}
Eggers, JG. 1993.
\newblock {\em Phys. Rev. Lett}, {\bf 71}, 3458.

\bibitem[261]{Sorkin:2004qq}
Sorkin, Evgeny. 2004.
\newblock {\em Phys.Rev.Lett.}, {\bf 93}, 031601.

\bibitem[262]{Figueras:2012xj}
Figueras, Pau, Murata, Keiju, and Reall, Harvey~S. 2012.
\newblock {\em JHEP}, {\bf 1211}, 071.

\bibitem[263]{Dias:2009iu}
Dias, Oscar~J.C., {et~al.} 2009.
\newblock {\em Phys.Rev.}, {\bf D80}, 111701.

\bibitem[264]{Figueras:2011gd}
Figueras, Pau, and Wiseman, Toby. 2011.
\newblock {\em Phys.Rev.Lett.}, {\bf 107}, 081101.

\bibitem[265]{Tanaka:2002rb}
Tanaka, Takahiro. 2003.
\newblock {\em Prog.Theor.Phys.Suppl.}, {\bf 148}, 307--316.

\bibitem[266]{Emparan:2002px}
Emparan, Roberto, Fabbri, Alessandro, and Kaloper, Nemanja. 2002.
\newblock {\em JHEP}, {\bf 0208}, 043.

\bibitem[267]{Son:2007vk}
Son, Dam~T., and Starinets, Andrei~O. 2007.
\newblock {\em Ann.Rev.Nucl.Part.Sci.}, {\bf 57}, 95--118.

\bibitem[268]{DeWolfe:2013cua}
DeWolfe, Oliver, {et~al.} 2013.
\newblock [arXiv:1304.7794].

\bibitem[269]{Chesler:2008hg}
Chesler, Paul~M., and Yaffe, Laurence~G. 2009.
\newblock {\em Phys.Rev.Lett.}, {\bf 102}, 211601.

\bibitem[270]{Chesler:2011ds}
Chesler, Paul~M., and Teaney, Derek. 2011.
\newblock [arXiv:1112.6196].

\bibitem[271]{Chesler:2010bi}
Chesler, Paul~M., and Yaffe, Laurence~G. 2011.
\newblock {\em Phys.Rev.Lett.}, {\bf 106}, 021601.

\bibitem[272]{Bjorken:1982qr}
Bjorken, J.D. 1983.
\newblock {\em Phys.Rev.}, {\bf D27}, 140--151.

\bibitem[273]{Luzum:2008cw}
Luzum, Matthew, and Romatschke, Paul. 2008.
\newblock {\em Phys.Rev.}, {\bf C78}, 034915.

\bibitem[274]{Buchel:2012gw}
Buchel, Alex, Lehner, Luis, and Myers, Robert~C. 2012.
\newblock {\em JHEP}, {\bf 1208}, 049.

\bibitem[275]{Buchel:2013lla}
Buchel, Alex, {et~al.} 2013.
\newblock {\em JHEP}, {\bf 1305}, 067.

\bibitem[276]{Buchel:2013gba}
Buchel, Alex, Myers, Robert~C., and van Niekerk, Anton. 2013.
\newblock {\em Phys.Rev.Lett.}, {\bf 111}, 201602.

\bibitem[277]{Hubeny:2011hd}
Hubeny, Veronika~E., Minwalla, Shiraz, and Rangamani, Mukund. 2011.
\newblock [arXiv:1107.5780].

\bibitem[278]{VanRaamsdonk:2008fp}
Van~Raamsdonk, Mark. 2008.
\newblock {\em JHEP}, {\bf 0805}, 106.

\bibitem[279]{Carrasco:2012nf}
Carrasco, Federico, {et~al.} 2012.
\newblock {\em Phys.Rev.}, {\bf D86}, 126006.

\bibitem[280]{Adams:2013vsa}
Adams, Allan, Chesler, Paul~M., and Liu, Hong. 2013.
\newblock [arXiv:1307.7267].

\bibitem[281]{Bantilan:2012vu}
Bantilan, Hans, Pretorius, Frans, and Gubser, Steven~S. 2012.
\newblock {\em Phys.Rev.}, {\bf D85}, 084038.

\bibitem[282]{Green:2013zba}
Green, Stephen~R., Carrasco, Federico, and Lehner, Luis. 2013.
\newblock [arXiv:1309.7940].

\bibitem[283]{Berger:2002st}
Berger, Beverly~K. 2002.
\newblock {\em Living Rev.Rel.}

\bibitem[284]{Belinskii:1972sg}
Belinskii, V.A., Lifshitz, E.M., and Khalatnikov, I.M. 1972.
\newblock {\em Zh.Eksp.Teor.Fiz.}, {\bf 62}, 1606--1613.

\bibitem[285]{1979PhR....56..371B}
{Barrow}, J.~D., and {Tipler}, F.~J. 1979.
\newblock {\em Phys. Rept.}, {\bf 56}, 371--402.

\bibitem[286]{Berger:1998vxa}
Berger, B.K., {et~al.} 1998.
\newblock {\em Mod.Phys.Lett.}, {\bf A13}, 1565--1574.

\bibitem[287]{Garfinkle:2003bb}
Garfinkle, David. 2004.
\newblock {\em Phys.Rev.Lett.}, {\bf 93}, 161101.

\bibitem[288]{Lim:2009dg}
Lim, Woei~Chet, {et~al.} 2009.
\newblock {\em Phys.Rev.}, {\bf D79}, 123526.

\bibitem[289]{1990PhRvD..41.1796P}
{Poisson}, E., and {Israel}, W. 1990.
\newblock {\em \prd}, {\bf 41}, 1796--1809.

\bibitem[290]{1996PhRvD..53.1754O}
{Ori}, A., and {Flanagan}, {\'E}.~{\'E}. 1996.
\newblock {\em \prd}, {\bf 53}, 1754.

\bibitem[291]{Dafermos:2012np}
Dafermos, Mihalis. 2012.
\newblock [arXiv:1201.1797].

\bibitem[292]{Christodoulou93}
Christodoulou, D., and Klainerman, S. 1993.
\newblock {\em The Global Nonlinear Stability of the Minkowski Space}.
\newblock Princeton: Princeton University Press.

\bibitem[293]{friedrichds}
Friedrich, H. 1986.
\newblock {\em J. Geom. Phys.}, {\bf 3}, 101--117.

\bibitem[294]{Dias:2011ss}
Dias, Oscar~J.C., Horowitz, Gary~T., and Santos, Jorge~E. 2012.
\newblock {\em Class.Quant.Grav.}, {\bf 29}, 194002.

\bibitem[295]{Buchel:2013uba}
Buchel, Alex, Liebling, Steven~L., and Lehner, Luis. 2013.
\newblock {\em Phys.Rev.}, {\bf D87}, 123006.

\bibitem[296]{Dias:2012tq}
Dias, Oscar~J.C., {et~al.} 2012.
\newblock {\em Class.Quant.Grav.}, {\bf 29}, 235019.

\bibitem[297]{Maliborski:2013jca}
Maliborski, Maciej, and Rostworowski, Andrzej. 2013.
\newblock [arXiv:1303.3186].

\bibitem[298]{Callan:1992rs}
Callan, Curtis~G., {et~al.} 1992.
\newblock {\em Phys.Rev.}, {\bf D45}, 1005--1009.

\bibitem[299]{Ashtekar:2010hx}
Ashtekar, Abhay, Pretorius, Frans, and Ramazanoglu, Fethi~M. 2011.
\newblock {\em Phys.Rev.Lett.}, {\bf 106}, 161303.

\bibitem[300]{Ashtekar:2010qz}
Ashtekar, Abhay, Pretorius, Frans, and Ramazanoglu, Fethi~M. 2011.
\newblock {\em Phys.Rev.}, {\bf D83}, 044040.

\bibitem[301]{Ramazanoglu:2010aj}
Ramazanoglu, Fethi~M., and Pretorius, Frans. 2010.
\newblock {\em Class.Quant.Grav.}, {\bf 27}, 245027.

\bibitem[302]{Ashtekar:2008jd}
Ashtekar, Abhay, Taveras, Victor, and Varadarajan, Madhavan. 2008.
\newblock {\em Phys. Rev. Lett.}, {\bf 100}, 211302.

\bibitem[303]{Almheiri:2012rt}
Almheiri, Ahmed, {et~al.} 2013.
\newblock {\em JHEP}, {\bf 1302}, 062.

\bibitem[304]{Guth:2007ng}
Guth, Alan~H. 2007.
\newblock {\em J.Phys.}, {\bf A40}, 6811--6826.

\bibitem[305]{Aguirre:2009ug}
Aguirre, Anthony, and Johnson, Matthew~C. 2011.
\newblock {\em Rept.Prog.Phys.}, {\bf 74}, 074901.

\bibitem[306]{Kleban:2011pg}
Kleban, Matthew. 2011.
\newblock {\em Class.Quant.Grav.}, {\bf 28}, 204008.

\bibitem[307]{2012PhRvD..85h3516J}
{Johnson}, M.~C., {Peiris}, H.~V., and {Lehner}, L. 2012.
\newblock {\em \prd}, {\bf 85}(8), 083516.

\bibitem[308]{2013arXiv1312.1357W}
{Wainwright}, C.~L., {et~al.} 2013.
\newblock [arXiv:1312.1357].

\bibitem[309]{2012ARNPS..62...57B}
{Buchert}, T., and {R{\"a}s{\"a}nen}, S. 2012.
\newblock {\em Annual Review of Nuclear and Particle Science}, {\bf 62}(Nov.),
  57--79.

\bibitem[310]{Zhao:2009yp}
Zhao, Xinghai, and Mathews, Grant~J. 2011.
\newblock {\em Phys.Rev.}, {\bf D83}, 023524.

\bibitem[311]{2013PhRvL.111p1102Y}
{Yoo}, C.-M., {Okawa}, H., and {Nakao}, K.-i. 2013.
\newblock {\em Physical Review Letters}, {\bf 111}(16), 161102.

\bibitem[312]{2014arXiv1404.1435Y}
{Yoo}, C.-M., and {Okawa}, H. 2014.
\newblock {\em ArXiv e-prints}, Apr.

\bibitem[313]{2014PhRvD..89f1503E}
{East}, W.~E., {Ramazano{\v g}lu}, F.~M., and {Pretorius}, F. 2014.
\newblock {\em \prd}, {\bf 89}(6), 061503.

\bibitem[314]{Barausse:2012da}
Barausse, Enrico, {et~al.} 2013.
\newblock {\em Phys.Rev.}, {\bf D87}, 081506.

\end{thebibliography}

\end{document}